\documentclass[%
    aip,
    prl,
     amsmath,amssymb,
     reprint,%
]{revtex4-2}

\usepackage{amsthm}
\usepackage{graphicx}
\usepackage{dcolumn}
\usepackage{bm}
\usepackage{comment}

\usepackage[utf8]{inputenc}
\usepackage[T1]{fontenc}
\usepackage{mathptmx}
\usepackage{etoolbox}

\makeatletter
\def\@email#1#2{%
 \endgroup
 \patchcmd{\titleblock@produce}
  {\frontmatter@RRAPformat}
  {\frontmatter@RRAPformat{\produce@RRAP{*#1\href{mailto:#2}{#2}}}\frontmatter@RRAPformat}
  {}{}
}%
\makeatother

\usepackage[margin=2cm]{geometry}
\allowdisplaybreaks

\usepackage{subfig}

\usepackage{hyperref}
\usepackage[dvipsnames]{xcolor}
\usepackage{amsmath}
\usepackage{amsfonts,amssymb}

\usepackage{tikz}
\usetikzlibrary{quantikz2}

\usepackage{soul}  
\usepackage{pgfplots}
\pgfplotsset{compat=1.18}

\usepackage{mathbbol}
\usepackage{xifthen}

\usepackage[titletoc,title]{appendix}

\usepackage{multirow}

\usepackage{esvect}

\colorlet{ygray}{gray!20}

\newcommand{\aeq}{\begin{equation}}
\newcommand{\eeq}{\end{equation}}
\newcommand{\aeqn}{\begin{eqnarray}}
\newcommand{\eeqn}{\end{eqnarray}}
\newcommand{\aeqns}{\begin{eqnarray*}}
\newcommand{\eeqns}{\end{eqnarray*}}

\newcommand{\yi}{\mathrm{i}}

\newcommand{\oO}{\mathcal{O}}

\theoremstyle{definition}
\newtheorem{theorem}{Theorem}
\newtheorem{definition}[theorem]{Definition}

\newcommand*\diff{\mathop{}\!\mathrm{d}}

\newcommand{\itimes}[2]{\langle #1|#2 \rangle}

\def\ed{e}
\def\es{\overline{e}}
\def\ea{\overline{\overline{e}}}

\def\ytheta{\theta}
\def\yphi{\phi}
\def\yTheta{{\theta'}}
\def\yPhi{{\phi'}}

\def\Vtheta{V^\ytheta}
\def\Vphi{V^\yphi}
\def\VTheta{V^\yTheta}
\def\VPhi{V^\yPhi}

\def\yFou{\mathcal{F}}
\newcommand{\yMM}[1]{\tilde{M}^{#1}}
\renewcommand{\vv}[1]{\boldsymbol{#1}}

\def\LTS{\mathrm{LTS}}

\def\dim{d}
\def\iltsindex{\ell}
\def\iindex{i}
\def\jindex{j}
\def\kindex{k}

\def\rindex{r}

\def\Nsingular{{n_s}}

\def\tcoord{t}
\def\xcoord{x}
\def\ycoord{y}
\def\zcoord{z}

\def\totalCoord{\zeta }
\def\varyCoord{\xi }
\def\constantCoord{\eta }
\def\spectralCoord{ \kappa }
\def\BtotalCoord{\boldsymbol{\totalCoord} }
\def\BconstantCoord{\boldsymbol{\constantCoord}}
\def\BvaryCoord{{\boldsymbol{\varyCoord}}}
\def\BspectralCoord{\boldsymbol{\spectralCoord}}

\newcommand\innerprod[2]{\left<#1,\,#2\right>}

\newcommand\dual[1]{#1^*}

\def\Sd{\mathsf{d}}

\def\FFT{\mathcal{F}}
\def\CC{\mathcal{C}}

\def\EvolutionOperator{U}
\def\FlowTransf{\Psi}

\def\LieDeriv{\mathfrak{L}}
\def\PathOrder{\mathcal{P}}

\def\Chart{\mathcal{R}}
\def\Manifold{M}

\def\Reals{\mathbb{R}}

\def\RealLine{\Reals}
\def\Sphere{S}
\def\Torus{T}
\def\Circle{\Sphere^1}

\def\LebesgueSpace{\mathcal{L}}
\def\HilbertSpace{\mathcal{H}}

\def\SchwartzSpace{\mathcal{S}}
\def\SchwartzDual{\dual{\SchwartzSpace}}
\def\SmoothFunc{\mathcal{C}}

\def\Bx{\mathbf{\xcoord}}

\def\By{\mathbf{\ycoord}}
\def\Bz{\mathbf{\zcoord}}
\def\BV{\mathbf{\Vflow}}

\def\Bpdg{\mathbf{\pdg}}
\def\Bpdf{\mathbf{\pdf}}

\def\Baction{\mathbf{\action}}
\def\Btheta{\boldsymbol{\vartheta}}
\def\Bphi{\boldsymbol{\phi}}
\def\Bpsi{\boldsymbol{\psi}}
\def\BOmega{{\boldsymbol{\Omega}}}

\def\Vflow{V}
\def\Jac{{\mathcal J}}
\def\scalar{s}

\def\pdf{f}
\def\pdg{g}

\def\waveFunc{\psi}

\def\Lag{L}
\def\Ham{H}

\def\Bpdf{\boldsymbol{\pdf}}
\def\BwaveFunc{\boldsymbol{\waveFunc}}

\def\Ham{H}
\def\Kgen{H}
\def\action{J}

\def\VolDensity{\nu}
\def\VolForm{\boldsymbol{\nu}}

\usepgfplotslibrary{fillbetween}

\begin{document}
\preprint{LLNL-...}


\title[]{The integrable Lie-Trotter-Suzuki decomposition for nonlinear dynamics:  \\
Efficient Hamiltonian simulation of the Stepanoff flow}

\author{Ilon Joseph}
\email{joseph5@llnl.gov}
\affiliation{Lawrence Livermore National Laboratory, Livermore, California 94550, USA}
\author{Ivan Novikau}
\affiliation{Lawrence Livermore National Laboratory, Livermore, California 94550, USA}
\author{Michael R. Montgomery}
\affiliation{Department of Mathematics, Dartmouth College, 
Hanover, New Hampshire 03755, USA}
\author{Joanna Slawinska}
\affiliation{Department of Mathematics, Dartmouth College, 
Hanover, New Hampshire 03755, USA}
\author{Dimitrios Giannakis}
\affiliation{Department of Mathematics, Dartmouth College, 
Hanover, New Hampshire 03755, USA}


\begin{abstract}
We propose an efficient unitary numerical time integration strategy for nonlinear dynamical systems: \emph{the integrable Lie-Trotter-Suzuki decomposition (I-LTS)}.
First, express the generator as a sum of integrable parts.
Second, approximate the full evolution operator using a Lie-Trotter-Suzuki (LTS) decomposition of the desired approximation order in time step.
The I-LTS construction ensures that the full evolution operator can be represented as a product of terms that each have an efficient explicit tensor product factorization that can be handled efficiently with tensor network and/or quantum block-encoding methods. 
Next, we apply this method to an important non-integrable point example of a dynamical system: the Stepanoff flow on the torus.
First, we derive an efficient unitary quantum algorithm for simulating the Stepanoff flow that can be thought of as a unitary classical version of a quantum map.
Then, we provide a detailed description of the encoding of the two-dimensional system into an efficient quantum circuit, including an efficient block-encoding for all required operators.
An efficient Hamiltonian simulation algorithm is constructed using a sequence of alternating quantum Fourier transforms (QFT) and Quantum Signal Processing (QSP) circuits.
Finally, the circuit is simulated using a numerical emulator of fault-tolerant quantum computers, and the results are found to be consistent with several alternative classical computational methods.
While efficient, the need for Trotterization introduces fundamental complexity in requiring small enough time step for an accurate approximation and solving for the transfer operator requires one to confront the linear scaling of the advection operator with grid size.
\end{abstract}
\maketitle

\section{Introduction}\label{sec:introduction}

\subsection{Unitary representations for nonlinear dynamics}
Quantum computers have been predicted to be quite powerful and a key advantage 
is their capacity to encode exponentially large amounts of information.
This capability can potentially be leveraged to solve differential equations with high spatial and temporal resolution.
As a result, quantum computers are starting to emerge as promising tools for simulating nonlinear dynamical systems of practical importance \cite{Joseph20, Liu21, Krovi23, Joseph23, Novikau25KvN, Novikau25global}.

However, due to the no-cloning theorem~\cite{Wootters82}, they are limited in their ability to perform nonlinear operations by the axioms of quantum mechanics.
Any iterative nonlinear quantum algorithm must incur an exponential cost in memory as the number of iterations increases~\cite{Leyton08, Brustle2025quantum}.
Thus, many researchers have been working to develop efficient techniques for simulating nonlinear dynamics on quantum computers using approaches that do not require iterative nonlinear operations\cite{Joseph23}.

The key to the Koopman--von Neumann approach \cite{Joseph20, Joseph23, Novikau25KvN} and the closely related Carleman embedding approach~\cite{Liu21, Krovi23, Joseph23, Novikau25global} is the realization that one can 
generate intrinsically linear representations of 
the system by simulating the conservation of probability in phase space.
These methods focus attention on the various forms of the advection equation for scalars, probability density functions (PDFs), and wavefunctions.
While there are unitary finite difference and finite element representations for these equations, they tend to have rather poor numerical properties~\cite{Lin22Koopman} and are rarely used in applications without adding some form of dissipation to the evolution~\cite{Novikau25KvN}.

Hence, in this work, we determine an alternate unitary approximation for the Koopman-von Neumann (KvN) evolution operator, based on a Lie-Trotter-Suzuki (LTS) decomposition\cite{Suzuki93, Childs21} for the generator, that is much better behaved while still being efficient to block-encode.
The \emph{integrable-LTS (I-LTS) decomposition} results from representing the generator  as a sum of integrable parts and applying an LTS product decomposition of a given approximation order.
In Sec.~\ref{sec:ILTS}, we prove that there are many ways to represent an arbitrary vector field on a manifold as a sum of vector fields, each of which is integrable.
The I-LTS decomposition is then simply a product of evolution operators, each corresponding to an evolution operator for one of the integrable vector fields in the sum. 
Because the Lie bracket of any two vector fields simply generates another vector field, 
the LTS decomposition corresponds to the exact evolution operator of another nearby vector field that approximates the flow of interest.
Finally, we note that the I-LTS decomposition applies to quantum dynamics equally well, and, even in the context of quantum dynamics, it affords the same complexity reduction.

The I-LTS strategy of breaking any given Hamiltonian operator into a sum of completely integrable parts leads to an efficient algorithm for the time evolution operator based on the LTS decomposition.
This is because the eigenvalues and eigenfunctions for any integrable evolution operator have a simple tensor product form that can be determined with a coordinate transformation to constants of the motion and subsequent Fourier transformation.
Hence, each individual evolution operator in the LTS product has an explicit block-encoding that both classical tensor network and quantum algorithms can exploit.

We note that operator splitting methods have certainly been used for advection-type operators in the past. For example, a similar technique was recently proposed by Ref.~\onlinecite{Banjara2025operator} for approximating the Koopman evolution operator via series of nonlinear maps, where a different map is used for each component of the flow.
They also made valuable contributions by studying the impact of high-order LTS product formulae for the Lotka-Volterra, Van der Pol, and Lorenz systems.
Here, we focus more on the Koopman-von Neumann evolution operator (which is closer to the Perron-Frobenious transfer operator / Schr\"odinger picture).
We also emphasize that it is the integrability of the individual components that reduces the complexity, so that the minimal number of terms depends on determination of an optimal integrable decomposition.

\subsection{Key results for the Stepanoff flow}

In order to understand how the I-LTS decomposition performs in practice, we explore the key point example of a relatively simple yet non-integrable  example of a dynamical system: the Stepanoff flow on the torus.~\cite{Stepanoff36, Oxtoby53}
In fact, we fully develop, analyze, and numerically verify the implementation of a quantum Hamiltonian simulation algorithm for the Stepanoff flow.
Similar I-LTS Hamiltonian simulation algorithms can be developed for more complicated chaotic flows, such as the Lorenz attractor, e.g. based on the integrable decompositions given in Appendix~\ref{sec:Lorenz}.

We chose to focus on this system as a prototypical nonlinear dynamical system exhibiting non-trivial dynamical complexity, yet possessing a number of attractive features 
that aid the development and implementation of quantum and tensor network algorithms.
In particular, the Stepanoff flow is equivalent to a Hamiltonian system because it is generated by a smooth divergence-free vector field.
Moreover, it has a compact phase space and 
the velocity components are simply first degree trigonometric polynomials with a single fixed point (zero of the vector field) at the origin.
Finally, this system also has the minimal state space dimension, two, required for non-integrable smooth dynamics in continuous time.
Put together, these facts make the Stepanoff flow a natural testbed for the development of quantum and tensor network algorithms for simulation of classical nonlinear dynamics.

From a dynamical standpoint, the Stepanoff flow is topologically conjugate (equivalent) to a time-reparameterized rotation system on the torus, 
where the function that determines the reparameterization of time
becomes singular at the fixed point.
This feature creates dynamical complexity, including non-integrability and sensitive dependence on initial conditions.
At the same time, the divergence-free nature of the Stepanoff vector field means that the  generator of the flow is  
a self-adjoint Hamiltonian operator acting on the Hilbert space of square-integrable observables with respect to the ergodic Haar measure, i.e. the area measure, on the torus.
Furthermore, by virtue of the fact that the vector field coordinates are trigonometric polynomials, the Hamiltonian is represented by a sparse banded matrix in the Fourier basis of this Hilbert space with analytically computable elements.

Recent work \cite{Giannakis24,Giannakis25} employed the Stepanoff flow to test the prediction skill of quantum-inspired techniques for approximating the Koopman evolution of observables of measure-preserving ergodic flows using tensor networks and bosonic Fock spaces.
In this work, we develop an explicit gate-based quantum algorithm, featuring a block-encoding of the Stepanoff Hamiltonian at its core.

The I-LTS decomposition provides an efficient quantum encoding using a combination of mixed real-space and Fourier-space coordinates, which yields a diagonal form for each of the generators of the integrable flows in the decomposition.
The initialization of the two-dimensional problem is implemented using quantum circuits that compute simple trigonometric functions.
These circuits are then used by two Quantum Singular Value Transformation (QSVT)~\cite{Gilyen19, Martyn21} circuits to construct the initial two-dimensional probability distribution on the torus.
This distribution is evolved in time using a sequence of alternating Quantum Fourier Transform (QFT)~\cite{Shor94, Coppersmith02, Nielsen10} and Quantum Signal Processing (QSP)~\cite{Low17, Low19} circuits.
The resulting quantum state encodes the probability distribution after dynamical evolution by the Stepanoff flow at the specified time instant.
The circuit is constructed using a limited set of gates, such as Pauli gates, the Hadamard gate, the Phase gate, and the rotation gates $R_y$ and $R_z$.
All of these gates are allowed to be controlled by multiple qubits.
Although the proposed circuit is implemented using standard well-established quantum algorithms such as QSP, QSVT, QFT, amplitude amplification (AA), and amplitude estimation (AE) techniques, a significant contribution of this work is to provide the explicit combination of these methods for modeling a nontrivial flow such as Stepanoff.

The quantum circuit is simulated on a classical emulator of fault-tolerant quantum computers using the numerical QuCF framework~\cite{QuCF} based on the QuEST toolkit~\cite{Jones19}. 
The results of the QuCF simulations are compared with those obtained from direct numerical integration of the Stepanoff flow.
The numerical simulations show good agreement between the exact classical evolution and the approximate QHS evolution, both in terms of the direct comparison of the classical and QuCF signals, and in the comparison of spatially integrated functions computed using the standard AE procedure\cite{Brassard02}.
The scaling of the constructed quantum circuit is analyzed both analytically and numerically.
The primary factor limiting quantum speedup  
is the Courant–Friedrichs–Lewy (CFL) condition, which is reflected in the growth of the spectral norm of the generator and, consequently, in the increase of the number of time steps that must be simulated to achieve a specified time interval.

\subsection{Quantum algorithms for differential equations}
Over the past decade, numerous quantum algorithms have been proposed for time integration of both quantum and classical systems.
Quantum methods which perform QHS include the LTS decomposition\cite{Suzuki93, Childs21} as well as state-of-the-art block-encoding-based techniques\cite{Lin22} such as QSP\cite{Low17, Low19} and QSVT\cite{Gilyen19, Martyn21}, to name a few.
The latter two algorithms are universal for approximating functions of operators and are used to construct polynomials in either the real eigenvalues or the singular values of a matrix.
They can be employed to build modular components for designing more advanced algorithms aimed at modeling dissipative and nonlinear problems on quantum computers.
Recently, these methods have been extended to handle complex eigenvalues\cite{Low24}, enabling general quantum eigenvalue processing.
In addition, the first theoretical and practical principles have been proposed for working with multiple matrices within a single QSP circuit\cite{Rossi22, Rossi23, Mori2024, Rossi2025}, potentially enabling the encoding of multivariable functions.

Another possible quantum algorithmic approach to time integration is the time-marching method\cite{Fang23}, which also allows one to simulate the evolution of dissipative systems.
In this approach, similar to classical time integration, the time interval is discretized with $N_t$ time points.
Then, evolution of the simulated system at each time instant is computed via a simple matrix-vector multiplication, $u^{t+1} = A u^t$, where the generally nonunitary matrix $A$ depends on the spatial and temporal discretization steps.
This matrix is mapped onto a quantum circuit using block-encoding techniques such as the Linear Combination of Unitaries (LCU) or sparse-matrix encoding methods.
One of the main disadvantages of this approach is the rapid decline in the circuit's success probability.
This issue can potentially be mitigated by using compression gadgets, but  this still requires $\log_2{N_t}$ ancillary qubits.

To date, quantum algorithms have been tested and analyzed on a variety of classical initial-value problems.
For instance, the time-marching method was used to simulate the advection-diffusion equation\cite{Over24TimeMarching}, Burger's equation\cite{Esmaeil24}, and the Lorentz strange attractor\cite{Koukoutsis25Lorentz}, to name a few.
Hybrid classical-quantum variational solvers were proposed for modeling the Poisson equation\cite{Trahan23}, Laplace equation\cite{Ghisoni24}, Maxwell's equations with adaptive mesh refinement\cite{Fressart25}, the steady-state heat equation\cite{Arora25}, the advection-diffusion equation\cite{Rao24, Alipanah25, Bharadwaj25}, the incompressible Navier-Stokes equations\cite{Song25}, and even for calibrating large-scale radio telescopes\cite{Renaud24}.
Quantum methods attracted significant attention as potential tools for modeling fluid systems\cite{Gaitan20, Gaitan21, Sanavio25, Malinverno25, Pfeffer25, Tennie25, Li25, Conde25, Bravyi25}, seismological processes\cite{Moradi18, Malley18, Greer20, Dukalski23, Rani25, Nguyen25}, electromagnetic wave phenomena\cite{Ram21, Vahala21, Novikau23, Vahala23, Koukoutsis23, Koukoutsis25}, general wave problems\cite{Schade24, Sato24, Lewis24}, and plasma systems\cite{Dodin20, Joseph23, Amaro23, Koukoutsis23plasma, Sundar25, Yifei25}.
For example, QSP and QSVT algorithms were explored for simulating initial-value fluid plasma\cite{Engel19, Novikau22} and boundary-value kinetic problems\cite{Novikau24-EVM, Goldfriend25}.
Because kinetic problems are high-dimensional, they are substantially more demanding than fluid problems and are therefore particularly interesting for potential acceleration by quantum computers\cite{Ameri23, Higuchi24, Toyoizumi24, Miyamoto24, Qin25, May25, Vaszary25}.
The efficient encoding of multidimensional problems is a separate issue that must be resolved to provide the optimal mapping of classical problems of practical interest onto quantum circuits, and, we turn to this issue next.

Efficient encoding of the generator of the evolution operator or of the matrix of a system of linear equations into a quantum circuit plays a central role in many fault-tolerant quantum algorithms\cite{Kan24, Ty24, Ollive25, Novikau24-EVM, Novikau25KvN, Goldfriend25, Novikau25Opt, Kharazi25, Kane25} and is crucial for ensuring quantum advantage over the best possible classical algorithms.
The core idea of this procedure is to construct a quantum circuit whose unitary matrix representation contains the target nonunitary matrix as a sub-block.
Moreover, the resulting circuit, also referred to as a block-encoding oracle, must scale polylogarithmically with the matrix size to maintain efficiency.
Otherwise, quantum algorithms that rely on an oracle to represent the problem generator can lose their quantum speedup.
Thus, the potential to achieve an exponential speedup factor in many quantum methods, including those based on QSP, strongly depends on finding an efficient block-encoding.

A key issue is that an efficient encoding is not guaranteed for arbitrary problems.
In fact,  an efficient block encoding does not generically exist for a generic problem specification; e.g. it would not exist for a Hermitian operator with otherwise random entries.
However, many physics problems are typically described by well-structured sparse matrices that are well-suited to efficient block-encoding.
This is one of the key reasons why even classical fluid and plasma problems are  
promising candidates for simulation  
with quantum algorithms.

Block-encoding can be considered a discrete optimization problem, where a numerical algorithm, potentially an artificial intelligence and machine-learning (AI/ML) based algorithm, must extract patterns in the positions and values of matrix elements.
Once these patterns are identified, the matrix can be mapped onto a quantum circuit by encoding the patterns themselves, rather than encoding each matrix element individually.
For an arbitrary matrix whose nonzero elements are either randomly distributed or that vary seemingly randomly without a promise of smoothness or any other discernible pattern, efficient block-encoding is most likely not possible.
In such cases, the number of gates in the oracle typically scales linearly with the number of nonzero elements in the matrix; i.e. exponentially in the number of qubits, which is why any quantum speedup would then be reduced to polynomial at best.

Significant effort is currently being devoted to developing advanced numerical frameworks for constructing efficient block-encoding oracles\cite{Clader22, Camps22, Camps23, Harrigan24, Kuklinski24, Sunderhauf24, Lemieux24, Simon25, Liu25BE, Yang25BE, Li25BE}.
The encoding may involve describing the matrix structure, i.e., the positions of nonzero elements in a matrix, and computing polynomials\cite{Rossi22, Conde24polynomial, Rosenkranz25, Guseynov25, OBrien25}, which, for example, describe the dependence of matrix elements on spatial coordinates.

\subsection{Outline}
This paper is organized as follows.
In Section~\ref{sec:ILTS}, we introduce the I-LTS decomposition for dynamical systems and prove that this technique can be applied to arbitrary vector fields in a local contractible region of space.
For example, Appendix~\ref{sec:Lorenz} lists several examples of integrable decompositions for the chaotic Lorenz 1963 flow.
We also show that the I-LTS decomposition has an explicit tensor product structure that is relatively easy for tensor network and quantum algorithms to exploit.
In Section~\ref{sec:QI-LTS}, we derive and simulate a quantum algorithm for the Stepanoff flow based on the I-LTS decomposition that is exponentially efficient in complexity per time step. 
We prove that this exponential savings holds even for an explicit end-to-end algorithm that includes the measurement of a fixed number of physical observables.
Finally, Section~\ref{sec:conclusion} provides concluding remarks.

\section{The Integrable Lie-Trotter-Suzuki (I-LTS) Decomposition \label{sec:ILTS}}

\subsection{Overview of the I-LTS Decomposition \label{sec:ILTS-overview}}
We now derive an efficient strategy for developing a unitary numerical discretization of the evolution operator and its generator, based on the Lie-Trotter-Suzuki (LTS) decomposition. 
The key idea of the integrable LTS (I-LTS) decomposition is to represent the generator as a sum of completely integrable parts~\cite{Arnold1992odes,Olver1993applications,Lichtenberg2013book} and the evolution operator as an LTS product of evolution operators corresponding to each of the integrable generators.
Each of the integrable parts generates a relatively efficient evolution that can be simplified with separation of variables and solved for in terms of constants of the motion.
Hence, the spectrum of each integrable operator completely factorizes into a tensor product form, e.g. of one-dimensional spectra, and the set of eigenfunctions can be formed from a tensor product of constants of the motion.
Thus, it is relatively cheap to apply the I-LTS evolution operator because applying each integrable evolution simply requires a coordinate transformation, a Fourier transform, and a phase shift, as well as the inverse of these transformations. 

Finding a unitary approximation is valuable because each generator is a type of advection operator, i.e. a first-order differential operator, and such operators are difficult to treat numerically.
Unfortunately, unitary 
discretizations of first-order differential  operators have poor numerical properties due to a near degeneracy of high and low frequency spatial harmonics.
This degeneracy is exact for the central-difference operator on a rectangular grid with constant mesh size and approximately holds for any anti-Hermitian finite difference or finite element approximation.
While some of the issues can partially be cured with the use of pseudo-spectral methods, there is no general theory of how to use such methods to guarantee convergence.

Another issue is that there is no guarantee that a numerical approximation preserves structural properties of the exact solutions.
For example, because the generator is a first-order differential operator, it has the property that the product of any two eigenfunctions must again be an eigenfunction whose total eigenfrequency is the sum of the two individual eigenfrequencies.
This gives the spectrum of the evolution operator a group structure that can potentially be exploited by tensor network and/or quantum algorithms \cite{Giannakis25}.
However, it is also known that a \emph{complete factorization} of all eigenfunctions into a tensor product form only occurs for the case of completely integrable systems.
This is because it is only in the integrable case that there is a sufficiently large symmetry group to solve the system with separation of variables. 
In this case, the orbit of the action of the symmetry group on points of the manifold is a measurable neighborhood of phase space.
Then, the Hilbert space decomposes into a tensor product factorization of separable degrees of freedom for each coordinate direction.

We shall prove that the I-LTS decomposition represents evolution by an approximate I-LTS generator, which is still a first-order differential operator.
Hence, it retains the same tensor product structure of products of eigenfunctions.
The I-LTS decomposition is most efficient when the individual components approximately commute with one another.
The symmetric case, where all components exactly commute, is the case where the first-order LTS approximation is exact.
Thus, the ``best'' I-LTS decompositions seek to determine an approximate symmetry group for the flow of interest.

The I-LTS decomposition takes an operator-theoretic point-of-view that allows one to blend the product of integrable evolution operators to form an approximate total evolution operator in a manner that is quite different from an analysis of individual trajectories.
Perhaps surprisingly, I-LTS constructs the approximate evolution operator from integrable generators that are not necessarily tangent to the flow of interest.
By choosing each of the parts to be completely integrable, each individual operator has a spectrum that is the product of the point spectrum along the flow and a continuous spectrum across the flow. 
While the geometry of the total flow may be quite complicated -- especially if it is chaotic -- analyzing the I-LTS evolution operator may be more straightforward because each of the operators in the product has a spectrum whose properties are well-understood.

Next, in Section~\ref{sec:evolution-operators}, we describe the many forms that the evolution operator can take for observables, probability densities, and wavefunctions.
Then, in Section~\ref{sec:ILTS-decomposition}, we derive the I-LTS decomposition and, in Section~\ref{sec:ILTS-exists}, we prove that I-LTS decompositions exist for any contractible region of a manifold.
Finally, in Sec.~\ref{sec:ILTS-complexity}, we consider different contributions to the complexity of the I-LTS-decomposition.

\subsection{ Evolution by any name: Koopman, Perron-Frobenius, Koopman-von Neumann \label{sec:evolution-operators}} 
In this work, we consider evolution operators that correspond to the motion along trajectories defined by a set of differential equations. 
While there are different notions of the evolution operator, they are all consistent with each other and, due to the uniqueness of solutions to differential equations, must agree as to which trajectory results from which initial condition.

\textbf{Flow of a Vector Field.} 
For simplicity, we will confine the discussion here to the concrete case of a flow, $\FlowTransf$, defined by a set of ordinary differential equations (ODEs) on a manifold, $\Manifold$.
The flow map, $\FlowTransf_\tau:\Manifold\rightarrow \Manifold$, where, here, time is parameterized by the generalized coordinate $\tau$, is generated by an autonomous vector field, $\Vflow=\BV\cdot \nabla$.
In any local chart, with coordinates $\Bz=\{\zcoord^\jindex\}$, the ODEs are defined to satisfy 
\begin{align} \label{eq:ODE_definition}
    d\Bz/d\tau=\BV(\Bz).
\end{align}
In fact, the equations of motion are more clearly stated as PDEs 
\begin{align} \label{eq:ODE-PDE_definition}
\left.\partial_\tau\Bz(\tau,\Bz_0)\right|_{\Bz_0}=\BV(\Bz(\tau,\Bz_0))
\end{align}
where the initial conditions are used as coordinates.
In order to clearly indicate the time at which the coordinates are evaluated, we will often use a subscript notation, where we define
\begin{align} \label{eq:z-tau_definition}
     \Bz_\tau(\Bz_0):=\Bz(\tau,\Bz_0)=\FlowTransf_\tau(\Bz_0)
\end{align}
where $\FlowTransf_\tau$ represents the flow map at at time $\tau$.

This formulation is general because, if the velocity is not autonomous, i.e. independent of the physical time coordinate, $\tcoord$, one can append time to the overall set of coordinates $\Bz'= \{\tcoord,\Bz\}$ and consider an evolution on the extended phase space   $\FlowTransf_\tau:\Reals\times \Manifold  \rightarrow\Reals\times\Manifold$, where $\Vflow'=\BV'\cdot\nabla'$ where $\BV'= \Vflow^\tcoord\{1, \BV\}$, when parameterized by $\tau$.

The flow $\FlowTransf_\tau$ represents a nonlinear transformation of the coordinates of the manifold.
Solving the set of ODEs requires applying the generally nonlinear velocity field to the nonlinear transformation of the coordinates.
To take advantage of computational tools that are optimized for linear operations, such as tensor networks and quantum algorithms, let us now turn towards linear representations of the action of the flow, i.e. the evolution operator.

\textbf{Action on Scalar Functions.}
In order to understand the action of the flow on spaces of functions on a manifold, first consider the simplest form of this action: the advection equation for scalar functions, $\scalar:\Reals\times \Manifold\rightarrow\Reals$, that are conserved in time.
For simplicity, here, we consider scalar functions that are smooth $\scalar\in\SmoothFunc^1(\Reals\times\Manifold)$.
For the standard formulation of the advection equation, the scalar field, $\scalar(\tau,\Bz_\tau)$, is considered at fixed \emph{final position}, $\Bz_\tau$.
The advection equation for a conserved scalar function is
\begin{align} \label{eq:scalar-advection}
    \partial_\tau\scalar(\tau,\Bz_\tau)=-\Vflow \scalar(\tau,\Bz_\tau)= -\BV\cdot \nabla \scalar (\tau,\Bz_\tau)=-\BV\cdot\Sd \scalar (\tau,\Bz_\tau). 
\end{align}
in terms of the exterior derivative $\Sd$.

The components of the flow, $\BV=\{\Vflow^{z^j}\}$, are given by the action of the flow on coordinate $\Bz=\{\zcoord^j\}$, so that 
\begin{align}
    \Vflow^{z^j}:=\BV\cdot\nabla \zcoord^j.
\end{align}
When the coordinate system is clear, we will sometimes shorten this notation to $\Vflow^j:=\Vflow^{z^j}$.

{\bf Action on Scalar Observables.}
However, in considering ODEs such as Eq.~\ref{eq:ODE_definition}, a scalar observable, such as a coordinate in a coordinate system, is considered at fixed \emph{initial position}, 
\begin{align}
    \Bz_0(\tau,\Bz)=\FlowTransf^{-1}_{\tau}(\Bz)=\FlowTransf_{-\tau}(\Bz).
\end{align} 
Using the chain rule, 
\begin{align} 
\left. \frac{d}{d\tau} \Bz_0 \right|_{\Bz_\tau}=-\BV\cdot\nabla\Bz_0=- \left. \frac{d}{d\tau} \Bz_\tau \right|_{\Bz_0}=-\BV\cdot\nabla\Bz_\tau,
\end{align}
hence, the observable $\scalar(\tau,\Bz_0)$, satisfies an advection equation with the opposite sign of the flow
\begin{align} \label{eq:observable-advection}
   \partial_\tau\scalar (\tau,\Bz_0)=\Vflow \scalar (\tau,\Bz_0)=\BV\cdot \nabla \scalar (\tau,\Bz_0)=\BV\cdot\Sd \scalar (\tau,\Bz_0).  
\end{align}
This is the generator of the \textit{Koopman evolution operator}, also known as the \textit{composition operator}.
To express all other equations below at fixed initial condition, one simply reverses the sign of $\BV$ and takes the complex conjugate of the generator (for Schr\"odinger-like equations that are complex).

\textbf{Koopman Evolution Operator.} 
The spatial adjoint of the Perron-Frobenius generator is the generator for the evolution of scalar observables.
The corresponding evolution operator is known as the Koopman evolution operator.
The key difference is that the Koopman operator defines the evolution at fixed \emph{initial position}.
Using the chain rule, at fixed initial position, the spatial components of the velocity simply change sign from $\BV\rightarrow-\BV$, and, for the extended phase space generator $\Vflow\rightarrow\Vflow^\tcoord (1,-\BV)$.
Hence, this transforms the scalar advection equation to the form
\begin{align} \label{eq:scalar-observable}
 \left. \partial_\tau\scalar\right|_{\Bz_0} = +\BV \cdot \nabla  \scalar .
\end{align}
In quantum mechanics, expressing the evolution of observables in this manner is known as the ``Heisenberg picture''. 
It is obvious that one can also write the Liouville and KvN equations at fixed initial position \cite{Joseph23}, but since this is not commonly used, we will not write the explicit form here.

\textbf{Action on Tensors and Differential Forms.}
A connection defines the differential gradient operator $\nabla$ which defines parallel transport on a manifold.
The operator $\Vflow=\BV\cdot\nabla$ is the generator for the evolution of a conserved tensor.
The objects of most interest to us here are differential forms, $\omega$, e.g. that define probabilities in phase space.
Assuming that the connection has no torsion, then the natural action of the generator on a differential form is independent of the specific choice of connection.
In this case, the appropriate action is simply the Lie derivative, which can be expressed as
\begin{align}
    \LieDeriv_\Vflow \omega=\Vflow\cdot\Sd\omega+\Sd (\Vflow\cdot \omega)
\end{align}
in terms of the exterior derivative, $\Sd$, using Cartan's formula.

\textbf{Action on Volume Forms and PDFs.}
The space-time adjoint of the scalar observable equation is the Liouville equation which generates the evolution of the probability distribution function (PDF), $\Bpdf=\pdf \Sd\zcoord^1\wedge \Sd\zcoord^2\wedge  \dots\wedge\Sd\zcoord^\dim$, which is a volume-form on the manifold.
The evolution of the volume form at fixed final position is
\begin{align} \label{eq:volume-form-advection}
    \partial_\tau\Bpdf = - \Sd(\BV\cdot \Bpdf ) =-\Sd \star (\BV \pdf )
\end{align}
where, in the last equality, we introduce the Hodge $\star$ operator, which converts $\kindex$-forms to $(\dim-\kindex)$-forms and vice-versa.
The action on the PDF, $\pdf$, which is the component of the volume form, $\Bpdf$, is
\begin{align} \label{eq:pdf-advection}
     \partial_\tau\pdf =\Vflow^\dagger\pdf 
     =-\nabla\cdot ( \BV \pdf) 
\end{align}
where $\Vflow^\dagger \pdf=-\nabla\cdot( \BV\pdf)$ is the adjoint of $\Vflow$.
This is the generator of the \textit{Perron-Frobenius evolution operator} for probability measures at fixed final position.

\textbf{Action on Wavefunctions.} A wavefunction is the ``square root'' of a probability measure, 
\begin{align}
    \BwaveFunc=e^{i\theta}\sqrt{\Bpdf}=\waveFunc\sqrt{\Bpdf/\pdf}
\end{align} 
together with a choice of complex phase, $\theta$. 
This implies that it represents an object known as a \emph{half-form} in geometric quantization~\cite{Kirillov2001geometric,Hall2013quantum}  terminology  which is actually the square root of a volume form.
It can be proven~\cite{Joseph20} that a wavefunction with a conserved probability density must satisfy a \textit{generalized Koopman-von Neumann} (KvN) equation.
At fixed final position, the KvN equation is
\begin{multline}
\label{eq:KvN-advection} 
   \partial_\tau\waveFunc-i\Lag\psi 
   = \tfrac{1}{2}(\Vflow^\dagger-\Vflow)\waveFunc
  \\ =  -\tfrac{1}{2}\left(\BV\cdot\nabla+\nabla\cdot   \BV\right)\waveFunc
   =-\tfrac{1}{2}\left\{\BV \cdot,\nabla\right\}\waveFunc
\end{multline} 
where $\Lag$ is a real phase space function. 
The KvN equation is essentially a ``semiclassical version'' of \textit{the Schr\"odinger equation} for nonlinear dynamics. 
For \textit{Hamiltonian flows}, there is a unique choice: $\Lag$ must be the Lagrangian (divided by Planck's constant) in order to represent the semiclassical limit of a generalized Schr\"odinger equation \cite{Joseph23JPA}.

\textbf{Relation between Evolution Operators.} 
Using the uniqueness of the trajectories for each initial condition, all of the evolution operators corresponding to the generators above are simply related to each other. 
Aside from the question of which representation is used (Koopman vs. KvN vs. Perron-Frobenious), the only difference is that the different phase space objects: scalars, wavefunctions, and PDFs, transform as different powers of the Jacobian of the transformation.  
The Hermitian part of the generator of all of these evolution equations has the form of scalar advection driven by a source
\begin{align} \label{eq:general_advection}
\rho^{-1} d \rho/dt  = \rho^{-1} \left(\partial_t+\BV\cdot\nabla \right) \rho/dt 
=-\alpha\nabla\cdot\BV 
\end{align}
where $\alpha = 0,1/2,1$ corresponds to the different cases of interest for scalar advection, KvN, and PDF advection.
This equation is satisfied by different powers of the Jacobian of the transformation between the initial coordinates, $\Bz_0$, and the final coordinates, $\Bz_\tau$,
\begin{align}
    \Jac^0_\tau=(\Jac^\tau_0)^{-1} :=\det{\left(\frac{\partial\Bz_0}{\partial \Bz_\tau}\right)} . 
\end{align}
Hence, the general solution can be expressed as
\begin{align}
    \rho_\tau(\tau,\Bz_\tau) =\rho_0(\Bz_0)(\Jac^0_\tau)^\alpha 
\end{align}
where $\rho_0(\Bz_0):=\rho(\tau=0,\Bz_0)$ is a conserved scalar function.

For example, a scalar function that is conserved along the flow must have the solution
\begin{align} \label{eq:scalar_eigenfunc}
\scalar (\tau,\Bz)=\scalar (0,\Bz_0).
\end{align}
This generates a corresponding solution for the probability density, $\pdf (\tau,\Bz)$, via
\begin{align} \label{eq:pdf_eigenfunc}
    \pdf (\tau,\Bz)= \scalar (\tau,\Bz) \Jac^0_\tau =  \scalar (0,\Bz) \Jac^0_\tau.
\end{align}
It also generates a corresponding KvN wavefunction $\waveFunc(\tau,\Bz)$, via
\begin{align} \label{eq:KvN_eigenfunc}
    \waveFunc(\tau,\Bz)= \scalar(\tau,\Bz) \left(\Jac^0_\tau \right)^{1/2} =  \scalar(0,\Bz) \left(\Jac^0_\tau\right)^{1/2}.
\end{align}
For a wavefunction, one must be careful to match the branch of the square root of the Jacobian across points where the Jacobian becomes  singular in an appropriate sense.
See \onlinecite{Joseph23JPA} for the derivation and for a detailed discussion of how these choices should be made in order to match semiclassical theory.

\textbf{Hilbert Space Inner Product.} 
Consider defining an inner product on phase space through integration over the volume of an orientable manifold.
There is a natural pairing between volume forms and scalar functions
 \begin{align} \label{eq:scalar-volform_pairing}
    \left<\scalar,\mathbf{\pdf}\right>:=\int  \scalar^*(\Bz)\mathbf{\pdf}[\Bz]=\int  \scalar^*(\Bz)\pdf(\Bz)d^\dim\Bz .
\end{align}
This can be considered a pairing between the scalar function, $\scalar\in C^\infty$, and the PDF of the volume form, $\pdf\in  \LebesgueSpace^1$, with the definition that
 \begin{align} \label{eq:scalar-pdf_pairing}
    \left<\scalar, \pdf \right>:=\int  \scalar^*(\Bz)\pdf(\Bz)d^\dim\Bz .
\end{align}
holds in any coordinate system.
However, because there is an asymmetry in the geometric objects in this pairing, it represents a duality between two different function spaces rather than a standard Hilbert space inner product.

Now consider the pairing between two half-forms, $\Bphi$ and $\Bpsi$, defined via
\begin{align}\label{eq:waveform_innerprod}
        \left<\Bphi,\Bpsi\right>:=\int  \Bphi^*(\Bz)\Bpsi(\Bz)=\int  \phi^*(\Bz)\psi(\Bz)d^\dim\Bz .
\end{align}
As an inner product for the wavefunctions, the definition is
\begin{align}\label{eq:wavefunc_innerprod}
        \left<\phi,\psi\right>:=\int  \phi^*(\Bz)\psi(\Bz)d^\dim\Bz 
\end{align}
in any coordinate system. 
Hence, in this case, we can consider the wavefunctions, $\phi$ and $\psi$, i.e. the components of the half-forms, to be elements of $\HilbertSpace^1=\LebesgueSpace^2$ with the standard volume for any coordinate system.

A key point is that since each of the Eqs.~\ref{eq:scalar-volform_pairing}-\ref{eq:wavefunc_innerprod} supply a volume form, the result transforms correctly to any coordinate system.
In particular, if the fields are Lie dragged via the appropriate advection equation, then the inner product is constant in time.
This is because the fields are simply advected scalar functions multiplied by a Jacobian factor that supplies the required change in volume element.
An explicit calculation shows that the integrand in Eq.~\ref{eq:scalar-pdf_pairing} is dynamically invariant
\begin{multline} \label{eq:scalar-pdf_invariant}
        \scalar(\tau,\Bz_\tau)\pdf(\tau,\Bz_\tau)d\Bz_\tau = \scalar(0,\Bz_0)\pdf(0,\Bz_0)\Jac^\tau_0 d\Bz_\tau \\
    = \scalar(0,\Bz_0)\pdf(0,\Bz_0)  d\Bz_0.
\end{multline}
Similarly, if the wavefunctions satisfy the  KvN equation, the phase factors precisely cancel and the inner product is constant in time.
Hence, the integrand in Eq.~\ref{eq:wavefunc_innerprod} is dynamically invariant
\begin{multline}
        \phi^*(\tau,\Bz_\tau)\psi(\tau,\Bz_\tau)d\Bz_\tau \\
        = e^{iS^*(\tau,\Bz_\tau}\phi^*(0,\Bz_0)e^{-iS(\tau,\Bz_t)}\psi(0,\Bz_0)|\Jac^\tau_0| d\Bz_\tau \\
    = \phi^*(0,\Bz_0)\psi(0,\Bz_0)  d\Bz_0.
\end{multline}

A related pathway to defining an inner product for the $\LebesgueSpace^2$ function space is to first choose a preferred
volume form, $\VolForm$, where, in any coordinate chart, one can express the volume form as
$\VolForm[\Bz]:=\nu(\Bz) (\wedge d\Bz)^\dim=\nu(\Bz) d\Bz^\dim$ where the component 
$\VolDensity(\Bz)\in \LebesgueSpace^1$. 
In particular cases with additional structure, such as Hamiltonian or volume-preserving dynamics, there may  be natural or distinguished choices of volume form that are of interest, typically because this represents a dynamically invariant volume form, and, hence, can be used to define a dynamically invariant inner product.
An ergodic measure defined by a volume form is another case of interest.
However, in these settings, it is usually the case that the Jacobian is constant in time, in which case, it is less important to keep track of these Jacobian factors.

For two scalar fields, $r,s$, this yields an inner product of the form
\begin{align} \label{eq:scalar_inner-product}
    \left<r,\scalar\right>:=\int  r^*(\Bz)\scalar(\Bz)\VolForm[\Bz] =\int  r^*(\Bz)\scalar(\Bz) \VolDensity(\Bz)  d^\dim\Bz.
\end{align}
In this case, one should assume that the scalar functions are elements of $\HilbertSpace^1=\LebesgueSpace^2$ with respect to this volume form.
This inner product is constant in time for any two scalar functions that satisfy the advection equation (without sources), as long as the volume form is also dynamically invariant. The proof is the same as that given in Eq.~\ref{eq:scalar-pdf_invariant}.

If there is a preferred choice of volume form, $\VolForm[\Bz]$, then  any other volume form can be expressed as
\begin{align}
    \Bpdf=\pdf(\Bz)\VolForm[\Bz]/\VolDensity(\Bz)=\pdf(\Bz) d^\dim\Bz.
\end{align}
In this form the Hodge dual of the PDF, $(\star \Bpdf[\Bz])=\pdf(\Bz)/\VolDensity(\Bz)$ appears as the component.
Hence, a natural inner product is
\begin{align} 
    \left<\Bpdg,\Bpdf\right>:=\int (\star \Bpdg[\Bz])^*\Bpdf[\Bz]=\int  \pdg(\Bz)^*\pdf(\Bz)\frac{\VolForm[\Bz] }{|\VolDensity(\Bz)|^2}.
\end{align}
This can also be written as an inner product for the PDFs
\begin{align} \label{eq:pdf_inner-product}
    \left<\pdg,\pdf\right>:=\int  \pdg^*(\Bz)\pdf(\Bz)\frac{\VolForm[\Bz] }{|\VolDensity(\Bz)|^2}
    =\int  \pdg^*(\Bz) \pdf(\Bz)\frac{  d^\dim\Bz}{ \VolDensity(\Bz)  } .
\end{align}
Here, it should be assumed that the PDFs are elements of $\HilbertSpace^1=\LebesgueSpace^2$ with respect to the volume form
$\VolForm[\Bz] /\VolDensity^2(\Bz)=d^\dim\Bz /\VolDensity(\Bz)$.
This inner product is constant in time if all three volume forms are dynamically invariant.

The same construction can be developed for half-forms. In this case, one can formally write any half-form as
\begin{align}
    \Bpsi=\waveFunc \left(\VolForm/\VolDensity\right)^{1/2}=\waveFunc \left(d^\dim\Bz\right)^{1/2}.
\end{align}
so that the scalar component is now $\waveFunc/ \VolDensity^{1/2}$.
Now the inner product becomes
\begin{align}
    \left<\Bphi,\Bpsi\right>:=\int  \phi^*(\Bz)\psi(\Bz)\frac{\VolForm[\Bz] }{ |\VolDensity(\Bz)| }
    =\int  \phi^*(\Bz) \psi(\Bz)   d^\dim\Bz .
\end{align}
This shows that the result is the same as  Eq.~\ref{eq:wavefunc_innerprod} and, hence, is independent of the choice of preferred volume form, $\VolForm$. 
Again, for KVN solutions, the inner product is constant in time.

Even when the solutions satisfy the appropriate advection equations, if the choice of volume form, $\VolForm$, is not dynamically invariant, then the scalar inner product  \ref{eq:scalar_inner-product}  and the PDF inner product  \ref{eq:pdf_inner-product}  are not constant in time.
In order to force the inner product to be constant in time, one would need to change the volume form so that it becomes dynamically invariant.
To do this, one should choose the volume form to be the same as it was at an initial point in time, given by $\Bz_0(\Bz)=\FlowTransf_{-\tau}(\Bz)$.
For the scalar case in Eq.~\ref{eq:scalar_inner-product},  this yields
\begin{align}
\VolForm[\Bz_0]=\Jac^0_\tau \VolForm[\Bz]
\end{align}
which requires multiplying by the inverse Jacobian $\Jac^0_\tau=\det \partial\Bz_0/\partial\Bz$.
For the PDF case in Eq.~\ref{eq:pdf_inner-product}, this yields
\begin{align}
 \VolForm[\Bz_0]  /\VolDensity^2 (\Bz_0)  = \Jac^\tau_0  \VolForm[\Bz]/\VolDensity^2(\Bz).
\end{align}
which requires multiplying by the Jacobian $\Jac^\tau_0 =\det \partial\Bz/\partial\Bz_0$.
Due to the explicit reference to the Jacobian from initial to final coordinates, neither of these conditions are as natural as the duality pairing \ref{eq:scalar-pdf_pairing}, which corresponds to the KvN inner product \ref{eq:wavefunc_innerprod}.   
Yet, if the volume form, $\VolForm$, is already dynamically invariant, then these compensating factors are correctly supplied by Eqs.~\ref{eq:scalar_inner-product} and ~\ref{eq:pdf_inner-product}, and there is no need introduce additional Jacobian factors.

\textbf{Unitary Evolution.} 
If the flow is divergence-free, then the Jacobian remains unity for all time.
In this case, all three generators are skew-Hermitian, and, with an appropriate choice of BCs, all three generators are skew-adjoint.
Hence, all three evolution operators are unitary and, therefore, equivalent.

If the flow is not divergence-free, then the KvN evolution operator is the \emph{only} skew-Hermitian form for the generator with the standard Hilbert space inner product.
However, the other forms simply lead to a different rescaling of the solutions with the Jacobian, by a factor $\left(\Jac^0_\tau\right)^{\pm 1/2}$, and, hence, of the natural inner product by the inverse factor, $\left(\Jac^0_\tau\right)^{\mp 1}$.
Hence, while there can be sources and sinks of particles, there is no loss in information flow.
In other words, the generalized eigenvalue problem posed by Eq.~\ref{eq:general_advection} for $\psi=(\Jac^\tau_0)^{1/2}\scalar=(\Jac^0_\tau)^{1/2}\pdf$ is skew-Hermitian because it has the KvN form of with $\alpha=1/2$.
Relative to the appropriate density, the three forms are skew-Hermitian, and, again, an appropriate choice of boundary conditions can make the operators skew-adjoint.

Another way of stating this fact is that the Jacobian itself allows one to define a  dynamically invariant, though typically non-autonomous, measure on the phase space.
For large classes of differential equations, the Jacobian and, hence, the dynamically invariant measure it produces, is well-defined for any finite time. 
This is because, if the differential equations are continuous of order $\CC^\kindex$, then the solution to Eq.~\ref{eq:general_advection} implies that the Jacobian is $\CC^\kindex$ for finite times.

\textbf{Evolution Operator.}
If boundary conditions are chosen that ensure the KvN generator is skew-adjoint,~\cite{Stengl2024existence} then the KvN evolution operator is unitary and the spectral theorem holds.
In this case, the KvN evolution operator has an eigenfunction/distribution expansion of the form
\begin{align}
\EvolutionOperator_\tau (\Bx,\By) &= \int_\Reals\mu[\lambda] e^{-i\lambda\tau} \psi_\lambda (\Bx) \psi_\lambda^*(\By) 
\end{align}
where $\mu[\lambda]$ is the appropriate spectral measure over the eigenvalues, indexed by $\lambda$.
If we define corresponding scalar eigenfunctions, $\scalar_\lambda =(\Jac^0_\tau)^{1/2} \psi_\lambda$, and $\dim$-form eigendensities, $\pdf_\lambda =(\Jac^\tau_0)^{1/2} \psi_\lambda$, then we can also define a similarity transformation of the KvN evolution operator for scalar functions
\begin{align}
\EvolutionOperator^{(0)}_\tau (\Bx,\By)&= \int\mu[\lambda] e^{-i\lambda\tau}  \scalar_\lambda (\Bx) \pdf_\lambda^*(\By)
\\
&=\int\mu[\lambda] e^{-i\lambda\tau}  \scalar_\lambda (\Bx) \scalar_\lambda^*(\By) \Jac^0_\tau(\By)
\end{align}
and a similarity transformation of the KvN evolution operator for $\dim$-form densities
\begin{align}
\EvolutionOperator^{(\dim)}_{\tau} (\Bx,\By)&= \int\mu[\lambda] e^{-i\lambda\tau} \pdf_\lambda(\Bx) \scalar^*_\lambda (\By)
\\
&= \int\mu[\lambda] e^{-i\lambda\tau} \pdf_\lambda(\Bx) \pdf^*_\lambda (\By) \Jac^\tau_0(\By)
.
\end{align}
The fact that they are consistent follows precisely from the different weights of the Jacobian factor that each form carries.
In fact, these operators are similarity transformations of one another, since
\begin{align}
\EvolutionOperator_{\tau} (\Bx,\By)&=\left(\Jac_\tau^0(\Bx)\right)^{1/2}\EvolutionOperator^{(0)}_{\tau} (\Bx,\By) \left(\Jac_\tau^0(\By)\right)^{-1/2} \\
&=\left(\Jac^\tau_0(\Bx)\right)^{1/2}\EvolutionOperator^{(\dim)}_{\tau} (\Bx,\By) \left(\Jac^\tau_0(\By)\right)^{-1/2} 
.
\end{align}
Note that these operators differ from the usual Koopman and Perron-Frobenious evolution operators by the presence of these Jacobian factors.
The Jacobian factors ensure that these operators are similar to the KvN operator.

There are also multiple equivalent forms for the resolution of the identity
\begin{align}\label{eq:resolution-identity}
\delta(\Bx-\By) &= \int\mu[\lambda]  \psi_\lambda (\Bx) \psi_\lambda^*(\By) \\
&= \int\mu[\lambda]  \scalar_\lambda (\Bx) \pdf_\lambda^*(\By)
=\int\mu[\lambda] \scalar_\lambda (\Bx) \scalar_\lambda^*(\By) \Jac^0_\tau(\By)
\\
&= \int\mu[\lambda] \pdf_\lambda(\Bx) \scalar^*_\lambda (\By)
=\int\mu[\lambda] \pdf_\lambda(\Bx) \pdf^*_\lambda (\By)  \Jac^\tau_0 (\By)
.
\end{align}
Here, the second line is for scalar functions and third line is for probabilty densities.
These forms explicitly illustrate the duality of 0-form and $\dim$-form eigenfunctions. 

\textbf{Eigenfunctions.} 
As we shall see, for the completely integrable systems defined below, the unitary part of the evolution operator is determined by a transformation to a new set of coordinates, $\BtotalCoord=\{\BvaryCoord ,\BconstantCoord \}$, that are either constants of the motion $d\BconstantCoord/d\tau=0$ or that evolve uniformly in the time parameter, i.e. they are \textit{isochronous coordinates}~\cite{Mauroy2013isostables,Williams2015data} that satisfy $d\BvaryCoord/d\tau=\BOmega(\BconstantCoord)$, where $\BOmega$ is constant along each trajectory, i.e. it is only allowed to depend on the constants of the motion.
Clearly, in these coordinates, the trajectories simplify to $\BconstantCoord_\tau=\BconstantCoord_0$ and $\BvaryCoord_\tau=\BvaryCoord_0+\BOmega(\BconstantCoord)\tau$. 
For the KvN/Liouville/Schr\"odinger picture of evolution at fixed final position, the phase depends on $\BvaryCoord_0=\BvaryCoord_\tau-\BOmega\tau$.
In contrast, for the dual KvN/Koopman/Heisenberg picture of evolution at fixed initial position, the phase depends on 
$\BvaryCoord_\tau=\BvaryCoord_0+\BOmega\tau$.
Thus, for the respective eigenfunctions, the only difference is a change in the sign of the phase evolution with time parameter $\tau$.

The importance of the eigenfunctions of the Koopman operator~\cite{Mauroy2013isostables,Williams2015data} and their relation to isochronous coordinates has found several important uses in data-driven methods for dynamical systems.
In this context, Ref.~\onlinecite{Mauroy2013isostables}, makes the important distinction between \emph{ischronal coordinates} for phase angles and \emph{isostable coordinates} for Lyapunov functions that are analogous to action variables in Hamiltonian dynamics.
The distinction between these types of variables is crucial for understanding the topology that is natural for the eigenfunctions.
In the following, we refer to this as the dynamically invariant \emph{isochronous topology} because it is the topology that is intrinsic to the dynamical system itself.
Hence, the distinction between isostables and isochrons is induced by the dynamically invariant isochronous topology.

\subsection{ The I-LTS Decomposition  \label{sec:ILTS-decomposition}} 
\subsubsection{ I-LTS Decomposition of the Evolution Operator \label{sec:ILTS-evolution-operator}} 
The key assumption that allows one to develop an explicit factorization of the evolution operator with favorable properties is the fact, in a locally contractible region of a manifold, a general non-integrable vector field can always be expressed as a linear combination of \emph{integrable flows}.
For each of the integrable flows, one can explicitly find a symmetry group that is large enough to perform separation of variables and solve for the eigenfunctions explicitly over a given region of phase space. 
The full evolution operator can then be constructed efficiently using the I-LTS decomposition which expresses the full evolution as a product of the individual evolution operators for each of the integrable flows.
While a vector field can also be written as a sum of non-integrable flows, this does not typically reduce the complexity of constructing an approximate evolution operator.

Assume that the generator of the KvN evolution operator can be represented as a sum of \textit{completely integrable} terms:
\begin{align}
\Kgen&=\tfrac{i}{2}\left(\Vflow^\dagger-\Vflow\right)
=\sum_\iltsindex \Kgen_\iltsindex \\
& =\tfrac{i}{2}\sum_\iltsindex \left(\Vflow_\iltsindex^\dagger-\Vflow_\iltsindex \right)
=-\tfrac{i}{2}\sum_\iltsindex \BV_\iltsindex\cdot\nabla+\nabla\cdot\BV_\iltsindex
,
\end{align}
where each $\Kgen_\iltsindex =i(\Vflow_\iltsindex^\dagger -\Vflow_\iltsindex)$ is completely integrable.
Then, the exact evolution operator,
\begin{align}
U=e^{-i\Kgen \tau}=e^{(\Vflow^\dagger-\Vflow)\tau/2}=e^{-(\nabla\cdot\BV+\BV\cdot\nabla)\tau/2}
,
\end{align}
can be approximated efficiently using the Lie-Trotter-Suzuki (LTS) decomposition.
For example, choosing an arbitrary ordering of the individual factors 
\begin{align}
\EvolutionOperator_\LTS(\Delta \tau)=\prod_\iltsindex e^{-i \Kgen_\iltsindex \Delta \tau} 
= \prod_\iltsindex e^{(\Vflow^\dagger_\iltsindex-\Vflow_\iltsindex)\Delta \tau/2}
\end{align}
leads to an approximation, $\EvolutionOperator_\LTS\simeq U$, that is first order accurate in time step, $\Delta \tau$.
Higher order LTS decompositions \cite{Childs21} use more complicated operator product ordering with specific time splittings to achieve any desired order of convergence with time step. 
Of course, if each of the individual generators $\Kgen_\iltsindex$, commutes with all of the others, then this approximation is exact, so that $\Kgen_\LTS=\Kgen$ and $\EvolutionOperator_\LTS=e^{-i\tau\Kgen_\LTS }=\EvolutionOperator=e^{-i\tau\Kgen }$.

We note that, when there are only two operators in the sum, the second order LTS is particularly efficient because it uses the same formula as first order LTS for the bulk of the calculation, but only takes a 1/2 time step for the initial and final operators.
Thus, it achieves second order convergence with only first order complexity.

If the vector fields in the I-LTS decomposition are smooth $C^\infty$ vector fields and the evolution operator acts on smooth $C^\infty$ functions, then for small enough time steps, $\Delta \tau$, the operator logarithm of the scalar evolution operator converges~\cite{AgracevGamkrelidze79,Strichartz1987campbell, Blanes2009magnus}
\begin{align}
     -i\Kgen_{\LTS }^{(0)} (\Delta \tau)&:= \frac{ 1}{\Delta \tau} \log {\left(\EvolutionOperator_\LTS^{(0)}(\Delta \tau)\right)}\\
     &=  \Vflow_\LTS= \BV_{\LTS }(\Delta \tau)\cdot\nabla.
\end{align} 
Moreover, the second line indicates that it is equal to a nearby first order differential operator $V_{\LTS } (\Delta \tau)$, up to an approximation error that vanishes in a particular limit, discussed further below.  
The propagator for $n$ time steps, i.e. for time $\tcoord=n\Delta \tcoord$, is simply defined by the $n$-fold product of single time step evolution operators, so that 
\begin{align}
\EvolutionOperator_\LTS(\tau, \Delta \tau) :=\EvolutionOperator_\LTS(n\Delta\tau, \Delta \tau) = \EvolutionOperator^n_\LTS(\Delta \tau).
\end{align}
Thus, with respect to integer time slices, the propagator satisfies the simple 1-parameter group relation
\begin{align}
    \EvolutionOperator_\LTS(\tau, \Delta \tau  )=e^{-i   \tau \Kgen_\LTS (\Delta \tau) }:=e^{-  \tau\Vflow_\LTS (\Delta \tau) }.
\end{align}
Note, however, that, if there are nontrivial commutators, then the generator of the evolution operator, $\Kgen_\LTS(\Delta \tau)=-  i\BV_\LTS(\Delta \tau) \cdot \nabla$, still explicitly depends on $\Delta \tau$ and, again, the approximation only becomes exact in the limit $\Delta \tau \rightarrow 0$.

Determination of the first order differential operator $\Kgen_\LTS=-i\Vflow_\LTS$ that reproduces $U_{LTS}$ at discrete time steps requires certain assumptions in order to be made mathematically rigorous.  
Moreover, in general, no such first order differential operator exists since, even for $M=S^1$ (see warning 6 from Ref.~\onlinecite{Milnor83}), the exponential map fails to be locally onto from the space of smooth sections of vector fields, viewed as sections of the tangent bundle, $\Gamma(TM)$, to the diffeomorphism group on the manifold, ${\rm Diff}(M)$, which includes flows that arise from dynamical systems. 
Instead of an exact implementation of $U_{LTS}$, we can generate a first order differential operator that reproduces $U_{\LTS}$ up to an arbitrarily small error.~\cite{AgracevGamkrelidze79}

The Magnus expansion is a continuous analog of the Baker-Campbell-Hausdorff (BCH) formula that represents the formal solution for $-i\Kgen_{\LTS}$.~\cite{Blanes2009magnus} 
Convergence of both the Magnus and BCH series  are difficult to prove due to the combinatorial growth in the number of terms that appear~\cite{AgracevGamkrelidze79,Strichartz1987campbell} unless the operators in the expansion form a finite-dimensional Lie algebra.~\cite{Strichartz1987campbell}
The Dyson series for the propagator $dU/d\tau=-iHU$ is a Volterra series that has the form of a time-ordered product and the Magnus expansion seeks to take a formal logarithm of this Volterra series.
Following Ref.~\onlinecite{AgracevGamkrelidze79}, which refers to the manipulation of such time-ordered products as ``chronological calculus,'' the action of $U_{LTS}(\Delta \tau)$ can be approximated in the topology of uniform convergence with respect to all derivatives by the exponential of truncations of the Volterra-Dyson series. 
Proposition 4.1 of Ref.~\onlinecite{AgracevGamkrelidze79} states conditions for convergence.
Using notation that closely follows \onlinecite{AgracevGamkrelidze79}, define the vector field 
\begin{align}
Y_{ \tau} :=\sum_\ell V_\ell \Theta(\tau_\ell-\tau)\Theta(\tau-\tau_{\ell-1})
\end{align}
which is piecewise continuous in time, where $\tau_\ell=\ell \Delta\tau/n$ for $n$ I-LTS components and $\Theta$ is the unit step function.
Then, form the vector field
\begin{align}
\Vflow_{LTS}^M(\Delta\tau) := \sum_{m=1}^M \Vflow_{LTS}^{(m)}(\Delta\tau, Y_{ \tau})
\end{align}
where
\begin{multline}
    \Vflow_\LTS^{(m)}(\Delta\tau, Y_{ \tau})  = \int_0^{\Delta\tau} d\tau_1 \int_0^{\tau_1} d\tau_2 \cdots \\
    \int_0^{\tau_{m-1}} d\tau_m g_m(Y_{\tau_1},\cdots,Y_{\tau_m})/\Delta\tau
\end{multline}
and the $g_m$'s are polynomials in terms of the Lie bracket. 
In this case, the action of the evolution operator $e^{\Vflow_{LTS}^M \Delta \tau}\phi$ converges to $U^{(0)}_\LTS(\Delta \tau)\phi$ provided
\begin{align}
\int_0^{\Delta \tau} \lVert Y_{ \tau} \rVert_{s+m} d\tau < 1 
\end{align}
where $\lVert\cdot \rVert_{s+m}$ is a norm defined as the maximum absolute value of the vector field and any of its partial derivatives up through order $s+m$.~\cite{AgracevGamkrelidze79}
Clearly, in order for this definition to remain finite, all the vector fields $\{V_\ell\}$ involved, as well as the state $\phi$ they act on via $V_\ell(\phi)=\BV_\ell\cdot\nabla(\phi)$, must be $C^\infty$ smooth.

Higher order terms in the Magnus series, $g_m$, and higher order LTS approximations include corrections due to commutators of the individual $\Kgen_\iltsindex$ terms.
From a structural point of view, these commutators always have the form of another vector field 
\begin{align}
C_{\iltsindex \kindex}=\left[V_\iltsindex,V_\kindex\right]= 
\sum_{jk} V^\iindex_\iltsindex 
(\partial_\iindex V^\jindex_\kindex) \partial_\jindex
- V^\iindex_\kindex(\partial_\iindex V^\jindex_\iltsindex) \partial_\jindex,
\end{align}
where the operators in parenthesis only act on terms within the parenthesis.
By induction, all commutator terms always simply add corrections to generate a nearby vector field.
Thus, the solution to $dU_{\LTS}/d\tau = -Y_\tau U_{\LTS}$ has the form
\begin{align}
    \Kgen_\LTS(\Delta \tau)= -i\BV_{\LTS,M}(\Delta \tau)\cdot\nabla:= \frac{i}{\Delta \tau} \log_{M}^{Volt}{\left(\EvolutionOperator_\LTS(\Delta \tau)\right)},
\end{align} 
where $\log_M^{Volt}$ is the truncation of the formal logarithm defined by the Volterra series. 
Thus, the I-LTS approximation proposed here is equivalent to determining the exact flow for a modified vector field, up to an arbitrarily small error.
With this understanding, we assume that one can take $\Delta \tau$ small enough that convergence is guaranteed and that the error vanishes in the limit $M\rightarrow \infty$.

Other relevant commutators also produce the same $V_\LTS$ because they have the same commutator structure aside from the order in which the partial derivatives are taken.
For example, the term
\begin{align}
D_{\iltsindex \kindex}=\left[V_\iltsindex^\dagger,V_\kindex^\dagger\right]= \sum_{jk} \partial_{\iindex} (V^\iindex_\iltsindex \partial_\jindex V^\jindex_\kindex) 
-
\partial_\iindex (V^\iindex_\kindex\partial_\jindex V^\jindex_\iltsindex),
\end{align}
has the form of the divergence of a vector field and has the exact same index structure as $C_{\ell k}$.
The only change is the ordering with respect to the placement of the partial derivatives and the velocity components.
Hence, the solution to $dU_{\LTS}/d\tau =  Y_\tau^\dagger U_{\LTS}$ can be expressed as $U_{\LTS}=e^{V_{LTS}^\dagger\Delta \tau}$ using the same $V_{\LTS}$, assuming it converges.
Aside from derivative ordering, the commutators $\left[V_\iltsindex,V_\kindex^\dagger\right]$ and $\left[V_\iltsindex^\dagger,V_\kindex\right]$, also have the same index structure.
Thus, the solution to the Koopman-von Neumann equation  $dU_{\LTS}/d\tau = ( Y_\tau^\dagger-Y_\tau) U_{\LTS}$ has a similar form of the solution  $U_{\LTS}(\Delta \tau)=e^{(V_{LTS}^\dagger-V_{LTS})\Delta \tau}$ in terms of the same vector field, $V_{\LTS}$.

The approximate Koopman eigenfunctions for $\EvolutionOperator_\LTS(\Delta \tau)$ are simply the eigenfunctions of the new Koopman generator $\Kgen_\LTS(\Delta \tau)$. Thus, the approximate Koopman eigenfunctions still retain the group structure discussed in \onlinecite{Giannakis24}.
Since $\Kgen_\LTS$ is a first order differential operator, any product of its eigenfunctions is also an eigenfunction. 
Thus, the tensor product factorization properties employed in \onlinecite{Giannakis25} are retained by this approximation.
Moreover, if the integrable vector field decomposition is chosen to respect the symmetry group of the original vector field, then the tensor product structure will also be determined by the symmetry group.

Our approach of expressing the generator for an arbitrary evolution in terms of a sum of integrable parts can be extended to any evolution operator, whether it be classical, quantum, or stochastic, or any combination thereof.
While operator splitting is a standard tool in applied mathematics, PDE theory, and numerical simulation, and the LTS decomposition is a standard tool in group theory, quantum mechanics, and quantum algorithms, \emph{here, we emphasize that the class of {\bf integrable decompositions} lead to the least computational complexity} when tensor network and/or quantum block-encoding techniques are employed.
This is because each term in the I-LTS operator product has a simple tensor product factorization that allows for very efficient time integration methods, called ``fast-forwarding'' algorithms in the quantum algorithms community.
This implies that each evolution operator in the I-LTS product can be fast-forwarded for any choice of time step, so that each term in the product is a single unitary.
Of course, the overall number of Trotter time steps must still be set by the desire to achieve a given accuracy.

Moreover, a key advantage of this decomposition is that if the flow admits a symmetry group that is not large enough to render the flow integrable, this symmetry group can still be used to significantly reduce the complexity of the I-LTS decomposition.
The symmetry group defines constants of the motion, $\BconstantCoord=\{\constantCoord^\jindex\}$, that satisfy $\BV\cdot\nabla \constantCoord^\jindex=0$.
Hence, using $\BconstantCoord$ as coordinates allows separation of variables to be employed to simplify the representation.
This then allows the number of terms in the I-LTS evolution operator product expansion to be significantly reduced.

The key step needed to prove that these methods are efficient is to determine an efficient transformation of bases between the different 
Hilbert spaces 
needed for each term in the I-LTS product.
In fact, the assumption that each evolution operator in the product is integrable implies that there is an efficient basis transformation that also possesses an efficient inverse.
For example, this transformation could be based either on the FFT, the Quantum Fourier Transform (QFT)~\cite{Coppersmith02, Nielsen10}, the quantized tensor network Fourier Transform (QTNFT)~\cite{Dolgov2012superfast, Chen2023quantum}, or through other efficient unitary transformations to bases of other special functions.
Thus, due to the integrability assumption, each transformation between Hilbert spaces that is needed to pass between terms in the I-LTS decomposition, as well as their inverses, are always efficiently computable.

Finally, when the total generator/evolution operator is not integrable, there will still be an inherent complexity in time evolution because the overall operator is not fast-forwardable.
Thus, one must integrate the system using an I-LTS product formula that is sufficiently small $\Delta \tau$, and, hence, a sufficiently large number of time steps, to achieve a given accuracy.
This inherent computational complexity will always remain for simulating chaotic dynamics.

\subsubsection{Completely Integrable Flows \label{sec:completely-integrable}}

Investigation of global complete integrability is an active area of research that combines differential geometry, topology, and foliation theory.~\cite{Fiorani2008geometrical,Zung2018conceptual}
The definition of \emph{complete integrability by quadratures} is that the equations of motion in any chart of the manifold can be determined by quadratures for some special set of coordinates.~\cite{Lichtenberg2013book,Olver1993applications,Carinena2016solvability,Grabowska2020solvable,Kozlov2024solvable}
While the Frobenious theorem implies that this is always true locally, the global definition also requires that the solutions can be globally extended in a well-defined manner.
Typically, this is only possible if a set of symmetries exists that allows one to smoothly generate families of new solutions from a given solution.

For example, at least locally, any vector field that is an element of a solvable Lie algebra that acts transitively on a given manifold is completely integrable by quadratures.~\cite{Carinena2016solvability,Kozlov2024solvable}
This implies that in any local neighborhood, one can find a (non-unique) canonical coordinate system, $\Bz=\{\zcoord^j\}$, in which the equations of motion have the general solvable quasilinear form~\cite{Grabowska2020solvable,Kozlov2024solvable}
\begin{align} \label{eq:solvable-eom}
    d\zcoord^j/d\tau&=\Vflow^j(\zcoord^j,\zcoord^{j+1},\dots,\zcoord^\dim) \\
    &=\zcoord^j c^j(\zcoord^{j+1},\dots,\zcoord^\dim)+b^j ( \zcoord^{j+1},\dots,\zcoord^\dim)
    .
    \label{eq:Grabowska-eom}
\end{align}
While this holds for complex solvable Lie algebras, for the real case, the structure in Eq.~\ref{eq:Grabowska-eom} applies to either 1 or 2 coordinates at a time, so that the $c^j(\BconstantCoord)$'s should be interpreted as either $1\times 1$ or $2\times 2$ matrices, respectively.~\cite{Carinena2016solvability}
For complete integrability to hold globally, the solvable form is required to smoothly agree across transition regions between local charts in a manner that is globally well-defined.
This allows one to solve for the equations of motion via a three stage procedure of first finding the canonical coordinate system in each chart, then solving the equations of motion locally, and finally gluing solutions across transition regions.

Yet, for many integrable systems of interest, compact invariant manifolds must be isomorphic to a torus.~\cite{Nekhoroshev2005types,Zung2018conceptual} 
The solution for the motion on the torus generically involves the multivalued angle coordinates on the torus.
To correctly account for the periodicity conditions on the angle variables, one should renormalize the corresponding coordinates in the solution so that they have the correct angular frequency of revolution,  $d\varyCoord^j/d\tau=\Omega^j$.
In principle, the frequency can potentially depend on arbitrary constants of the motion that satisfy the solvable form of the solution, $\Omega^j(\totalCoord_0^{j+1},\dots,\totalCoord_0^\dim)$.
In order to simplify the discussion of the eigenfunctions, we shall restrict our attention to the case $\Omega^j(\BconstantCoord)$ alone.

Hence, in practice, we consider a restricted notion of complete integrability that corresponds to the definition of broad integrability,~\cite{Zung2018conceptual}  also known as ``integrability in the broad sense,'' as first defined by Bogoyavlenskij.~\cite{Bogoyavlenskij1998extended}  
This significantly simplifies the derivation of the structure of KvN eigenfunctions, and, hence, of the eigenfunctions of the I-LTS decomposition.
We refer to this as \emph{abelian integrability} becausde it assumes an abelian Lie algebra of symmetries.
\begin{definition}[\bf $(p,q)$-Abelian Integrable System \onlinecite{Zung2018conceptual}]
\label{def:pq-integrable}
A  \emph{$(p,q)$-abelian integrable system} on a manifold, $M$, is defined by $p$ smooth global vector fields, $X_i$, in involution 
and $q$ smooth global constants of the motion, $\BconstantCoord=\{\constantCoord^j\}$, that satisfy
\begin{align}
    [X_i,X_j]&=0\\ 
    X_i\cdot d\constantCoord^j&=0.
\end{align}
In addition, assume the nondegeneracy conditions
\begin{align}
    X_1\wedge \dots \wedge X_p&\neq 0\\
    d\constantCoord^1\wedge \dots \wedge d\constantCoord^q&\neq 0
\end{align}
hold almost everywhere.
A level set, $N$, defined by $\BconstantCoord=constant$, is \emph{regular} if the nondegeneracy conditions hold throughout the set.
The map $\BconstantCoord:\Manifold\rightarrow \Reals^q$ is \emph{proper} if it is a proper topological map almost everywhere, and, hence the level sets are regular almost everywhere.
\end{definition}
Zung proves~\cite{Zung2018conceptual} that abelian integrable systems satisfy a version of the Liouville-Arnold theorem.
\begin{theorem}[\bf Liouville-Arnold $(p,q)$-Integrability ~\onlinecite{Zung2018conceptual}]
\label{thm:pq-Liouville}
Consider a abelian $(p,q)$ integrable system as defined above.
In any tubular neighborhood, $U(N)$ of a regular and proper compact level set, $N$, there is a free torus action
\begin{align}
    \rho: T^p\times U(N)\rightarrow U(N)
\end{align}
unique up to automorphisms of $T^p$.
Hence, $N$ is diffeomorphic to $T^p$, and $U(N)$ is diffeomorphic to $T^p\times B^q$.
Thus, there are $p$ periodic coordinates, $\BvaryCoord=\{\varyCoord_i\}$, in which $X_i=\sum_j X_i^j(\BconstantCoord)\partial_{\varyCoord^j}$.
Moreover, any vector field of the form 
\begin{align}
    \Vflow=\sum_j V^j(\BconstantCoord)X_j 
\end{align}
is completely integrable and can be expressed as
\begin{align}\label{eq:Vflow_broad-integrable}
\Vflow=\sum_j \Omega^j(\BconstantCoord)\partial_{\varyCoord^j}.
\end{align}
\end{theorem}
\begin{proof}
    Zung~\cite{Zung2018conceptual} proves the theorem for $\Vflow=X_1$.
    By direct computation, any smooth $\Vflow$ of the assumed form, $\Vflow=\sum_j V^j(\BconstantCoord)X_j$, satisfies the assumptions required of $X_1$.
    Hence, one can always choose $X_1=\Vflow$ of this form.
\end{proof}

For $(p,q)$-abelian integrable vector fields, 
the equations of motion $d\BvaryCoord/d\tau=\BOmega(\BconstantCoord)$ lead to the
the solution
\begin{align}
\varyCoord^j_\tau=\varyCoord^j_0+\Omega^j(\BconstantCoord) \tau.
\end{align}
The frequency is defined by the angular average of the velocity 
\begin{align}
   \frac{1}{\Omega^j(\BconstantCoord)}=\lim_{\Theta\rightarrow\infty} \frac{1}{2\Theta}\int_{-\Theta}^{\Theta}  \frac{d \theta^j }{V^{\theta^j}}
\end{align}
where $\theta^j$ is any choice of angular coordinates with the same winding numbers as $\varyCoord^j$. 
Similarly, 
for any coordinate $\xi^k$ that is not multivalued, it may still be useful to perform a similar normalization with respect to some reference coordinate and define a similar frequency $\Omega^k(\BconstantCoord)$.

For example, any $2\dim$-dimensional Hamiltonian system that is completely integrable in the Liouville-Arnold sense has a commuting Poisson algebra of invariants of type $\BconstantCoord$.~\cite{Lichtenberg2013book,Olver1993applications}
In this case, the solvable Lie algebra is abelian, as in the definition of broad $(p,q)$-integrability \cite{Zung2018conceptual}.
For simplicity, consider a system in canonical coordinates, $\Bz=\{\mathbf{q},\mathbf{p}\}$, where the Poisson bracket has the canonical form.
First, to normalize the solution, a set of action invariants $\Baction(\BconstantCoord)=\{J^j\}$ can be determined using the action integral
\begin{align}
 2\pi  \action^j=\left. {\mathcal S}^j  \right|^{2\pi}_0&&  {\mathcal S}^j = \int  d\theta^j  (\partial_{\theta^j} \mathbf{q})\cdot \mathbf{p}
\end{align}
where, here, the coordinate $\theta^j$ is any choice of coordinate such that $V^{\theta^j}$ is non-vanishing almost everywhere along the trajectory.
Assuming the Hamiltonian, $\Ham(\Baction)$, is autonomous, the frequency for coordinate $\varyCoord^j$, is determined by  
\begin{align}
    \Omega^j(\Baction)=\partial_{\action^j} H.
\end{align}
For every constant of the motion, $\action^j$, we can define a canonically conjugate  coordinate  by quadrature via 
\begin{align}
  \varyCoord^j=\Omega^j(\Baction)\int \frac{d \theta^j }{V^{\theta^j}(\zcoord^j,\Baction)}=\frac{\partial {\mathcal S}^j}{\partial {\action^j}} .
\end{align}
Thus, the solution is  $\xi^j=\xi^j_0+\Omega^j(\Baction)\tau$.

{\bf The Non-Compact Case.} In the Hamiltonian case, it has been proven ~\cite{Fiorani2008geometrical} that for non-compact level sets $N$ with complete orbits for all $X_i$, the solution manifold is of the form $\Torus^{p-r}\times\Reals^{r}$ for some $0\leq  r\leq p$.
In a tubular neighborhood of $N$, where the leaves of $U(N)$ are diffeomorphic, then there is an abelian action of $\Torus^{k}\times\Reals^{\dim-k}$ on $U(N)$, and, $U(N)$ is diffeomorphic to $\Torus^{p-r}\times\Reals^{r}  \times B^q$. 
One expects a generalization of this result to apply to more general non-Hamiltonian abelian-integrable systems.

To be precise, for noncompact domains, one should split the coordinates into $\BvaryCoord=\{\Btheta,\Bx\}$, where $\Btheta$ are angles and $\Bx$ are noncompact coordinates.
When non-compact coordinates arise, $\xcoord^j\in \Reals$, it may also be convenient to normalize the motion by a constant velocity/frequency, $\Omega^j(\BconstantCoord)$.
This leads to the same structure for the vector field in Eq.~\ref{eq:Vflow_broad-integrable} and, hence, to the same structure for the solutions to the equations of motion. 
For notational simplicity, we will not distinguish between these two types of coordinates in the following.

\subsubsection{Eigendecomposition for Completely Integrable Flows \label{sec:ILTS_eigs}}

An eigenfunction with respect to parameter $\tau$ must satisfy 
\begin{align}
    i\partial_\tau\phi=\omega\phi,
\end{align}
and, hence, $\phi(\tau,\Bz)=\phi_0(\Bz)e^{-i\omega\tau}$.
According to the definition, a abelian integrable system has a set of canonical coordinates $\BtotalCoord=\{\BvaryCoord,\BconstantCoord\}$ in which the only nontrivial equations of motion  have the form
\begin{align}
    d\BvaryCoord/d\tau=\BOmega(\BconstantCoord)
\end{align}
and hence the solution has the form
\begin{align}
    \BvaryCoord_\tau=\BvaryCoord_0+\BOmega(\BconstantCoord)\tau && \BconstantCoord_\tau=\BconstantCoord_0.
\end{align}
Thus, nontrival eigenfunctions must have the form
\begin{align}
\scalar_{\BspectralCoord}(\BtotalCoord)=a(\BconstantCoord) e^{i\BspectralCoord\cdot\BvaryCoord_0} = a(\BconstantCoord) e^{i\BspectralCoord\cdot(\BvaryCoord_\tau-\BOmega  \tau) }=a(\BconstantCoord) e^{i\BspectralCoord\cdot \BvaryCoord_\tau-i\omega  \tau  }
\end{align}
where $\BspectralCoord$ represents a dual vector of Fourier harmonics in $\BvaryCoord$ and the total frequency is $\omega=\BspectralCoord\cdot\BOmega$.
Because the dependence on $\BconstantCoord$ is arbitrary, the spectrum is continuous in these coordinate directions. Hence, the KvN operator has eigendistributions
\begin{align}
\scalar_{\BspectralCoord,\BconstantCoord_0}(\BtotalCoord)=\delta(\BconstantCoord-\BconstantCoord_0) e^{i\BspectralCoord\cdot\BvaryCoord_0} = \delta(\BconstantCoord-\BconstantCoord_0)e^{i\BspectralCoord\cdot(\BvaryCoord_\tau-\BOmega  \tau) }.
\end{align}

For the I-LTS decomposition, each $\Vflow_{\iltsindex}$ is assumed to be completely integrable.
Hence, for each $\Vflow_{\iltsindex}$, one can find coordinates that rectify each flow, and, for each dynamically invariant region of phase space, there are coordinates, $\BtotalCoord_\iltsindex=\{\BvaryCoord_\iltsindex,\BconstantCoord_\iltsindex\}$, such that $\Vflow_\iltsindex(\totalCoord^\jindex_\iltsindex)=\BV_\iltsindex\cdot\nabla\totalCoord^\jindex_\iltsindex=\Omega^\jindex_\iltsindex(\BconstantCoord)$ is constant in time. 
For each phase space region covered by $\BtotalCoord_\iltsindex$, the trajectories are completely described by the relations
\begin{align}
    \totalCoord^\jindex_{\iltsindex,\tau} =  \totalCoord^\jindex_{\iltsindex,0} + \Omega^\jindex_\iltsindex(\BconstantCoord)  \tau,
\end{align}
 where $\totalCoord^\jindex_{\iltsindex,\tau}$ is the $j$-th coordinate at time $\tau$.
Thus, for each  dynamically invariant  region for $\Vflow_\iltsindex$, the KvN eigendistributions for $\Vflow_\iltsindex$ can be written in terms of Fourier harmonics as
\begin{align}
\scalar_{\iltsindex,\BspectralCoord}(\tau,\BtotalCoord_\iltsindex) =e^{i\BspectralCoord\cdot \BtotalCoord_{\iltsindex,0}}= e^{i\BspectralCoord\cdot(\BtotalCoord_\iltsindex-\BOmega_\iltsindex \tau)}=e^{i\sum_\jindex \spectralCoord_\jindex(\totalCoord^\jindex_\iltsindex-\Omega^\jindex_\iltsindex \tau)}
\end{align}
where $\BspectralCoord=\{\spectralCoord_\jindex\}$ represents a vector of wavenumbers for the Fourier transformation.
Clearly, the frequency is $\omega_\iltsindex=\spectralCoord_\jindex \Omega^\jindex_\iltsindex$.

By Pontryagin duality, either a Fourier series or a Fourier transform is necessary for a complete set of eigenfunctions, depending on whether the range of $\totalCoord^\jindex_\iltsindex$ is compact or not. 
The latter case occurs when the spatial range is infinite or if the trajectory asymptotically approaches a fixed point in the past or future, e.g. if the trajectory lies on a separatrix, so that the temporal dynamics cannot be considered to be periodic except on an infinite domain.
This is precisely the point at which the dynamically invariant isochronous topology is needed to understand whether there are proper eigenfunctions or only eigendistributions.

Since both the coordinate transformations  $\BtotalCoord_\iltsindex$, and the frequencies  depend on the invariants, 
$\Omega_\iltsindex(\BconstantCoord_\iltsindex)$,
it is computationally advantageous to keep the dependence on the $\BconstantCoord_\iltsindex$ coordinates local in space rather than performing a Fourier transformation.
In this case, we label the generalized Koopman eigendistributions with the following notation
\begin{align}
\scalar_{\iltsindex,\BspectralCoord,\BconstantCoord_0}(\tau,\BtotalCoord_\iltsindex ) &= \delta(\BconstantCoord_\iltsindex-\BconstantCoord_0) e^{i\BspectralCoord\cdot(\BvaryCoord_\iltsindex-\BOmega_\iltsindex \tau)}\\
&=\delta(\BconstantCoord_\iltsindex-\BconstantCoord_0) e^{i\sum_\jindex \spectralCoord_\jindex(\varyCoord^\jindex_\iltsindex-\Omega^\jindex_\iltsindex \tau)}.
\end{align}
Thus, in this case, the eigendistributions are local Dirac delta-functions in $\BconstantCoord_\iltsindex$ and Fourier harmonics in $\BvaryCoord_\iltsindex$.

\subsubsection{Efficient Path-Ordered Evolution Operator \label{sec:ILTS_evolution-operator}}
Now, let $\Phi_\iltsindex: \Bz \rightarrow \BtotalCoord_\iltsindex=\{\BvaryCoord_\iltsindex,\BconstantCoord_\iltsindex\}$, denote the coordinate transformation needed to determine the integrable coordinate system for each $\Vflow_\iltsindex$, and let $\FFT_\iltsindex:\scalar(\BvaryCoord_\iltsindex,\BconstantCoord_\iltsindex)\rightarrow\scalar_{\BspectralCoord}(\BconstantCoord_\iltsindex)$, denote the Fourier transform in the non-isolating coordinates, $\BvaryCoord_\iltsindex$.
An $m$-stage LTS operator product expansion has the form of a path-ordered product of terms 
\begin{align}
\EvolutionOperator_\LTS &= e^{-i\tau_m \Kgen_m}\cdots e^{-i\tau_2 \Kgen_2}  e^{-i\tau_1 \Kgen_1}
=\PathOrder\prod_{\iltsindex}  e^{-i\tau_\iltsindex \Kgen_\iltsindex}.
\end{align}
Here, the path-ordering operator, $\PathOrder$, indicates that consecutive terms along the path are added to the left.
Each evolution operator in the product can be diagonalized with a simple series of transformations.
The complete expression for $\EvolutionOperator_\LTS(\Bz',\Bz):=\left<\Bz'|\EvolutionOperator_\LTS|\Bz\right>$ is 
\begin{multline}
\EvolutionOperator_\LTS(\Bz',\Bz) =  \Phi_m^{-1}  \circ \FFT_m^{-1}e^{-i\tau_m \Kgen_m}\FFT_m  \Phi_m \circ \cdots 
\\
\circ \Phi_2^{-1}  \FFT_2^{-1}  e^{-i\tau_2 \Kgen_2}\FFT_2  \Phi_2 \circ  \Phi_1^{-1}  \FFT_1^{-1}  e^{-i\tau_1 \Kgen_1}\FFT_1  \Phi_1
\end{multline}
which can be expressed as
\begin{align}
\EvolutionOperator_\LTS(\Bz',\Bz)& = \PathOrder\prod_\iltsindex \Phi_\iltsindex^{-1} \FFT_\iltsindex^{-1}e^{-i\tau_\iltsindex \Kgen_\iltsindex}\FFT_\iltsindex  \Phi_\iltsindex
.
\end{align}
Again, it is more computationally efficient to only apply the Fourier transformation to the non-isolating invariants, $\BvaryCoord_\iltsindex$, as described above, and to leave the remaining isolating coordinates, $\BconstantCoord_\iltsindex$, in a local representation.
We note that, because each coordinate transformation $\Phi_\jindex$ and inverse transformation $\Phi_\jindex^{-1}$ can be performed efficiently, transformations between coordinate systems $\Phi^\jindex_k=\Phi_k\circ \Phi_\jindex^{-1}$ can also be performed efficiently.
Similarly, since the required Fourier transformations $\FFT_\jindex$ and their inverses $\FFT_\jindex^{-1}$ can be performed efficiently, functional transformations between coordinate systems and their Fourier transformations can also can be performed efficiently.

The Fourier transformations (FTs), $\FFT^{\pm1}_\iltsindex$ in these equations are relatively inexpensive because of the FFT, which scales as $N\log(N)$ for a data vector of size $N$. 
Classically, an approximate FFT can be evaluated very inexpensively either using sparse FFTs or, if the data is intrinsically low-rank, by using a low-rank tensor network decomposition.
For example, quantized tensor network FT's (QTNFTs)   exist for classical computers that exploit the low-rank structure of the FFT in quantized tensor train format \cite{Dolgov2012superfast, Chen2023quantum}.
The QFT for quantum computers is an even more efficient approach that only scales as $[\log_2(N)]^2$ for an exact result \cite{Coppersmith02, Nielsen10} and $\log_2(N)\log_2(\log_2(N))$ for an approximate result.

\subsection{Existence of local I-LTS Decompositions  \label{sec:ILTS-exists}}
\subsubsection{Autonomous 1D Vector Fields are Integrable \label{sec:1D-ILTS}}
 
An explicit solution for the flow can be given for any autonomous vector field, $\BV(\zcoord)=\{\Vflow^\zcoord(\zcoord)\}$, in 1D.
As in previous sections, we use the coordinate $\tau$ to represent the parameterization of time and assume $\Vflow^\zcoord(\zcoord)$ is independent of $\tau$.
To treat the non-autonomous case where $\Vflow^\zcoord(\tau,\zcoord)$ has an explicit time dependence, one can use the extended phase space formalism, where $t$ is appended to the coordinate system.
In this case, the solution has the form given in the next section on general vector fields in one higher dimension.

\textbf{Solution of 1D ODE.}
For an autonomous 1D vector field, $\Vflow^\zcoord(\zcoord)$, the ODE to be solved is
\begin{align}
d\zcoord/d\tau&=\Vflow^\zcoord(\zcoord).
\end{align}
To solve this equation, we note that, when $\Vflow^\zcoord$ does not have an explicit dependence on $\tau$, then the solution can be obtained with a simple coordinate transformation from $\zcoord$ to an isochronous coordinate $\varyCoord$ defined by  
\begin{align}
 \varyCoord_\tau = \int_{\zcoord_0}^{\zcoord_\tau} d\zcoord/\Vflow^\zcoord(\zcoord).
\end{align}
In this new  coordinate system, the velocity $V^\varyCoord=1$ and, so the constant of the motion is simply
\begin{align}
\varyCoord_0=\varyCoord_\tau-\tau.
\end{align}

\textbf{Solution of 1D Advection.}
The scalar advection equation is simply
\begin{align}
    d\scalar/d\tau = \left(\partial_t + \Vflow^\zcoord(\zcoord)\partial_\zcoord\right) \scalar. 
\end{align}
This corresponds to advection with the velocity
\begin{align}
d\zcoord/d\tau&=\Vflow^\zcoord(\zcoord).
\end{align}
The evolution operator from time 0 to $\tau$, with the final position $\zcoord$ held fixed, is defined by
\begin{align}
    \scalar(\tau,\zcoord) = \EvolutionOperator_\tau^{(0)} \scalar(0,\zcoord) = \scalar(0,\zcoord_0),
    \label{eq:scalar_advection}
\end{align}
where $\zcoord_0:=\zcoord(\tau=0)$ is the initial position at time $\tau=0$.

\textbf{Isochronous Metric and Distance.}
In 1D, the isochronous 1-form, $d\xi$, defines an isochronous metric, $g_{\xi\xi}=1$, and, inverse metric, $g^{\xi\xi}=1$, for parallel transport along the flow.
Hence, the isochronous metric in the original $\zcoord$ coordinate is
\begin{align}
   g_{\zcoord \zcoord}&=g_{\xi\xi}(d\xi/d\zcoord)^2=1/(V^\zcoord)^2 \\
   g^{\zcoord \zcoord}&=g^{\xi\xi}(d\zcoord/d\xi)^2=(V^\zcoord)^2 
\end{align}
and the isochronous Jacobian is
\begin{align}
 \Jac(\zcoord)=\sqrt{g_{\zcoord \zcoord}}=d\xi/d\zcoord = 1/V^\zcoord(\zcoord).
\end{align}
Clearly, the isochronous metric defines an isochronous distance function 
\begin{align}
   \Delta \xi= \int d\xi =  \int \frac{d\xi}{d\zcoord} d\zcoord = \int \sqrt{g_{\zcoord \zcoord}} d\zcoord = \int \Jac d\zcoord.
\end{align}
Because the evolution of $\varyCoord$ is linear in time, the isochronous Jacobian plays a role that is similar to that of $\Jac^0_\tau$, discussed above.

\textbf{Isochronous Topology.}
Define the isochronous topology through the isochronous distance function.
In this topology, a 1D manifold breaks into multiple components depending on the number of fixed points of the vector field $V^\zcoord(\zcoord)$ and the topology of the original manifold.
If $V^\zcoord$ has no fixed points, then the manifold is homeomorphic to itself in the isochronous topology.

Any connected 1D real manifold must be homeomorphic to a circle, a line segment, or the real line.
To obtain an accurate count of the number of fixed points, assume that the original manifold is orbitally complete; i.e. that all orbits are contained within the manifold.
However, if $V^\zcoord$ has $\Nsingular$ isolated fixed points, then, in the isochronous topology, the manifold breaks into multiple components consisting of line segments and fixed points.
For example, if the original manifold is homotopic to a circle, the number of line segments equals $\Nsingular$, while, if the original manifold is a line segment, then the number of line segments is $\Nsingular-1$. 
Because there is no communication between these individual dynamically invariant components, they are disconnected in the isochronous topology and can be treated as separate sets.

\textbf{Isochronous Hilbert Space.}
To determine the Hilbert space of interest, one must first define the function spaces and their boundary conditions.
If the manifold in the isochronous topology is a circle or line segment, 
and periodic boundary conditions are imposed, then 
it is natural to use the standard Hilbert space of complex square-integrable functions, $\HilbertSpace^1=\LebesgueSpace^2$, on the respective spaces. 
For example, for the circle, $\Circle$, the inner product is defined via
\begin{align}\label{eq:inner-product_T1}
\innerprod{\phi}{\psi}=\oint_{\Circle} \phi^*(\xi)\psi(\xi) d\xi.
\end{align}
When the time interval for traversing the line segment is finite, then the Hilbert space is isomorphic to that of the circle.

However, if $\Vflow^\zcoord(\zcoord)$ has isolated nondegenerate fixed points, then, in the isochronous topology, the manifold is composed of disconnected subsets: the subset of discrete fixed points and the subset of line segments that are homeomorphic to the entire real line, each of which has infinite measure.
For each fixed point, $p$, the standard Hilbert space metric is simply $\innerprod{\phi}{\psi}_p=\phi^*(p)\psi(p)$, while for each line segment, $\Gamma$, the standard Hilbert space inner product is 
\begin{align}\label{eq:inner-product_R1}
\innerprod{\phi}{\psi}_\Gamma=\int_{\RealLine^1} \phi^*(\xi)\psi(\xi) d\xi.
\end{align}
Because these sets are disconnected in the isochronous topology, any weighted combination of these standard inner products can be used.

These considerations show that, whenever there are dynamically invariant sets that are homeomorphic to the entire real line, e.g. due to the presence of fixed points, then the infinite length of the line generates a continuous spectrum.
This can be handled with the well-equipped (fully-rigged) Hilbert space~\cite{Gelfand2014generalized} approach of nested function spaces.
In this setting, the appropriate set of test functions is the standard Schwartz space~\cite{Schwarts1950theorie} of rapidly decaying complex functions on the real line, $\SchwartzSpace(\RealLine^1)$.
The space that is dual to the test functions, $\SchwartzDual(\RealLine^1)$, under the inner product~\ref{eq:inner-product_R1} is the standard space of tempered distributions.
Hence, the Gelfand triple is the set of nested function spaces
\begin{align}
    \SchwartzSpace \subset H^1(\RealLine) \subset \SchwartzDual.
\end{align} 
For the problems of interest below, the dual basis consists of generalized eigendistributions.
As we shall see, these take the form of Fourier modes on the real line and Dirac point masses at the fixed points.

\textbf{Eigenfunctions / Eigendistributions.} 
For 1D spaces of finite measure, such as the circle and line, the advection equation has constant speed in the isochronous $\{\tau,\varyCoord\}$ coordinate system
\begin{align}
d\scalar/d\tau=(\partial_\tau+\partial_\varyCoord) \scalar = 0.
\end{align}
Because the constant flow is divergence free with respect to the isochronous measure, $d\xi$, the advection operator is skew-symmetric.
Choosing periodic boundary conditions (BCs) then yields a skew-adjoint operator.
Other types of BCs of interest can often be handled by determining the answer for periodic BCs and then using an appropriate linear combination of harmonics.

For periodic boundary conditions, the KvN eigenfunctions are simply Fourier harmonics 
\begin{align}
s_\spectralCoord(\tau,\xi)=e^{i\spectralCoord\varyCoord_0}=e^{i\spectralCoord(\varyCoord-\tau)}=e^{i\spectralCoord\varyCoord-i\omega \tau},
\end{align}
with wavenumber $\spectralCoord$ and eigenfrequency $\omega=\spectralCoord$.
For a circle or line segment, these are genuine eigenfunctions.
For the real line, these are eigendistributions.

For the fixed points themselves, the KvN equation is advection at zero speed. Assuming these are isolated fixed points, then the eigendistributions are Dirac delta functions at these points; i.e. for fixed point $\xi_\jindex$, the eigendistribution is $\delta(\xi-\xi_\jindex)$.

\textbf{Example: Linear Source/Sink.} 
Let us now examine the form of the solution near a source or sink, so that the velocity takes the form
\begin{align}
    \Vflow^\zcoord=\gamma \zcoord.
\end{align}
In this case, 
\begin{align}
\varyCoord_\tau-\varyCoord_0=\ln{(\zcoord/\zcoord_0)}=\gamma(\tau-\tau_0),
\end{align}
which can also be stated as
\begin{align}
\zcoord_\tau/\zcoord_0=e^{\varyCoord_\tau-\varyCoord_0}=e^{\gamma(\tau-\tau_0)}.
\end{align}
Given any initial scalar field $\scalar_0(\zcoord_0):=\scalar(\tau=0,\zcoord_0)$ we can determine the solution at the final time via
\begin{align}
\scalar(\tau,\zcoord_\tau) =\scalar( \tau=0,\zcoord_0)=s_0(\zcoord_0)=s_0(\zcoord_\tau e^{-\gamma \tau}).
\end{align}
Note, however, that because the transformation has a logarithmic singularity, there are four disjoint dynamically invariant sets that must be treated separately: the negative real line $\zcoord<0$, the fixed point at $\zcoord=0$, the positive real line $\zcoord>0$, and the fixed points at $\pm\infty$.

\textbf{Eigenfunctions in original coordinates.}
The same considerations also yield the solution to the KvN and Liouville equations.
In the original coordinates, the eigenfunctions for the Liouville and KvN equations carry powers of the Jacobian.
The key point is that, if the Jacobian of the transformation is 
\begin{align}
    \Jac(\zcoord)=d\varyCoord/d\zcoord=1/\Vflow^\zcoord(\zcoord),
\end{align}  
then the solutions to the Liouville equation for the PDF, $\pdf(\tau,\zcoord)$, equations is simply
\begin{align}
  \pdf(\tau,\zcoord) = \Jac(\zcoord) \scalar (\tau,\zcoord)= \Jac(\zcoord) \scalar (0,\zcoord_0),
\end{align}
where $\scalar(\tau,\zcoord)$ is a solution to the scalar advection equation~\ref{eq:scalar_advection}.
Using the same notation, the solution to the KvN equation,  $\psi(\tau,\zcoord)$, is simply
\begin{align}
\psi(\tau,\zcoord)=\left(\Jac(\zcoord)  \right)^{1/2} \scalar(\tau,\zcoord)=\left(\Jac(\zcoord)  \right)^{1/2} \scalar(0,\zcoord_0).
\end{align}
For example, near a source or sink, the Jacobian is $\Jac(\zcoord)=1/\Vflow^\zcoord(\zcoord)=1/\gamma \zcoord$.
Thus, the solutions and eigenfunctions for the KvN wavefunction and Liouville PDF simply have the form of a solution to the scalar advection equation that are then multiplied by $\zcoord^{-1/2}$ and $\zcoord^{-1}$, respectively.

More generally, assume there are $\Nsingular$ isolated nondegenerate singular points where the velocity either vanishes or blows up.
If one compactifies the real line, then this number should include the point(s) at infinity when it is also singular.
On the circle, there are $2\Nsingular$ invariant sets defined by the $\Nsingular$ singular points themselves and the $\Nsingular$ regions between the singular points.
One the extended real line, this increases to $2\Nsingular+1$ invariant sets because there are  $\Nsingular+1$ singular points.
The advection equation for each invariant set can be solved with a transformation to the new coordinate $\varyCoord$ in that particular region.
The Koopman eigenfunctions for each nontrivial invariant set is given by Fourier harmonics in coordinate $\varyCoord$.

The set of eigenfunctions always has the tensor product form discussed in \onlinecite{Giannakis24}.
However, since there are $2\Nsingular$ disjoint invariant sets, multiplying eigenfunctions from distinct regions yields a result that simply vanishes, so, in order to avoid over-counting, these should not be included.
In this case, $\Nsingular$ of the invariant sets are homeomorphic to $\Reals$, and starting from $\spectralCoord=1$, all eigenfunctions can be generated as powers of $(\scalar_1(\varyCoord))^\spectralCoord$ for $\spectralCoord\in\Reals$.
For the $\Nsingular$ fixed points, only the Dirac delta point mass is necessary, but considering fractional powers $0<\spectralCoord<1$ can still have an interesting interpretation (e.g., see the appendices of \onlinecite{Joseph23JPA}).
In fact, because an RKHS and an RKHA have well-defined reproducing kernels, fractional powers of such kernels are well-defined and have recently found important uses in developing efficient tensor network representations of the Koopman operator \cite{Giannakis24,Montgomery2025algebra}.

\subsubsection{Local I-LTS Decompositions on Manifolds \label{sec:local-ILTS}}
In this section, we prove a theorem showing that the I-LTS decomposition always exists for any autonomous vector field, $\Vflow(\zcoord)$, in any contractible region of a smooth  manifold.
In particular, this holds true for any region of $\Reals^\dim$ that is homeomorphic to the $\dim$-ball.
According to the lemma, the I-LTS decomposition requires at most $\dim$ components for $\dim$ spatial dimensions.
If the vector field is not autonomous, then we append time, $\tcoord$, to the coordinate system, $\zcoord=\{\tcoord,\Bz\}$, and solve for an autonomous flow in $\dim+1$ dimensions generated by the vector field  $V(\tau,\zcoord)=\Vflow^\tcoord(\zcoord)\{1,\BV(\zcoord)\}$.
Thus, a corollary of this lemma is that, in a contractible region of a smooth manifold, an I-LTS decomposition always exists for any non-autonomous vector field that requires at most  $\dim+1$ components for $\dim+1$ space-time dimensions.
The importance of this theorem is that an I-LTS decomposition can be much easier to work with than exact Koopman eigenfunctions.

\begin{theorem}[\bf Local I-LTS decompositions exist]
\label{thm:local_I-LTS}
Consider the flow of a vector field, $\Vflow=\BV\cdot \nabla$, on a real $\dim$-dimensional manifold, $M$.
In any contractible local chart, $\Chart$, with coordinates $\Bz=\{\zcoord^\jindex \}$, the flow can be expressed in this coordinate basis as $V=\sum_\jindex V^{\zcoord^\jindex}\partial_{\zcoord^\jindex}$.
Hence, $V$ can be expressed as the sum of $\dim$ integrable vector fields $V=\sum_\iltsindex V_\iltsindex$, where 
\begin{align}
V_\iltsindex=\delta_{\iltsindex}^\jindex V^{\zcoord^\jindex}(\Bz)\partial_{\zcoord^\jindex}
\end{align}
only retains the single component $\jindex=\iltsindex$ (no sum here).
This integrable decomposition determines a family of I-LTS product expansions for the evolution operator that determines the flow.
\end{theorem}
\begin{proof}
Consider a general autonomous vector field in a contractible region of a $\dim$-dimensional real manifold.
Within each coordinate chart of the manifold, with coordinates, $\Bz=\{\zcoord^\jindex\}$, the generator of the evolution operator has the form
\begin{align}
    \Vflow=\BV\cdot\nabla :=\sum_{\jindex=1}^\dim V^{\zcoord^\jindex}\partial_{\zcoord^\jindex}
    =:
    i\sum_\iltsindex \Kgen_\iltsindex.
\end{align}
In this case, for any choice of coordinates, one can determine an I-LTS decomposition by breaking the advection operator into the $d$ different individual parts that correspond to the different components of the vector field
\begin{align}
\Vflow_\iltsindex = i\Kgen_\iltsindex = \delta^\jindex_\iltsindex V^{\zcoord^\jindex}\partial_{\zcoord^\jindex}
\end{align}
(no sum intended).
Using the results for the 1D vector field above, for each individual term of the form $\Kgen_\iltsindex$, the solution is given by a coordinate transformation to 
\begin{align}
\varyCoord^\jindex_{\iltsindex,\tau} =\varyCoord^\jindex_{\iltsindex,0} - \delta^\jindex_\iltsindex  \int_0^{\zcoord^\jindex} d\zcoord^\jindex/V^\jindex(\tau,\zcoord)  =\varyCoord^\jindex_{\iltsindex,0} + (\tau-\tau_0) \delta^\jindex_\iltsindex .
\end{align}
Hence, for each coordinate system, we can identify the time-varying coordinate $\varyCoord_{\iltsindex,\tau}=\varyCoord^\iltsindex_{\iltsindex,\tau}$ and the constant coordinates, $\BconstantCoord_{\iltsindex,\tau }=\BconstantCoord_{\iltsindex,0 }=\{\varyCoord^{\jindex\neq \iltsindex}_{\iltsindex,\tau}\}=\{\varyCoord^{\jindex\neq \iltsindex}_{\iltsindex,0}\}$.
Thus, for each vector field $\Vflow_\iltsindex$, the solution in these coordinates is of the form  $\BtotalCoord_{\ell,\tau}=\{\varyCoord_{\ell,\tau},\BconstantCoord_{\ell,0}\}$

For a unitary LTS-decomposition, use any Lie-Trotter-Suzuki decomposition based on the individual factors, $\EvolutionOperator_\iltsindex(\Delta\tau)=e^{(\Vflow_\iltsindex^\dagger-\Vflow_\iltsindex)\Delta \tau/2}$. 
For example, for the first order LTS expansion,  choose $\EvolutionOperator_\LTS(\Delta\tau)=\prod_\ell \EvolutionOperator_\ell(\Delta\tau)$.
\end{proof}

\textbf{Scalar advection.}
In the new coordinate system given by $\BtotalCoord_\iltsindex=\{\varyCoord^\jindex_\iltsindex\}$, the solution to the scalar advection equation for each $\Vflow_\iltsindex$ is simply
\begin{align}
\scalar_\iltsindex(\tau,\BtotalCoord_{\iltsindex,\tau}) = \EvolutionOperator_{\iltsindex}^{(0)}(\tau) \scalar (0,\BtotalCoord_{\iltsindex,0}) = \scalar( 0,\BtotalCoord_{\iltsindex,0}) .
\end{align}
This can also be restated in the original coordinate system as
\begin{align}
\scalar_{\iltsindex }(\tau,\Bz_\tau) =\EvolutionOperator_{\iltsindex}^{(0)}(\tau) \scalar_\iltsindex(0,\Bz_0)=\scalar_{\iltsindex}\left(0,\Bz_0(\BtotalCoord_{\iltsindex,0})\right).
\end{align}

\textbf{Local Eigendecomposition.}
For each integrable flow $\Vflow_\iltsindex$, the building blocks of the Koopman eigenfunctions are simply Fourier harmonics in the $\BtotalCoord$ coordinates
\begin{align}
\scalar_{\iltsindex,\BspectralCoord}(\tau,\BtotalCoord) 
&=\scalar_{\iltsindex,\BspectralCoord}(0,\BtotalCoord_{\iltsindex,0})
=e^{i \BspectralCoord \cdot \BtotalCoord_{\iltsindex,0}}\\
&=e^{i \BspectralCoord \cdot \BtotalCoord_{\iltsindex,\tau} -i\spectralCoord_\iltsindex \tau}
=
e^{i \BspectralCoord \cdot \BtotalCoord_{\iltsindex,\tau} -i\omega^\iltsindex \tau},
\end{align}
where, for each operator $\Kgen_\iltsindex$, the local eigenfrequencies are simply $\omega^\jindex_\iltsindex=\delta^\jindex_\iltsindex \spectralCoord_\jindex$.
As discussed in Sec.~\ref{sec:ILTS-decomposition}, 
it is actually more convenient to use the eigendistribution representation
\begin{align}
\scalar_{\iltsindex,\spectralCoord,\BconstantCoord_0}(\tau,\BtotalCoord) 
&=\delta^{\dim-1} (\BconstantCoord_\iltsindex - \BconstantCoord_0) e^{i \delta^\jindex_\iltsindex\spectralCoord_\jindex  \varyCoord_{\iltsindex,0}^\jindex  }\\
&=\delta^{\dim-1} (\BconstantCoord_\iltsindex - \BconstantCoord_0) e^{i  \spectralCoord_\iltsindex (\varyCoord_{\iltsindex,\tau}^\iltsindex -  \tau) }
.
\end{align}

\textbf{Global Eigendecomposition.} In order to determine a global Koopman eigenmode over the entire manifold, one must use linear combinations of these local eigenmodes to match boundary conditions over the transition regions between charts and to satisfy the global boundary conditions. 
For example, satisfying global boundary conditions will require satisfying some form of periodicity around closed loops  in phase space, etc.

For any completely integrable flow, this can be performed with relatively little computational cost thanks to the simple expression of the eigenmodes in terms of Fourier harmonics.
Thus, for the I-LTS decomposition, this computation can be performed for each of the integrable subflows, $\Vflow_\iltsindex$, with relatively little cost.

However, since the original flow, $\Vflow$, and the approximate flow $\Vflow_\LTS$, are both generically chaotic, determining their eigenmode decomposition is much more challenging because the spectrum of a chaotic flow is continuous.
For example, a chaotic flow has a dense set of periodic orbits, and, each periodic orbit defines a valid generalized eigendistribution. 
Yet, the I-LTS decomposition implies that there is an efficient computational algorithm for applying the evolution operator to a vector and, hence, an efficient computational method for determining the eigenmodes using standard numerical methods.

\subsection{Complexity of the I-LTS Decomposition \label{sec:ILTS-complexity}}
The LTS operator product expansion of the propagator indicates that there are multiple forms of complexity.
First, there are different integrable components, $\{\Vflow_\iltsindex\}$, that sum to form the generator.
Second, an ordering must be chosen for the individual integrable evolution operators in the LTS product formula, $\{e^{-\tau\Vflow_\iltsindex}\}$. 
Third, there are different regions, $\{\Chart_\rindex\}$, that may need to be treated separately.
Fourth, boundary conditions must be applied globally in order to determine global eigenfunctions.

In fact, the coordinate transformations, $\BtotalCoord_\iltsindex =
\{\BvaryCoord_\iltsindex,\BconstantCoord_\iltsindex\}=\Phi_\iltsindex(\Bz)$, present another important source of complexity.
An integrable system can have countably many distinct dynamically invariant regions that must be treated separately.
For example, an autonomous Hamiltonian system possesses separatrices that separate the maxima and minima of the Hamiltonian function. 
Each dynamically invariant region, whose boundary is defined by the separatrices between the different regions, must be handled with a separate coordinate transformation.
This gives the Koopman, KvN, and Perron-Frobenious operators a direct sum structure that represents a fundamental source of complexity.
From this perspective, if one attempts to approximate a chaotic dynamical system with a series of integrable dynamical systems, it is the self-similar structure of the dynamically invariant regions that represents the fundamental computational challenge because they are knotted together with ever-increasing complexity as the winding numbers are allowed to increase. 

\section{Quantum I-LTS Algorithm for the Stepanoff Flow}\label{sec:QI-LTS}
In this section, we derive and simulate a quantum algorithm for the Stepanoff flow based on the I-LTS decomposition that is exponentially efficient in complexity per time step. 
In fact, this exponential savings holds even for an explicit end-to-end algorithm that includes the measurement of a fixed number of physical observables.
First, in Sec.~\ref{sec:Stepanoff-theory}, we introduce the Stepanoff flow and  reformulate the system through an integrable decomposition to simplify its encoding into a quantum circuit.
Then, in Sec.~\ref{sec:QHS} we present an efficient quantum circuit for performing QHS of the Stepanoff flow based on the I-LTS decomposition.
Notably, the complexity of the I-LTS decomposition only grows quadratically with the number of qubits per dimension, while the complexity of a standard Pauli decomposition grows exponentially fast, in fact, faster than linearly with the number of states.
Finally, in Sec.~\ref{sec:ae} amplitude estimation is used to determine the expectation values for observables over the probability density function after evolution by the Stepanoff flow.

\begin{figure}[!t]
\centering
\subfloat[]{\includegraphics[width=0.25\textwidth]{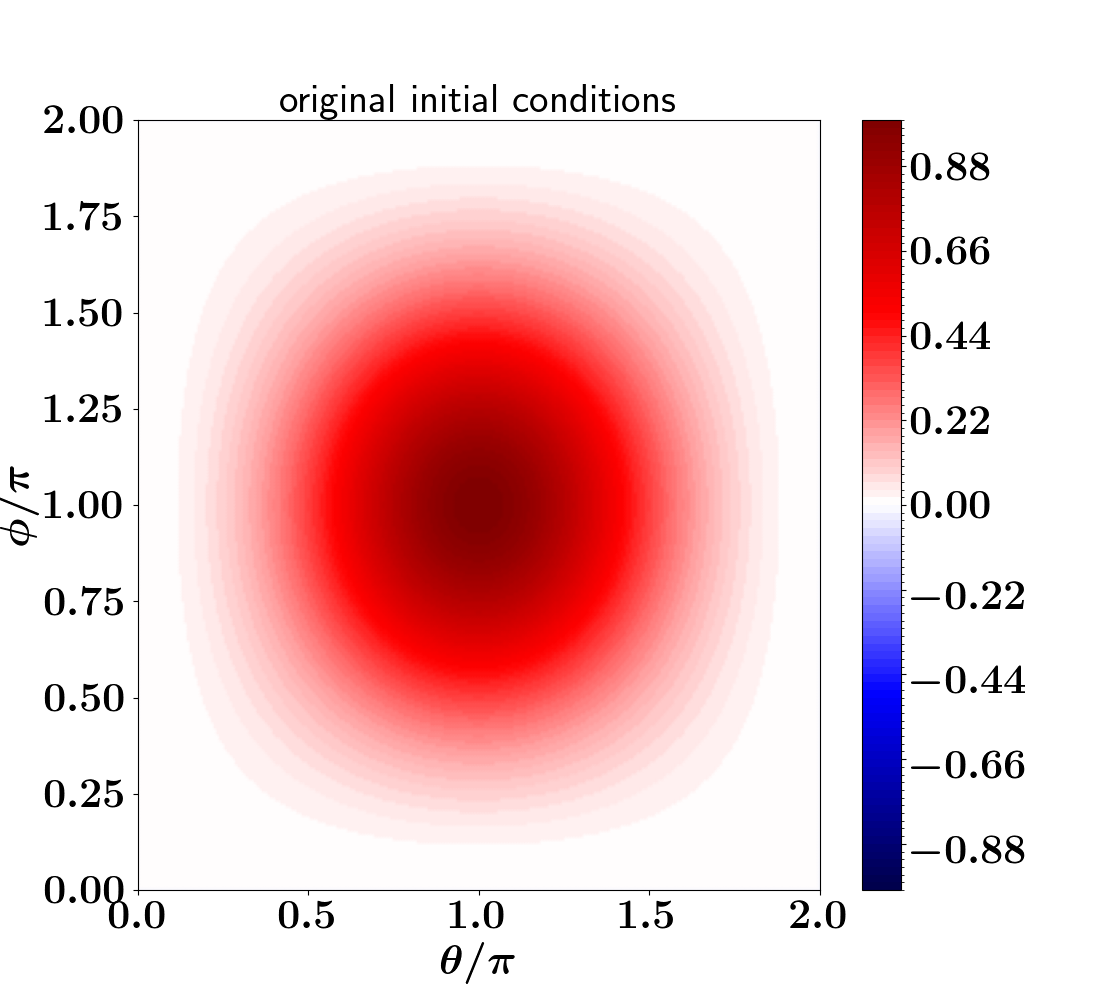}}
\subfloat[]{\includegraphics[width=0.25\textwidth]{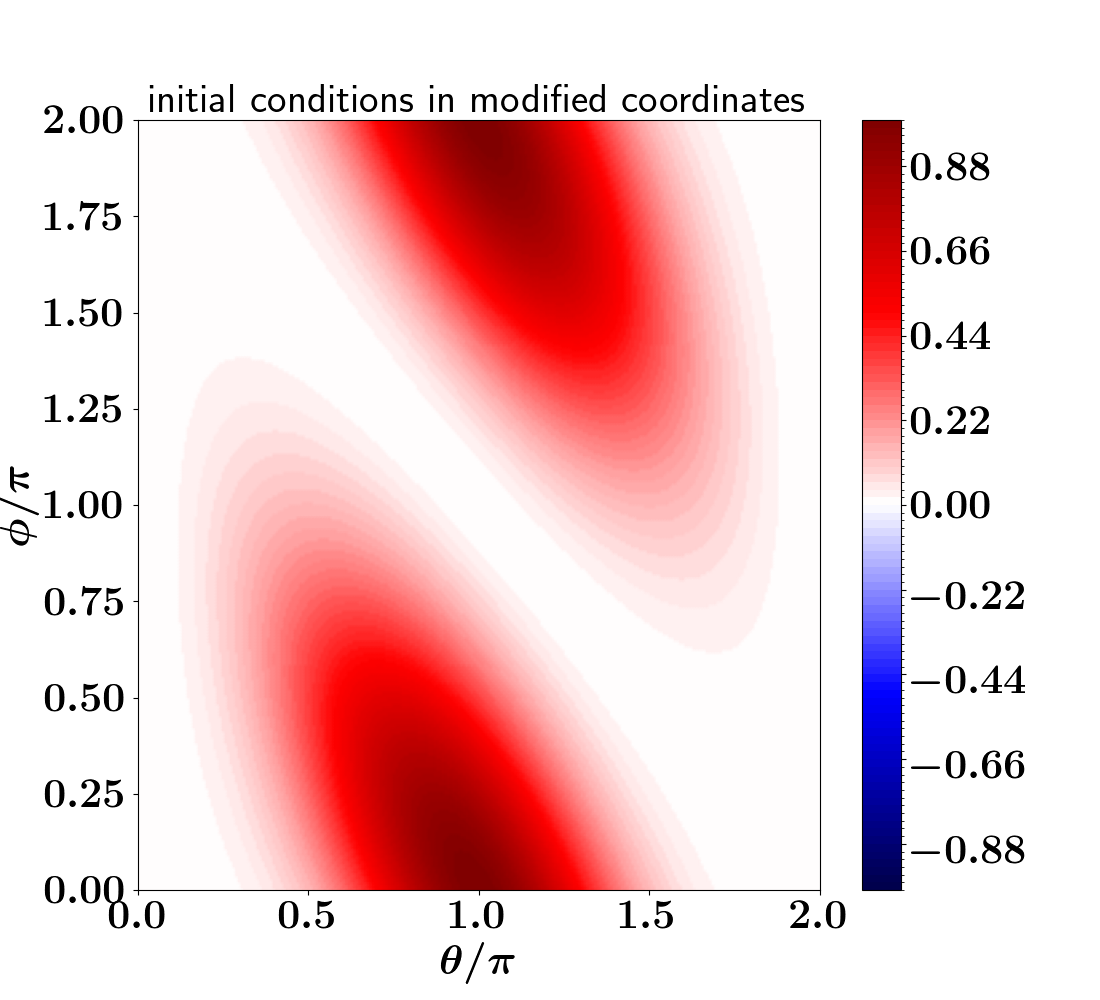}}

\caption{
    \label{fig:initial-conditions}
    (a) The initial condition ~\eqref{eq:init} for $\psi(\ytheta,\yphi,t=0)$.
    (b) The initial condition ~\eqref{eq:init-mod} for $\psi(\yTheta,\yPhi,t=0)$ in the modified coordinates ~\eqref{eq:theta-new} where $\yTheta=\ytheta+\yphi$ and $\yPhi=\yphi$.
}
\end{figure}

\subsection{The Stepanoff Flow}\label{sec:Stepanoff-theory}

\subsubsection{Real space representation}
\label{sec:Stepanoff_real-space}
We consider the temporal evolution of the distribution $\psi(\ytheta, \yphi, t)$ under the Stepanoff flow $\vv{V}$ on the torus~\cite{Stepanoff36, Oxtoby53, Giannakis24},
\begin{subequations}\label{eq:sys-lagrangian}
\begin{eqnarray}
    &&\frac{\diff \ytheta}{\diff t} = \Vtheta,\quad \frac{\diff \yphi}{\diff t} = \Vphi,\\
    &&\Vtheta = \Vphi + \beta (1 - \cos\yphi),\\
    &&\Vphi = \alpha [1 - \cos(\ytheta-\yphi)].
\end{eqnarray}
\end{subequations}
This can also be described by the Hamiltonian
\begin{equation}
    H =  
    \alpha \left[\ytheta - \yphi - \sin{(\ytheta-\yphi)} \right]
    -\beta \left[\yphi - \sin{\yphi}\right]
\end{equation}
with the Poisson bracket $\left\{ \yphi,\ytheta\right\}=1$, so that $\ytheta$ is the momentum conjugate to $\yphi$.
For simplicity, in the following, we consider the case where $\beta = 1 - \alpha$.

The Eulerian linear form of the Stepanoff flow, i.e. the Koopman -- von Neumann (KvN) embedding\cite{Joseph20}, can be written as
\begin{equation}\label{eq:sys-KvN}
     \partial_t \psi  = -\BV\cdot\nabla\psi,\quad \BV\cdot\nabla = \Vtheta \partial_\ytheta + \Vphi \partial_\yphi.
\end{equation}
As an initial state of the KvN wavefunction, we take the following localized periodic function:
\begin{subequations}\label{eq:init}
\begin{eqnarray}
    &&\psi(\ytheta, \yphi, 0) = \frac{g(\ytheta)}{g(\pi)} \frac{g(\yphi)}{g(\pi)},\\
    &&g(\ytheta) = \exp\left(\kappa \sin(\ytheta/2)^2 \right) - 1.\label{eq:init-1D}
\end{eqnarray}
\end{subequations}
The parameters $\alpha = \sqrt{20}$ and $\kappa = 1$ are used in all numerical simulations.
These initial conditions are shown in Fig.~\ref{fig:initial-conditions}(a).

\begin{figure}[!t]
\centering
\subfloat{\includegraphics[width=0.49\textwidth]{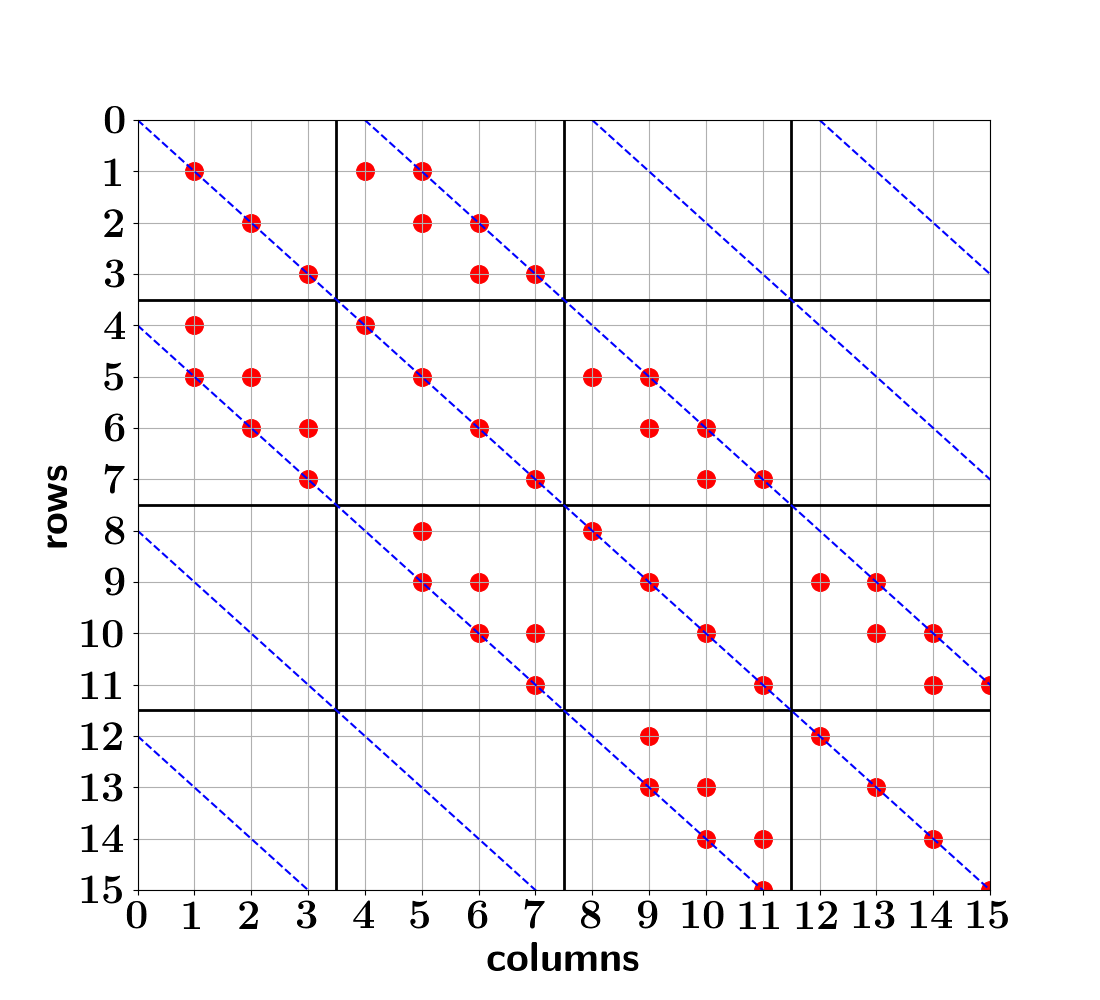}}
\caption{
    \label{fig:matrix-M}
    Structure of the matrix $\hat{K}$, Eq.~\eqref{eq:M}, showing the positions of the matrix nonzero elements in the case of $N = 4$.
    The black lines indicate the $N\times N$ blocks, while the blue lines highlight the main diagonals within each block.
}
\end{figure}

\subsubsection{Fourier space representation} 
\label{sec:Stepanoff_Fourier-space}
An alternative approach proposed in Ref.~\onlinecite{Giannakis24} is to simulate the system in Fourier space.
The real space coordinates, $(\ytheta, \yphi)$, are discretized using $N$ grid points in each dimension, and an integer index, $x$, is used to arrange the resulting $N^2$ elements of the discretized density $\psi_x$ into the form of a vector:
\begin{subequations}\label{eq:discr}
\begin{eqnarray}
    &&\psi_x(t) \equiv \psi(\ytheta_{x_1}, \yphi_{x_2}, t),\quad x = x_2 N + x_1,\quad N = 2^n,\\
    &&x_1 = 0, 1, \dots (N-1),\quad  x_2 = 0, 1, \dots (N-1),\\
    &&\ytheta_{x_1} = \Delta\ytheta x_1,\quad \Delta\ytheta = 2\pi/(N-1),\\
    &&\yphi_{x_2} = \Delta\yphi x_2,\quad \Delta\yphi = 2\pi/(N-1).
\end{eqnarray}
\end{subequations}
The corresponding generator of the dynamical system in Fourier space is
\begin{equation}\label{eq:Fourier}
    \hat{K}_{rc} =\langle e^{ \yi \vv{r}\cdot \vv{\ytheta}} | \yi \vv{V}\cdot \nabla |e^{\yi \vv{c}\cdot \vv{\ytheta}}\rangle
    =  \langle e^{-\yi \vv{r}\cdot \vv{\ytheta}} \yi \vv{V}\cdot \nabla e^{\yi \vv{c}\cdot \vv{\ytheta}} \rangle_{\theta,\phi} ,
\end{equation}
where $\vv{\ytheta} = (\ytheta,\yphi)$ and $\langle\cdot\rangle_{\theta,\phi}$ indicates the average over both angles. 
The vectors $\vv{r} = (r_1, r_2)$ and $\vv{c} = (c_1, c_2)$ store row and column indices, respectively, to indicate positions in the discretized Fourier space.
The scalar row and column indices, $r$ and $c$, are
\begin{subequations}\label{eq:r}
\begin{eqnarray}
    &&r = (r_2 + N/2 )N + (r_1 + N/2),\\
    &&c = (c_2 + N/2) N + (c_1 + N/2),
\end{eqnarray}
\end{subequations}
where $r_1, r_2, c_1, c_2 = -N/2, - N/2+1,\dots , 0, 1,\dots, N/2-1$.
As a result, the discretized Stepanoff flow in Fourier space is described by the $N^2\times N^2$ matrix
\begin{align}\label{eq:M}
    \hat{K}_{r,c} &= \ed_{c_1, c_2}\delta_{r_1, c_1}\delta_{r_2, c_2} \nonumber\\
    &+\es_{c_1, c_2} \bigg(\delta_{r_1, c_1 + 1}\delta_{r_2, c_2-1} + \delta_{r_1, c_1-1}\delta_{r_2, c_2+1} \bigg)\nonumber\\
    &+\ea_{c_1} \delta_{r_1, c_1} \bigg(\delta_{r_2, c_2-1} + \delta_{r_2, c_2+1}\bigg),
\end{align}
where the following discrete functions are used to describe the matrix elements:
\begin{subequations}
\begin{eqnarray}
    &&\ed_{c_1, c_2} = -(c_1 + \alpha c_2),\\
    &&\es_{c_1, c_2} = \frac{\alpha}{2}(c_1 + c_2),\\ 
    &&\ea_{c_1} = \frac{1-\alpha}{2} c_1.
\end{eqnarray}
\end{subequations}
The dynamics of this system is described by 
\begin{equation}\label{eq:hs-fourier}
    \psi(t) = \yFou^{-1} e^{-\yi\hat{K}t}\yFou\psi(0),
\end{equation}
where $\yFou$ denotes the Fourier transform with respect to both $\ytheta$ and $\yphi$.
As seen from Fig.~\ref{fig:matrix-M}, the matrix $\hat{K}$ can be split into $N\times N$ blocks where each block contains $N\times N$ elements.
The generator is Hermitian and sparse. 
This means that system~\eqref{eq:hs-fourier} can potentially be simulated using a standard QSP circuit.
However, the matrix~\eqref{eq:M} poses challenges for quantum circuit encoding due to its non-diagonal elements, which depend on the monotonically increasing indices $c_j$.

\begin{figure*}[!t]
\centering
\subfloat{\includegraphics[width=0.33\textwidth]{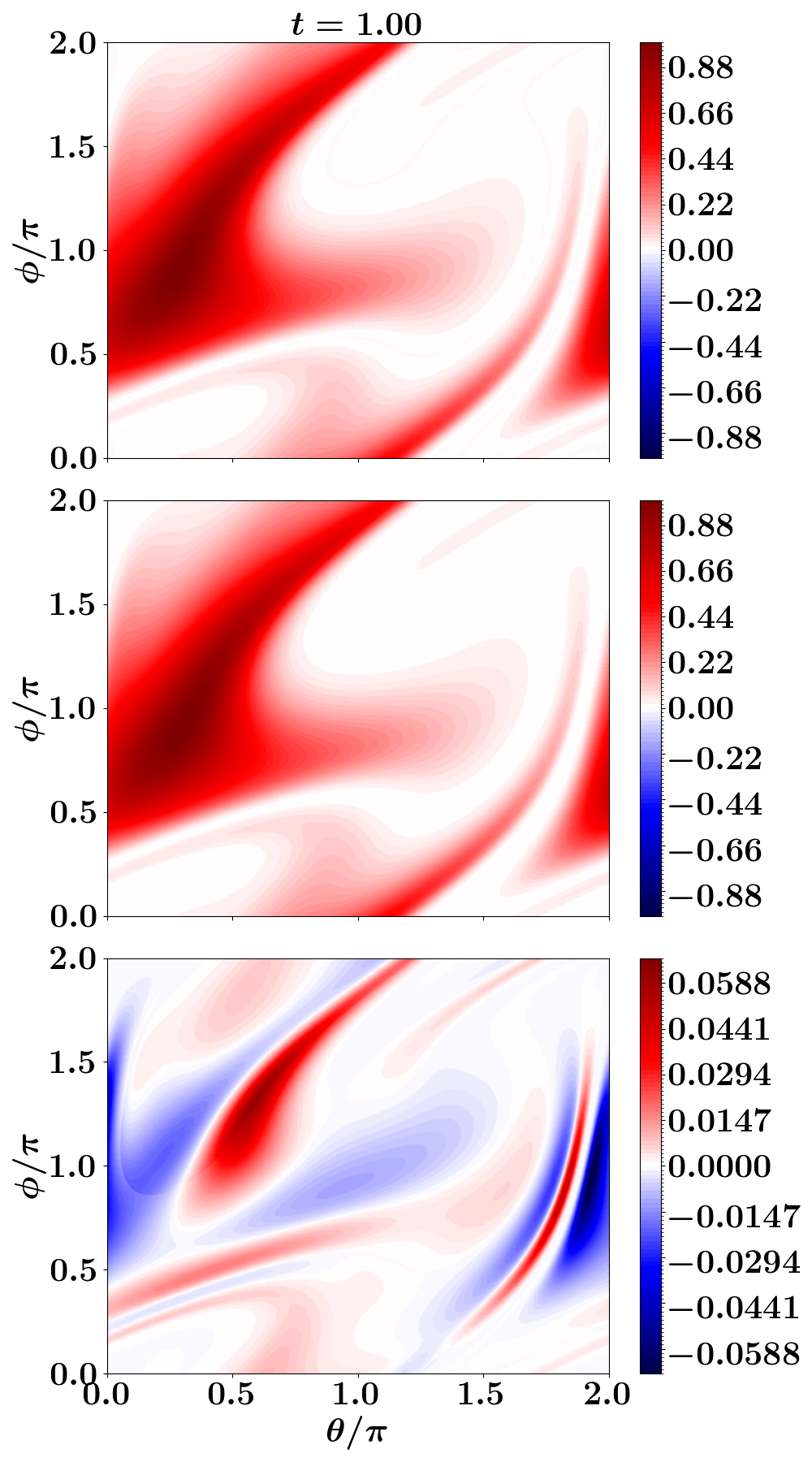}\label{fig:comp-cl-dt001-t1}}
\subfloat{\includegraphics[width=0.33\textwidth]{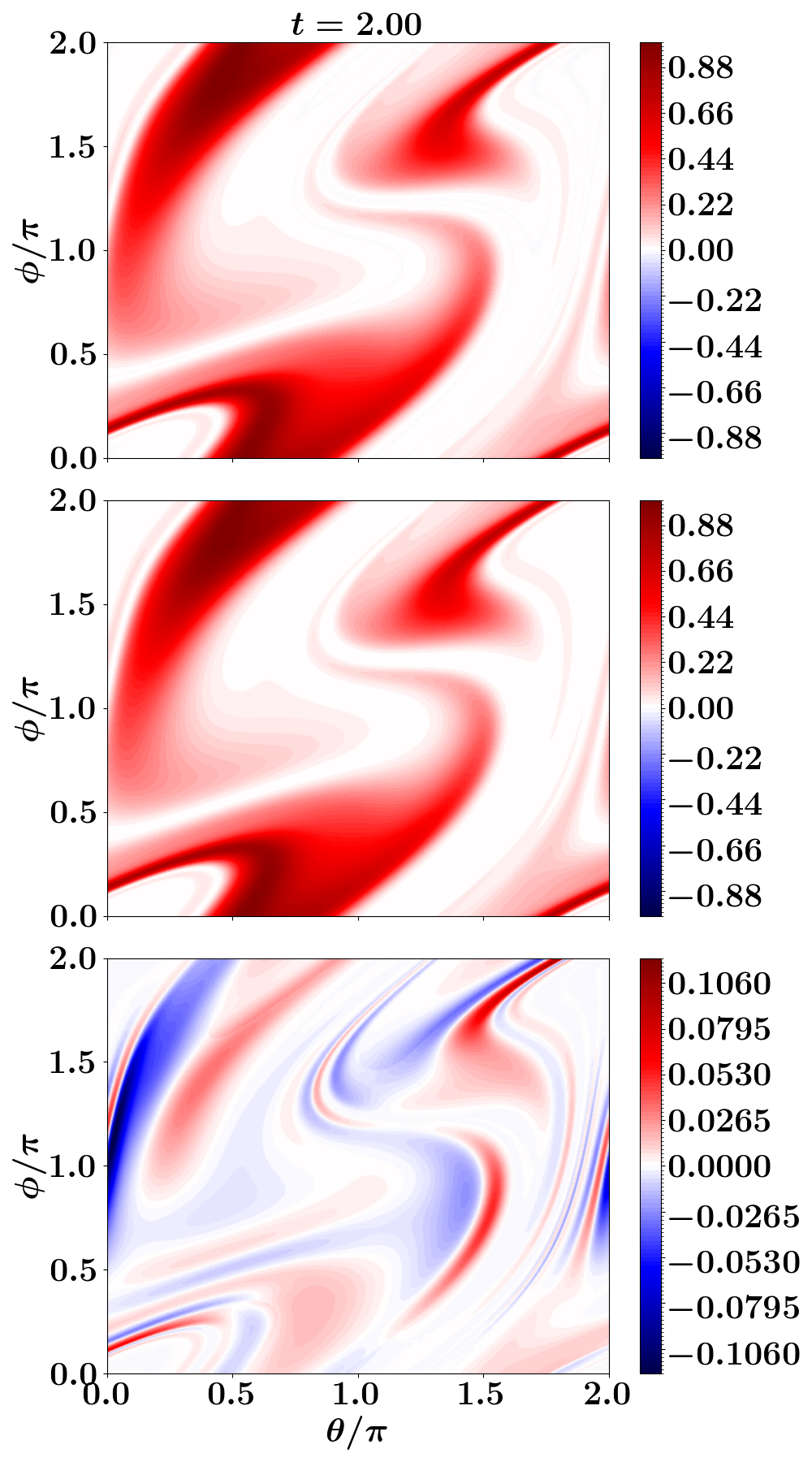}\label{fig:comp-cl-dt001-t2}}
\subfloat{\includegraphics[width=0.33\textwidth]{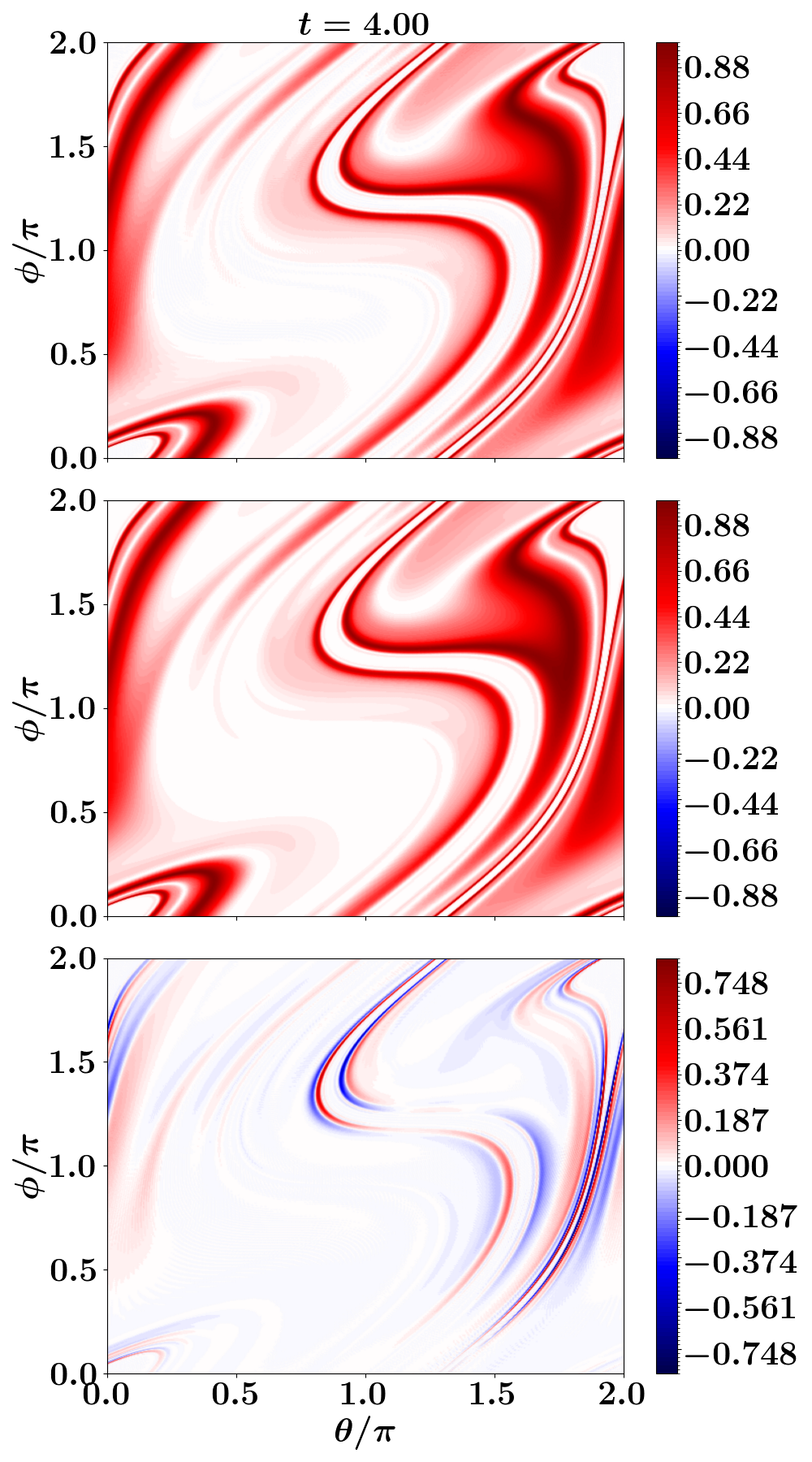}\label{fig:comp-cl-dt001-t4}}
\caption{
    \label{fig:comp-cl-dt001}
    Top row: the distribution $\psi(\ytheta, \yphi, t)$ at various time points $t$, obtained from classical simulations in real-Fourier space, Eq.~\eqref{eq:hs-real-fourier}, with $\Delta t = 0.01$. 
    The distribution is normalized to one.
    Middle row: results from classical simulations in real space, Eq.~\eqref{eq:sys-lagrangian}.
    Bottom row: the difference between the classical simulations in real space and real-Fourier space.
    In all simulations, the parameters $n = 9$, $\kappa = 1$, and $\alpha = \sqrt{20}$ are used.
}
\end{figure*}

\subsubsection{Restoring symmetry with a coordinate transformation} \label{sec:Stepanoff_mod-space}
To simplify the encoding, we perform the following coordinate transformation:
\begin{equation}\label{eq:theta-new}
    \yTheta = \ytheta - \yphi,\quad \yPhi = \yphi.
\end{equation}
This allows for localizing the dependence on each angle within each flow component: 
\begin{subequations}\label{eq:V1V2-new}
\begin{eqnarray}
    && \frac{\diff \yTheta}{\diff t} = \VTheta = \Vtheta - \Vphi = \beta (1 - \cos\yPhi),\\
    && \frac{\diff \yPhi}{\diff t} = \VPhi = \alpha (1 - \cos\yTheta).
\end{eqnarray}
\end{subequations}
This can be derived from the Hamiltonian
\begin{equation}
    H=\alpha\left(\yTheta-\sin{\yTheta}\right)
    -\beta\left(\yPhi-\sin{\yPhi}\right)
\end{equation}
with the Poisson bracket $\{\yPhi,\yTheta\}=1$, so that $\yTheta$ is the momentum canonical to $\yPhi$.
Again, for simplicity, we only consider the case $\beta=1-\alpha$ below.

In the new coordinates~\eqref{eq:theta-new}, the KvN equation is simply
\begin{equation}
 \partial_t \psi =-\vv{V}\cdot \nabla \psi,
\quad \vv{V}\cdot\nabla=\VTheta \partial_\yTheta \psi + \VPhi \partial_\yPhi \psi 
\end{equation}
The matrix elements in Fourier space are recast as
\begin{equation}
    \hat{M}_{rc} 
    = \langle e^{ \yi \vv{r}\cdot \vv{\yTheta}} | \yi \vv{V}\cdot \nabla | e^{\yi \vv{c}\cdot \vv{\yTheta}} \rangle
    =\langle e^{-\yi \vv{r}\cdot \vv{\yTheta}}  \yi \vv{V}\cdot \nabla e^{\yi \vv{c}\cdot \vv{\yTheta}} \rangle_{\theta',\phi'},
\end{equation}
where 
$\vv{\yTheta} = (\yTheta, \yPhi)$ and
\begin{align}
    \vv{V}\cdot \nabla e^{\yi \vv{c}\cdot \vv{\yTheta}}
    &= \VTheta\, \frac{\partial e^{\yi \vv{c}\cdot \vv{\yTheta}} }{\partial\yTheta}        
         + \VPhi\, \frac{\partial e^{\yi \vv{c}\cdot \vv{\yTheta}} }{\partial\yPhi} \nonumber\\
    &= \yi c_1 \VTheta e^{\yi \vv{c}\cdot \vv{\yTheta}} + \yi c_2 \VPhi e^{\yi \vv{c}\cdot \vv{\yTheta}}.
\end{align}
Using Eqs.~\eqref{eq:V1V2-new}, one obtains
\begin{subequations}
\begin{eqnarray}
    &&c_1 \VTheta = -(1-\alpha) c_1 + \frac{1-\alpha}{2} c_1 \left(e^{\yi\yPhi} + e^{-\yi\yPhi}\right),\\
    &&c_2 \VPhi = -\alpha c_2 + \frac{\alpha}{2} c_2 \left(e^{\yi\yTheta} + e^{-\yi\yTheta}\right).
\end{eqnarray}
\end{subequations}
The generator in the modified discretized Fourier space takes the form
\begin{align}\label{eq:M-fourier-mod}
    \hat{M}_{rc} = 
        &-\left((1-\alpha)c_1 + \alpha c_2\right)\delta_{r_1, c_1}\delta_{r_2, c_2} \\
        &+ \frac{1-\alpha}{2} c_1 \delta_{r_1, c_1} (\delta_{r_2, c_2+1} + \delta_{r_2, c_2-1})\\
        &+ \frac{\alpha}{2} c_2 (\delta_{r_1, c_1+1} + \delta_{r_1, c_1-1})\delta_{r_2, c_2}.
\end{align}
In modified coordinates~\eqref{eq:theta-new}, the initial condition~\eqref{eq:init} becomes
\begin{subequations}\label{eq:init-mod}
\begin{align}
     \psi'(\yTheta, \yPhi, 0) &= \psi(\yTheta+\yPhi, \yPhi, 0)
     =\frac{g(\yTheta+\yPhi)}{g(\pi)} \frac{g(\yPhi)}{g(\pi)},\\
    g(\yTheta+\yPhi) &=\exp\left(\kappa \sin\left[\frac{\yTheta + \yPhi}{2}\right]^2 \right) - 1,\label{eq:init-1D-mod}
\end{align}
\end{subequations}
where $g(\cdot)$ is defined in Eq.~\eqref{eq:init-1D}.

As seen from Eq.~\eqref{eq:M-fourier-mod}, the matrix still contains non-diagonal elements that are difficult to encode into a quantum circuit.
Recently, Ref.~\onlinecite{Low25} introduced straightforward methods for encoding this type of behavior using additional ancillary registers and logical comparison operators. 
Unfortunately, this approach can be expensive and, depending on context, it should be advantageous to eliminate this step~\cite{Novikau25Opt}.

\subsubsection{I-LTS representation in mixed real space \& Fourier space}
\label{sec:Stepanoff_mixed-space}
The Stepanoff flow can be represented as the sum of the two components $\VTheta(\yPhi)$ and $\VPhi(\yTheta)$ given in Eq.~\ref{eq:V1V2-new}, each of which is completely integrable.
Each flow is well-defined over the entire phase space, so only one I-LTS decomposition is required in this case.
As discussed in Section~\ref{sec:ILTS_evolution-operator}, this leads to a simple LTS decomposition if we write each operator in real space for one coordinate and in Fourier space for the canonically conjugate coordinate:

\begin{subequations}
\begin{align}
     \yMM{\yTheta}_{r_1, x_2, c_1, y_2}
     &= 
     \langle e^{ \yi r_1 \yTheta} |\yi \VTheta\partial_{\yTheta} | e^{\yi c_1\yTheta} \rangle      
     \\&= \langle e^{-\yi r_1 \yTheta} \yi \VTheta\partial_{\yTheta}e^{\yi c_1\yTheta} \rangle_\yTheta,
    \\
      \yMM{\yPhi}_{x_1, r_2, y_1, c_2} &= 
     \langle e^{\yi r_2 \yPhi}|  \yi \VPhi\partial_{\yPhi} |e^{\yi c_2\yPhi} \rangle \\
    &= \langle e^{-\yi r_2 \yPhi}  \yi \VPhi\partial_{\yPhi} e^{\yi c_2\yPhi} \rangle_\yPhi
\end{align}
\end{subequations}
where, now, $\langle\cdot\rangle_\yTheta$ indicates an average over  angle $\yTheta$, and similarly for $\yPhi$.
In these expressions, $r_j,c_j$ refer to the Fourier harmonics for the $j$'th angle while $x_j,y_j$ refer to the position index for the $j$'th angle.
Taking into account Eqs.~\eqref{eq:V1V2-new}, one has
\begin{equation}\label{eq:M1-real-fourier}
    \yMM{\yTheta}_{r_1, x_2, c_1, y_2} = -(1 - \alpha) c_1 \left(1 - \cos\yPhi_{x_2}\right)
        \delta_{r_1, c_1}\delta_{x_2,y_2},
\end{equation}
and
\begin{equation}\label{eq:M2-real-fourier}        
    \yMM{\yPhi}_{x_1, r_2, y_1, c_2} = -\alpha c_2  \left(1 - \cos\yTheta_{x_1}\right)
        \delta_{x_1,y_1}\delta_{r_2, c_2}.
\end{equation}
In this way, the generator is split into a sum of two distinct diagonal matrices, each operating in its own independent subspace. 
\emph{Thus, the I-LTS decomposition has eliminated the off-diagonal elements that appear in both the real space and Fourier space form of the operators.
}

To perform Hamiltonian simulation using these disjoint generators, we simply discretize the time interval $t$ into $N_t$ steps of length $\Delta t$, and apply LTS operator splitting to simulate the system dynamics in the form
\begin{subequations}\label{eq:hs-real-fourier}
\begin{eqnarray}
    &&\psi'(t) = \left(\prod^{N_t}_{j=1} U_{\Delta t}\right) \psi'(0),\\
    &&U_{\Delta t} = \yFou_{\yPhi}^{-1} e^{-\yi\yMM{\yPhi} \Delta t} \yFou_{\yPhi}\yFou_{\yTheta}^{-1} e^{-\yi\yMM{\yTheta}\Delta t}\yFou_{\yTheta},
\end{eqnarray}
\end{subequations}
where $\yFou_{\yTheta}$ and $\yFou_{\yPhi}$ denote the Fourier transforms with respect to $\yTheta$ and $\yPhi$, respectively.

\begin{figure*}[!t]
\centering
\includegraphics[width=0.80\textwidth]{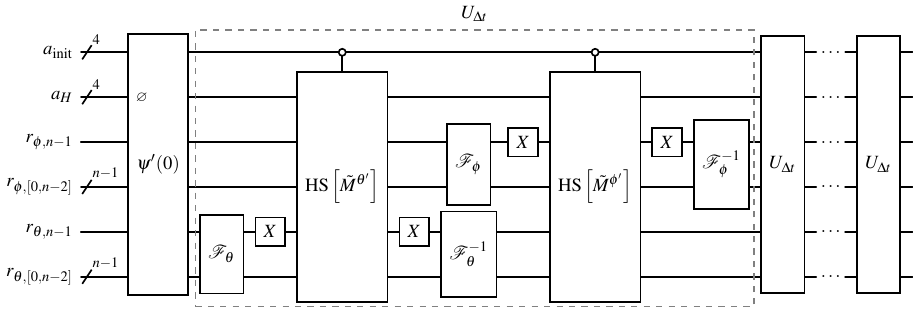}
\caption{
    \label{circ:scheme-mix}
    Schematic circuit performing QHS of the Stepanoff flow in real-Fourier space, Eq.~\eqref{eq:hs-real-fourier}.
    The symbol $\varnothing$ indicates that the corresponding operator does not act on that register.
    The circuit for the initialization operator $\psi'(0)$ is shown in Fig.~\ref{circ:init}.
    The circuit for the QHS operator HS$\left[\yMM{k}\right]$ is shown in Fig.~\ref{circ:enc-diag-M1}.
}
\end{figure*}

\subsubsection{Classical simulation results}
\label{sec:Stepanoff_classical-results}
Figure~\ref{fig:comp-cl-dt001} compares classical simulations of the evolution of a KvN wavefunction using two different representations.
In all cases, the wavefunction is assumed to start with the initial conditions in~\eqref{eq:init} and the evolution is simulated until time $t=1,2,4$.
The top row shows the evolution simulated using the linear mixed real-Fourier space representation, Eq.~\eqref{eq:hs-real-fourier}, derived in Sec.~\ref{sec:Stepanoff_mixed-space}.
In this case, $n = 9$ so that there are $N=2^9=512$ Fourier modes for each angle, and the time step is $\Delta t = 0.01$.
The middle row shows the evolution simulated using the standard transfer operator approach discussed in Sec.~\ref{sec:Stepanoff_real-space} (middle row) on a fine spatial grid with $N=512$ points in each direction.
The bottom row displays the difference between solutions. 
The explicit implementation for these cases can be found in ~\onlinecite{StepanoffSource}.
A similar comparison between the real-space transfer operator approach and the Fourier-space transfer operator approach of Sec.~\ref{sec:Stepanoff_Fourier-space} is shown in \onlinecite{Giannakis24} (see their Fig.~X).

The standard transfer operator approach~\cite{Giannakis24} works in the following manner.
Since the value of the wavefunction is constant in time, to find the value at any final position, one can simply integrate the ODEs backwards in time to find the initial position. 
Then, the value is found by evaluating the initial conditions in Eq.~\eqref{eq:init} at the initial position.
In this case, the Dormand-Prince $5$th order ODE solver is used for time integration.
While the transfer operator is linear, this calculation relies on nonlinear operations for the integration of the ODEs.
While both methods yield a unitary transfer operator, this intrinsically nonlinear method is more accurate than the linear method because the motion of the trajectories is not limited by the spatial resolution of the grid.
Since the wavefunction is well-resolved at this relatively high resolution of $N=512$, the bottow row essentially shows the error of the classical linear method. 

\subsection{Quantum Hamiltonian Simulation (QHS) Algorithm for the Stepanoff Flow}\label{sec:QHS}

\subsubsection{Overview of the QHS circuit}
To derive a quantum algorithm for Hamiltonian simulation of the Stepanoff flow, we employ the linear approach of the mixed real space-Fourier space representation with a first-order Trotter-Suzuki decomposition of the generator in Eq.~\eqref{eq:hs-real-fourier}.
A schematic circuit for performing QHS of this model is shown in Fig.~\ref{circ:scheme-mix}.
The circuit illustrates the principle steps of Eq.~\eqref{eq:hs-real-fourier}, including initializing the system in real space, performing QHS for each time step by transforming to Fourier space for one coordinate while keeping the other in real space, and, so on, until the desired time is reached.

In this circuit, the register $a_{\rm init}$ contains all ancillary qubits required for state initialization, while the register $a_H$ contains all ancillary qubits required to perform QHS.
The registers $r_{\ytheta}$ and $r_{\yphi}$ are used to encode the position information in either real or Fourier space for the coordinates $\ytheta$ and $\yphi$, respectively.
The operator denoted as $\psi'(0)$ computes the initial state vector defined in Eq.~\eqref{eq:init-mod}.
The construction of $\psi'(0)$ is described in Sec.~\ref{sec:be-init}.
The operator HS$\left[\yMM{k}\right]$, which performs QHS with the generator $\yMM{k}$ over the time interval $\Delta t$, is described in Sec.~\ref{sec:enc-Hamiltonian}.
The additional $X$ gates, used in conjunction with the QFT operators $\mathcal{F}_{\ytheta}$ and $\mathcal{F}_{\yphi}$, are necessary to arrange the Fourier spectrum to match the ordering of the matrix elements in $\yMM{k}$.

The QHS circuit was constructed and simulated using the QuCF numerical framework~\cite{QuCF}, which is based on the QuEST toolkit~\cite{Jones19}, and can be found in Ref.~\onlinecite{StepanoffSource}.


\begin{figure*}[!t]
\centering
\subfloat[]{\includegraphics[width=0.34\textwidth]{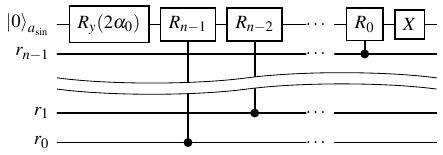}\label{circ:sin}}
\hspace{1em}
\subfloat[]{\includegraphics[width=0.30\textwidth]{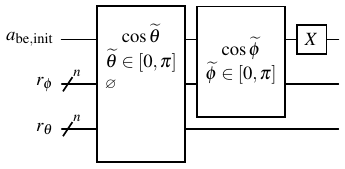}\label{circ:init-be}}
\hspace{1em}
\subfloat[]{\includegraphics[width=0.30\textwidth]{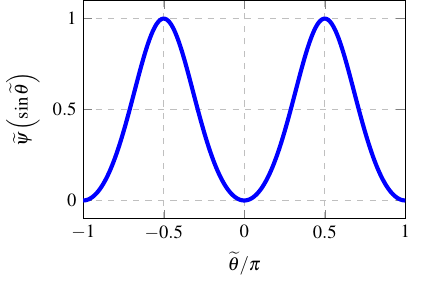}\label{fig:init-sin-full}}
\caption{
    \label{circ:sin-indermediate-function}
    (a): The circuit to encode $\sin(\widetilde{\ytheta}_j)$, where $j = 0,1,\dots 2^n-1$, into the amplitude of the state $\ket{0}_{a_{\rm sin}}\ket{j}_r$. 
    Here, $R_k = R_y(2\alpha_1/2^k)$. 
    The real parameters $\alpha_0$ and $\alpha_1$ are derived from Eq.~\eqref{eq:be-sin} to obtain the desired grid of the discretized angle $\widetilde{\ytheta}_j$.
    Without the Pauli $X$ gate at the end, the circuit encodes $\cos(\widetilde{\ytheta}_j)$ into the amplitude of the state $\ket{0}_{a_{\rm sin}}\ket{j}_{r}$.
    (b): The oracle encoding $\sin(\widetilde\theta_{j_{\theta}}+\widetilde\phi_{j_{\phi}})$ with $\widetilde\theta_{j_{\theta}}, \widetilde\phi_{j_{\phi}} \in [0, \pi]$ into the amplitude of the state $\ket{0}_{a_{\rm be, init}}\ket{j_{\phi}}_{r_{\phi}}\ket{j_{\theta}}_{r_{\theta}}$.
    The blocks $\cos\widetilde{\theta}$ and $\cos\widetilde{\phi}$ are implemented using circuit (a).
    (c): The function $\widetilde{\psi}(\sin\widetilde{\ytheta})$, Eq.~\eqref{eq:init-for Chebyschev},  with $\widetilde{\ytheta} = [-\pi,\pi]$.
}
\end{figure*}

\subsubsection{Encoding the initial condition}\label{sec:be-init}

To map the initial condition~\eqref{eq:init-mod} onto a quantum circuit, we begin by encoding Eq.~\eqref{eq:init-1D} using the QSVT algorithm.
Specifically, we compute the QSVT angles\cite{Dong21, Dong21code} to implement the function 
\begin{equation}\label{eq:init-for Chebyschev}
    \widetilde{\psi}(s) = \widetilde{g}(s)/\widetilde{g}(1),\quad \widetilde{g}(s) = \exp\left(\kappa s^2 \right) - 1,
\end{equation}
where $s \in [-1, 1]$ represents the singular values of the block-encoded matrix addressed as an oracle within the corresponding QSVT circuit.
The circuit for computing the sine function, Fig.~\ref{circ:sin}, serves as the block-encoding oracle.
If the register $r$ has $n$ qubits, then the unitary corresponding to this oracle embeds a diagonal matrix with $N$ elements $\sin(\widetilde{\ytheta}_j)$: 
\begin{equation}\label{eq:be-sin}
    \widetilde{\ytheta}_j = \alpha_0 + j \Delta\widetilde{\ytheta},\quad j = 0,1,\dots N-1,\quad \Delta\widetilde{\ytheta} = 2\alpha_1/N,
\end{equation}
where the real parameters $\alpha_0$ and $\alpha_1$ define the grid of the discretized angle $\widetilde{\theta}_j$.
As a result, the block-encoded diagonal matrix has eigenvalues equal to $\sin\widetilde{\ytheta}_j$.
This corresponds to the substitution $s = \sin\widetilde{\ytheta}$ in Eq.~\eqref{eq:init-for Chebyschev}.
The function $\widetilde{\psi}(\sin\widetilde{\ytheta})$ for $\widetilde{\ytheta} = [-\pi,\pi]$ is shown in Fig.~\ref{fig:init-sin-full}.
If we take
\begin{equation}\label{eq:be-sin-pars}
    \alpha_0 = 0,\quad \alpha_1 = \frac{\pi N}{2(N - 1)},
\end{equation}
then $\widetilde{\ytheta} = [0, \pi]$ and $\widetilde{\psi}(\sin\widetilde{\ytheta}_j) \equiv g(\ytheta_j)/g(\pi)$, with $j = 0,1,\dots N-1$, where $g(\ytheta_j)$ is the target function described in Eq.~\eqref{eq:init-1D}.
Thus, to prepare the function $g(\ytheta_j)/g(\pi)$, we use the QSVT circuit parameterized by the QSVT angles precomputed for constructing function~\eqref{eq:init-for Chebyschev}, and employ circuit~\ref{circ:sin} with the parameters~\eqref{eq:be-sin-pars} as the block-encoding oracle queried by the QSVT circuit.

To encode the initial condition~\eqref{eq:init-1D-mod}, we prepare a diagonal matrix with the elements $\sin(\widetilde{\ytheta} + \widetilde{\yphi})$ using circuit~\ref{circ:init-be}.
The circuit design relies on the fact that $\sin(\widetilde{\ytheta} + \widetilde{\yphi}) = \cos\widetilde{\ytheta}\sin\widetilde{\yphi} + \sin\widetilde{\ytheta}\cos\widetilde{\yphi}$.
There, the trigonometric functions are computed using circuit~\ref{circ:sin}.
In particular, the ancilla $a_{\rm be, init}$ in circuit~\ref{circ:init-be} serves as the ancilla $a_{\rm sin}$ in circuit~\ref{circ:sin}.
In the subcircuit $\cos\widetilde\theta$, the state register $r_{\phi}$ is not used, and the register $r_{\theta}$ corresponds to the register $r$ used as the control register in circuit~\ref{circ:sin}.
Similarly, the register $r_{\phi}$ in the subcircuit $\cos\widetilde\phi$ serves as the control register $r$.

\begin{figure}[!t]
\centering
\includegraphics[width=0.5\textwidth]{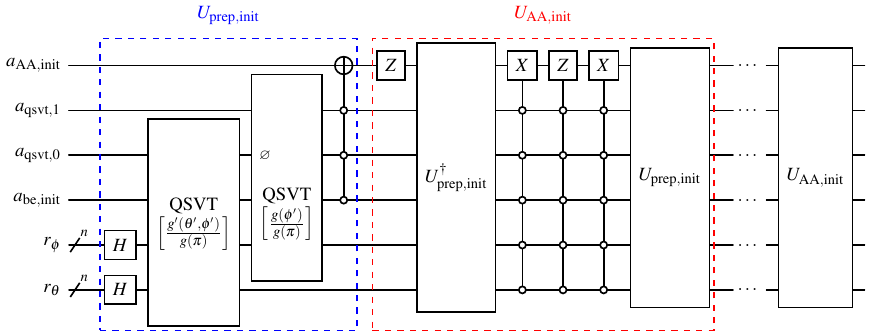}
\caption{
    \label{circ:init}
    Initialization circuit encoding the initial condition~\eqref{eq:init-mod}.
    The AA operator $U_{\rm AA, init}$ (red box) includes the preparation oracle $U_{\rm prep,init}$ (blue box) as a subcircuit and is repeated $N_{\rm AA, init}$ times, Eq.~\eqref{eq:N-AA}, within the initialization circuit.
    The four ancillary qubits used in this circuit correspond to the register $a_{\rm init}$ in Fig.~\ref{circ:scheme-mix}. 
}
\end{figure}

The final circuit for preparing the initial condition~\eqref{eq:init-mod} is shown in Fig.~\ref{circ:init}.
Circuit~\ref{circ:init-be} is used as the block-encoding oracle in the leftmost QSVT operator in the blue box.
This QSVT circuit computes $g'(\yTheta,\yPhi)/g(\pi)$ in Eq.~\eqref{eq:init-mod}.
The rightmost QSVT operator employs circuit~\ref{circ:sin} as the block-encoding oracle with $r_\phi\equiv r$ and $a_{\rm be, init}\equiv a_{\rm sin}$ and computes $g(\yPhi)/g(\pi)$ in Eq.~\eqref{eq:init-mod}.
Both QSVT circuits are parameterized by the same set of QSVT angles precalculated classically to construct function~\eqref{eq:init-for Chebyschev}.
The ancillae $a_{{\rm qsvt},0}$ and $a_{{\rm qsvt}, 1}$ serve as the target qubits for the sequence of gates $R_z$ parameterized by the QSVT angles for the left and right QSVT circuits, correspondingly.

The entire blue box in Fig.~\ref{circ:init} is the preparation oracle $U_{\rm prep, init}$, which computes the initial condition~\eqref{eq:init-mod} and encodes it into the unit-state amplitude of the ancilla $a_{\rm AA, init}$ using the closing controlled Pauli $X$ gate.
The Hadamard gates at the beginning of the preparation oracle allow the QSVT subcircuits to compute function~\eqref{eq:init-mod} for all angles $\yTheta \in [0, 2\pi]$ and $\yPhi\in [0, 2\pi]$ simultaneously.
Without these gates, the QSVT circuits would compute the function only at the point $(\yTheta, \yPhi) = (0,0)$.

The circuit $U_{\rm prep, init}$ is not fully-coherent and its success probability decreases with $N$.
To boost the success probability, the standard amplitude amplification (AA) procedure~\cite{Grover97, Grover98} is applied.
The AA operator $U_{\rm AA, init}$ is indicated by the red box in Fig.~\ref{circ:init}:
\begin{equation}\label{eq:AA-init}
    U_{\rm AA, init} = U_{\rm prep, init} U_{R,{\rm init}}  U_{\rm prep, init}^\dagger U_{R, {\rm prep}}.
\end{equation}
Here, $U_{R, {\rm prep}}$ is the reflector on the target state prepared by $U_{\rm prep, init}$.
This reflector changes the sign of the state, which encodes the initial condition~\eqref{eq:init-mod}, and keeps the perpendicular state unmodified.
This operator is implemented by a single Pauli $Z$ gate acting on the ancilla $a_{\rm AA, init}$.
The operator $U_{R,{\rm init}}$ changes the sign of the initial condition, which is the zero state of all qubits in the circuit.
This operator can be implemented by a combination of controlled Pauli $X$ and $Z$ gates as shown in Fig.~\ref{circ:init}.

After repeating the AA operator $N_{\rm AA, init}$ times, the probability of the initialization circuit is increased to a value close to $1$.
The number of these repetitions, $N_{\rm AA, init}$, is computed knowing the original probability $p_{\rm prep}$ of the state encoding the non-amplified initial condition returned by $U_{\rm prep, init}$ (i.e., the probability of the unit state of $a_{\rm AA, init}$):
\begin{equation}\label{eq:N-AA}
    N_{\rm AA, init} = \left\lceil  \frac{1}{2}\left(\frac{\pi}{2\arcsin{\sqrt{p_{\rm prep}}}} - 1\right)  \right\rceil.
\end{equation}
In the case of function~\eqref{eq:init-for Chebyschev} with $\kappa = 1$, $N_{\rm AA, init} = 2$ is enough for a wide range of $N$ to boost the success probability of the initialization circuit up to $0.999$.

The outcome of initialization circuit~\ref{circ:init} is the state
\begin{equation}
    \psi'_{\rm circ}(t = 0) = \eta_{\rm init} \psi'(0),
\end{equation}
where $\psi'(0)$ is the original initial condition~\eqref{eq:init-mod} which is used in Eq.~\eqref{eq:hs-real-fourier}.
The scalar coefficient $\eta_{\rm init}$ is 
\begin{equation}\label{eq:norm-init}
    \eta_{\rm init} = \left(\frac{\eta_{\rm QSVT}}{2^{n/2}}\right)^2 \eta_{\rm AA}.
\end{equation}
Here, the scalar $\eta_{\rm QSVT} < 1$ is used to guarantee the successful computation of the QSVT angles for the function~\eqref{eq:init-for Chebyschev} (in our case, $\eta_{\rm QSVT} = 0.99$).
The coefficient $2^{-n}$ appears due to the Hadamard gates in the preparation circuit $U_{\rm prep, init}$.
The scalar $\eta_{\rm AA} > 1$ depends on $N$ and appears due to the application of the AA procedure which balances the coefficient $2^{-n}$ in the initialized state.
The scalar $\eta_{\rm AA}$ can be computed as the ratio of the maximum amplitudes of the quantum states returned by the amplified initialization circuit and the preparation circuit $U_{\rm prep, init}$.

\subsubsection{Encoding Hamiltonian evolution}\label{sec:enc-Hamiltonian}
\begin{figure}[!t]
\centering
\includegraphics[width=0.49\textwidth]{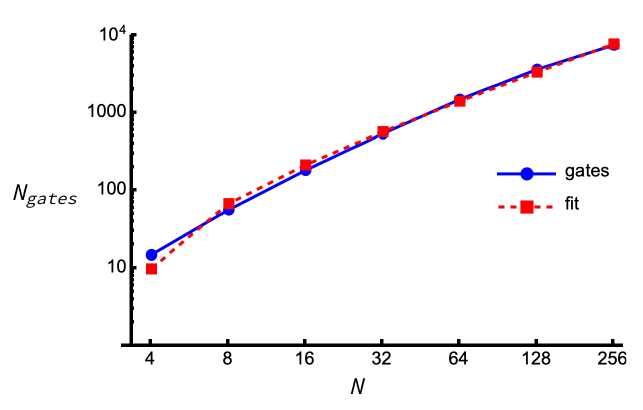}
\caption{
    \label{fig:scaling-HS}
    The number of gates in the standard Pauli decomposition of the matrix $\yMM{\yTheta}$ for a single Trotter step  vs. the number of grid points/Fourier modes $N$.
    The standard Pauli encoding is inefficient and eliminates the opportunity for exponential speedup because the complexity grows faster than $N$, as can be seen by comparing the actual gate count (solid blue circles) to a fit (dashed red squares) to the form $a N\log(N)+b N+c \log(N)+d$.
}
\end{figure}

The next step is to implement the unitary diagonal operators $e^{-\yi\yMM{\yTheta} \Delta t}$ and $e^{-\yi\yMM{\yPhi} \Delta t}$.
One possible approach is to decompose the matrices into a sequence of tensor products of Pauli matrices.
This allows for constructing $e^{-\yi\yMM{k} \Delta t}$ via a combination of CNOT and Pauli $Z$ gates.
However, in this case, as shown in Fig.~\ref{fig:scaling-HS}, the scaling of the resulting circuit is 
$\oO(N\log(N))$,  which completely eliminates any potential for exponential quantum advantage.

An alternative approach, entirely enabled by the I-LTS decomposition, is to use a sequence of QSP circuits.
The fact that the generators are diagonal allows the unitaries $e^{-\yi\yMM{\yTheta}\Delta t}$ and $e^{-\yi\yMM{\yPhi} \Delta t}$ to be encoded without fully constructing the block-encoding oracles for the matrices $\yMM{\yTheta}$ and $\yMM{\yPhi}$.
Let us start from considering the following transformation:~\cite{Georgeot01, Porter22}
\begin{align}\label{eq:linear-exp}
    \frac{1}{2^{n/2}}&\sum_{k=0}^{N-1} e^{-\yi\omega k}\ket{k} \nonumber\\
        &=\frac{1}{2^{n/2}}\sum_{k_0=0}^{1}\dots\sum_{k_{n-1}=0}^{1} e^{-\yi\omega \sum_{l=0}^{n-1} k_l 2^{l}}\ket{k_{n-1} k_{n-2} \dots k_0} \nonumber\\
        &=\frac{1}{2^{n/2}}\sum_{k_0=0}^{1}\dots\sum_{k_{n-1}=0}^{1} \bigotimes_{l = 0}^{n-1} e^{-\yi\omega k_l 2^{l}} \ket{k_l} \nonumber\\
        &=\frac{1}{2^{n/2}}\bigotimes_{l = 0}^{n-1}\sum_{k_l=0}^{1} e^{-\yi\omega k_l 2^{l}} \ket{k_l} \nonumber\\
        &=\frac{1}{2^{n/2}}\bigotimes_{l = 0}^{n-1}\left(\ket{0} + e^{-\yi\omega 2^{l}} \ket{1}\right),
\end{align}
where $\omega$ is a constant scalar and we assume that the integer $k$ is decomposed into an array of bits $k_l$, with the index $l = 0, 1, \dots (n-1)$, as
\begin{equation}
    k = k_{n-1} 2^{n-1} + k_{n_2} 2^{n-2} + \dots + k_{0} 2^0 = \sum_{l=0}^{n-1} k_l 2^{l}.
\end{equation}
Equation~\eqref{eq:linear-exp} prepares an exponential function depending on a linearly growing index $k$.
Starting from the zero state, the expression described in Eq.~\eqref{eq:linear-exp} can be prepared by a combination of Hadamard and phase gates.
Indeed, 
\begin{equation}
    H^{\otimes n}\ket{00\dots0} = \frac{1}{2^{n/2}}\bigotimes_{l = 0}^{n-1}\left(\ket{0} + \ket{1}\right),
\end{equation}
where the target register has $n$ qubits and one Hadamard gate $H$ is acting on each qubit in the register. 
Then, we need a phase gate with the angle $-\omega 2^l$ acting on the $l$-th qubit:
\begin{align}
    \frac{1}{2^{n/2}}&\bigotimes_{l = 0}^{n-1} P\left(-\omega 2^l\right)(\ket{0} + \ket{1}) \nonumber\\
        &= \frac{1}{2^{n/2}}\bigotimes_{l = 0}^{n-1} \left(\ket{0} + e^{-\yi\omega 2^l} \ket{1}\right).
\end{align}

We can extend this technique to the case where the exponential function takes a matrix $A$ as an argument:
\begin{equation}\label{eq:exp-w-A}
    \exp\left(-\yi k\omega A\right).
\end{equation}
The above operator can be considered as a Hamiltonian simulation with the generator $A$ in the time interval $k\omega$.
To encode this function, one can use $n$ QSP circuits acting on an additional register block-encoding the matrix $A$.
These QSP circuits should be also controlled by qubits in the register encoding the integer $k$.
Let us denote this register by $r_k$.
Each QSP controlled by the $l$-th qubit in $r_k$ performs Hamiltonian simulation with the generator $A$ in the time interval $\omega 2^{l}$, where $l = 0, 1,\dots, (n-1)$.
A somewhat similar approach was used in Ref.~\onlinecite{Hu24Sch} (see Fig. 12 there).

\begin{figure}[!t]
\centering
\subfloat[]{\includegraphics[width=0.20\textwidth]{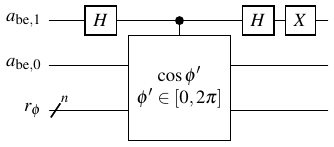}\label{circ:be-qhs-phi}}
\hspace{1em}
\subfloat[]{\includegraphics[width=0.20\textwidth]{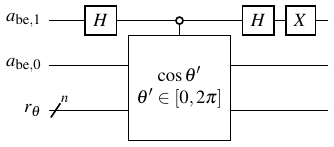}\label{circ:be-qhs-theta}}
\caption{
    \label{circ:be-qhs}
    (a): Circuit of the block-encoding oracle computing $(1 - \cos\yPhi)/2$ for the generator $\yMM{\yTheta}$, Eq.~\eqref{eq:M1-real-fourier}.
    (b): Circuit of the oracle computing $-(1 - \cos\yTheta)/2$ for the generator $\yMM{\yPhi}$, Eq.~\eqref{eq:M2-real-fourier}.
    The trigonometric functions $\cos\yPhi$ and $\cos\yTheta$ are computed using circuit~\eqref{circ:sin}.
}
\end{figure}

In the case of the matrix $\yMM{\yTheta}$, Eq.~\eqref{eq:M1-real-fourier}, the matrix $A$ is diagonal with the elements \begin{equation}\label{eq:M1-A-diag}
    \left(1 - \cos\yPhi_{x_2}\right)/2,
\end{equation}
where the integer $x_2$ is encoded by bitstrings in the register $r_{\yphi}$. 
This matrix $A$ can be encoded by using a combination of circuit~\ref{circ:sin} and a circuit of Linear Combination of Unitaries (LCU).
The resulting circuit is shown in Fig.~\ref{circ:be-qhs}.
The usage of the LCU circuit explains the division by $2$ in Eq.~\eqref{eq:M1-A-diag}.
Also, $k \equiv c_1$ in Eq.~\eqref{eq:exp-w-A}, i.e. $r_k \equiv r_{\ytheta}$, and $\omega = - 2 (1-\alpha) \Delta t$.
The multiplicative factor of $2$ in $\omega$ is necessary to account for the LCU block-encoding of the matrix $A$.

\begin{figure}[!t]
\centering
\includegraphics[width=0.49\textwidth]{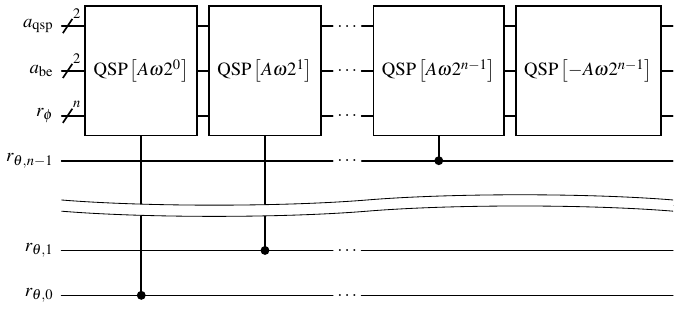}
\caption{
    \label{circ:enc-diag-M1}
    Circuit computing the unitary $e^{-\yi\yMM{\yTheta} \Delta t}$.
    Each QSP operator uses circuit~\ref{circ:be-qhs-phi} as an oracle to encode the dependence of the generator $\yMM{\yTheta}$ on the angle $\yPhi$.
    To prepare the unitary $e^{-\yi\yMM{\yPhi} \Delta t}$, the registers $r_{\yphi}$ and $r_{\ytheta}$ are swapped and circuit~\ref{circ:be-qhs-theta} is employed as a block-encoding oracle acting on the target state register $r_{\ytheta}$, while the QSP circuits are controlled by the register $r_{\yphi}$.
    In both cases, the combination of the ancillae $a_{\rm qsp}$ and $a_{\rm be}$ corresponds to the register $a_H$ used in circuit~\ref{circ:scheme-mix}.
}
\end{figure}

The resulting circuit for computing the unitary $e^{-\yi\yMM{\yTheta} \Delta t}$ is shown in Fig.~\ref{circ:enc-diag-M1}. 
There, the series of $n$ conditioned QSP circuits encodes the unitary $e^{-\yi\yMM{\yTheta}\Delta t}$ with the assumption that $c_2 = 0, 1, \dots N-1$.
However, the rightmost QSP circuit shifts $c_2$ by $-N/2$ to be in accordance with Eq.~\eqref{eq:r}.

For the matrix $\yMM{\yPhi}$, Eq.~\eqref{eq:M2-real-fourier}, the procedure is identical.
In this case, the diagonal matrix $A$ has the elements equal to $-\left(1 - \cos\yTheta_{x_1}\right)/2$, the index $k$ corresponds to the index $c_2$, i.e. $r_k\equiv r_{\yphi}$, and $\omega = 2 \alpha \Delta t$.

\begin{figure*}[!t]
\centering
\subfloat{\includegraphics[width=0.33\textwidth]{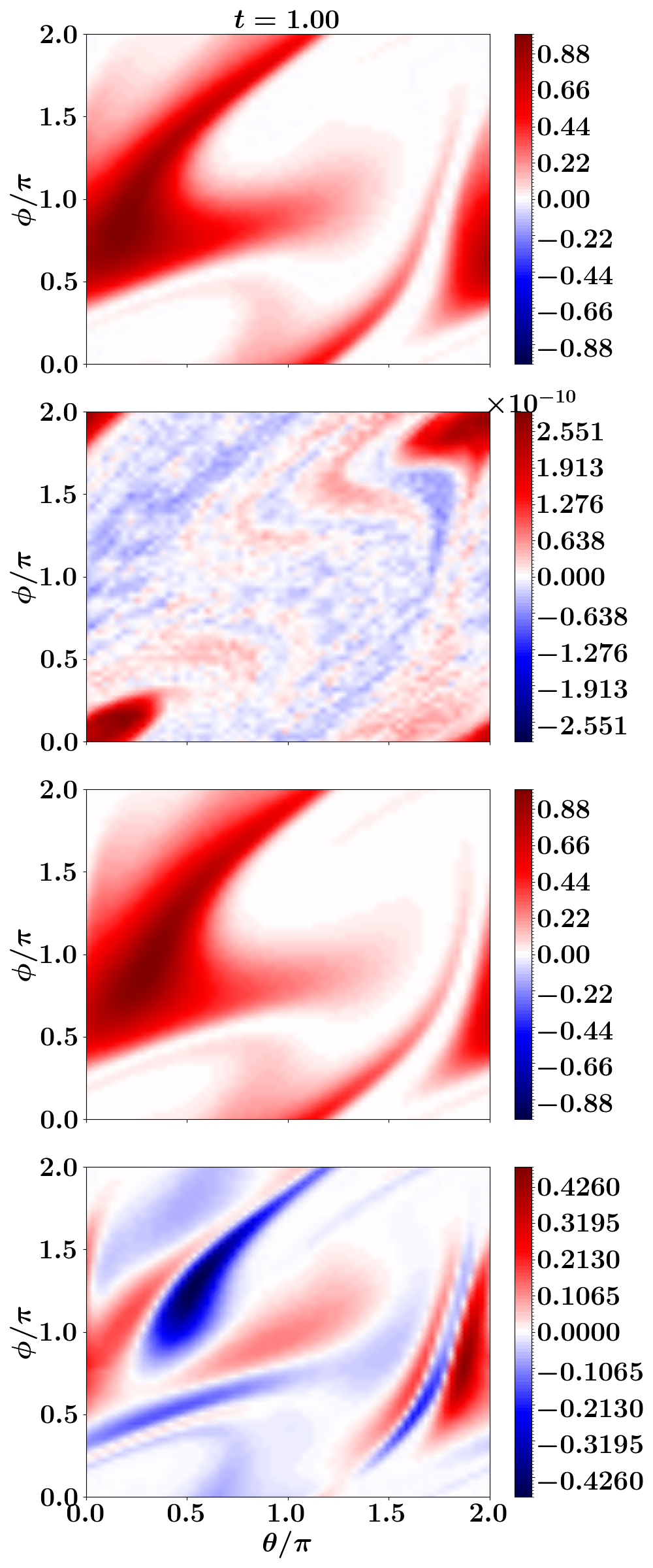}\label{fig:comp-t1}}
\subfloat{\includegraphics[width=0.33\textwidth]{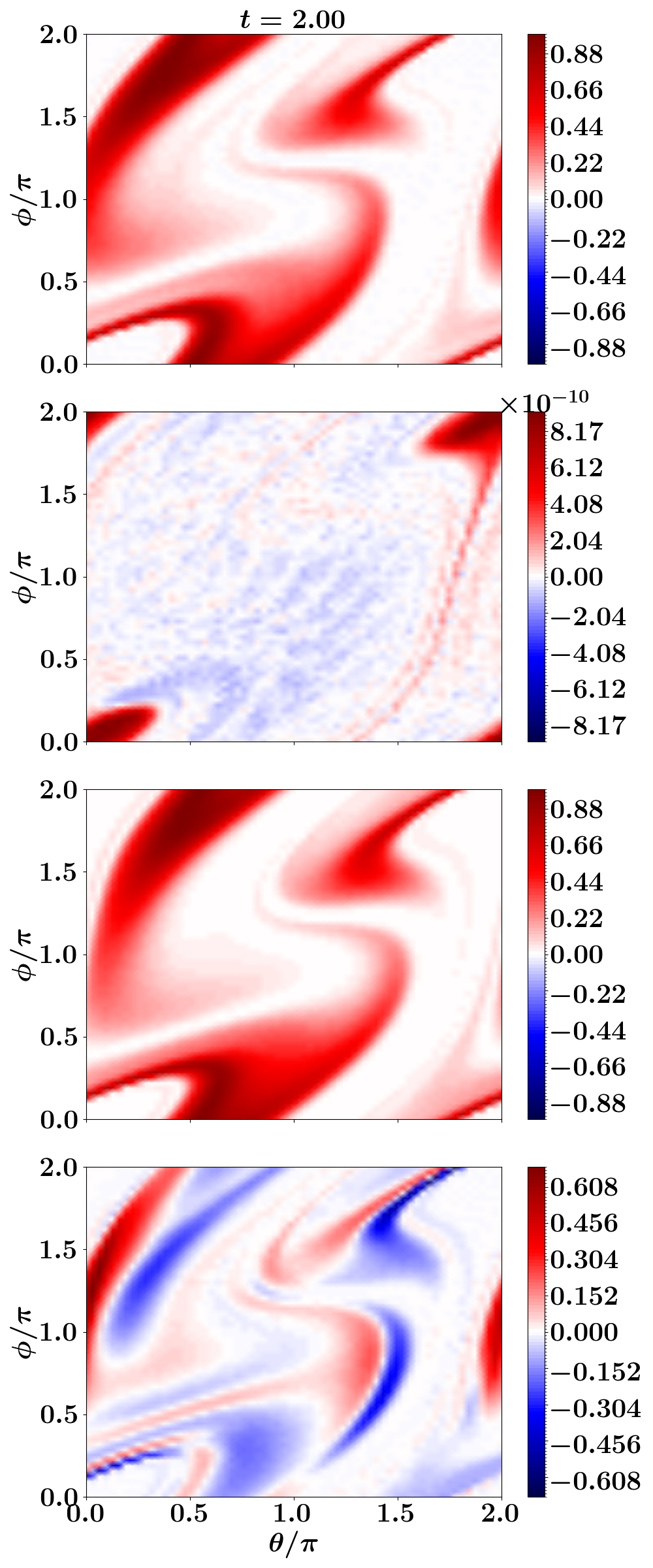}\label{fig:comp-t2}}
\subfloat{\includegraphics[width=0.33\textwidth]{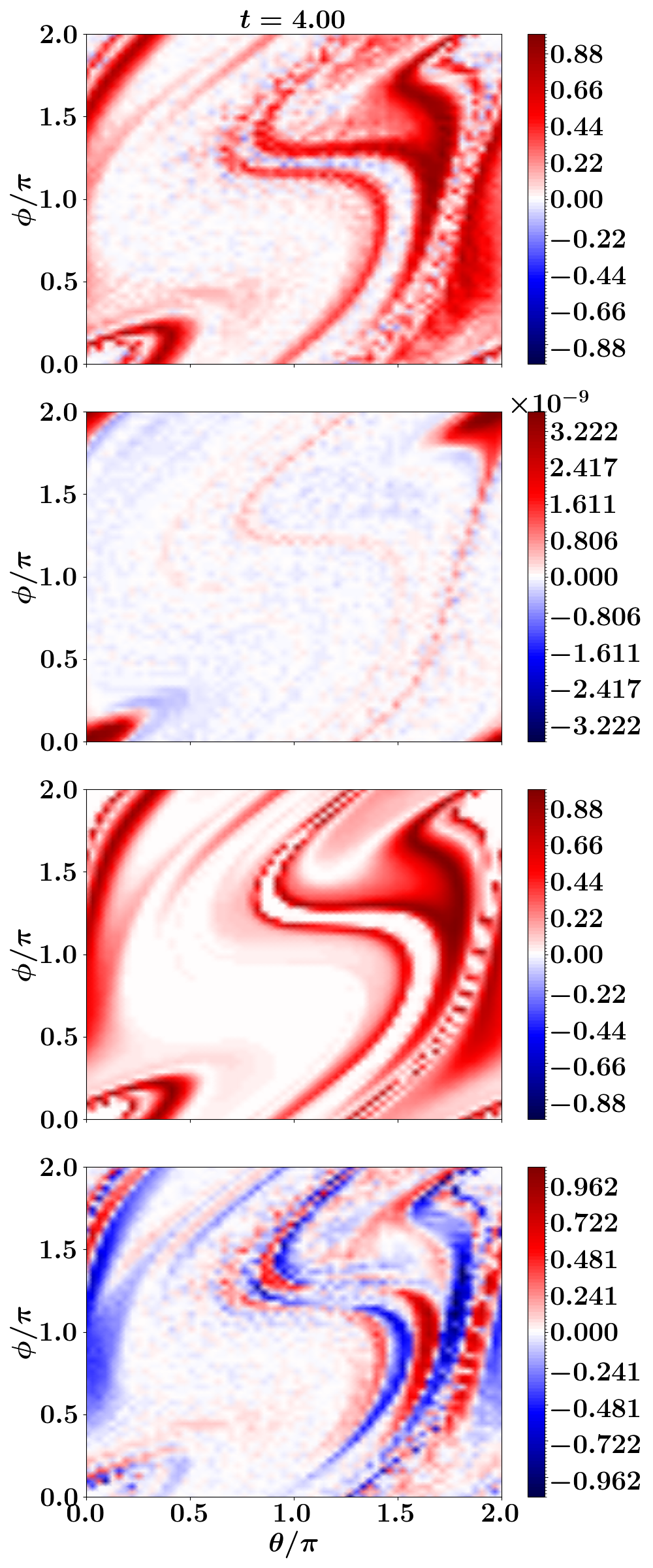}\label{fig:comp-t4}}
\caption{
    \label{fig:comp}
    Top row: the distribution $\psi(\ytheta, \yphi, t)$ for various time intervals $t$ in the QuCF simulations of the Stepanoff flow in real-Fourier space, Eq.~\eqref{eq:hs-real-fourier}, using circuit~\ref{circ:scheme-mix} with $\Delta t = 0.1$ and using the initial condition described in Eq.~\eqref{eq:init-mod}. The distribution is normalized to one.
    Second row: the difference between the QuCF and classical simulations in real-Fourier space.
    Third row: the distribution obtained in the classical simulations in real space, Eq.~\eqref{eq:sys-lagrangian}.
    Bottom row: the difference between the QuCF simulations in real-Fourier space and the classical simulations in real space.
    In all these simulations, the parameters $n = 6$, $\kappa = 1$, and $\alpha = \sqrt{20}$ are used.
    The source files of these simulations can be found in Ref.~\onlinecite{StepanoffSource}. 
}
\end{figure*}

\subsubsection{Quantum Hamiltonian simulation (QHS) results}
To evaluate the performance of the QHS algorithm,  circuit~\ref{circ:scheme-mix} was emulated using the QuCF~\cite{QuCF, StepanoffSource} framework.
Figure ~\ref{fig:comp} shows a comparison between the quantum simulation using cicuit~\ref{circ:scheme-mix} and the classical linear mixed real-Fourier space algorithm corresponding to Sec.~\ref{sec:Stepanoff_mixed-space}.
Again, the KvN wavefunction is initialized with the initial conditions~\eqref{eq:init}, shown in Fig.~\ref{fig:initial-conditions}, and then evolved until time $t=1,2,4$ with time step $\Delta t=0.1$.
In this case, $n=6$ to reduce the cost of emulating the QHS algorithm, so the number of Fourier modes per angle is $N=2^6=64$.

The top row shows the result of using the QHS circuit.
The middle row shows the difference between the results of the QHS circuit and the classical linear algorithm in Sec.~\ref{sec:Stepanoff_mixed-space}.
One can see that the quantum simulations reproduce the classical results with high precision, so that the error is on the order of $10^{-10}-10^{-9}$.
However, the linear simulation in mixed real-Fourier space is less accurate than the nonlinear classical transfer operator approach shown in the third row.
The bottom rows shows the difference between the linear QHS method and the nonlinear transfer operator method.
For $t=1,2$ and for most of the phase space for $t=4$, this can essentially be considered to be the error in the linear QHS method.
Using higher $N$ would reduce this error to manageable values, e.g. as shown for $N=9$ in Fig.~\ref{fig:comp-cl-dt001}.

\begin{figure*}[!t]
\centering
\includegraphics[width=01.0\textwidth]{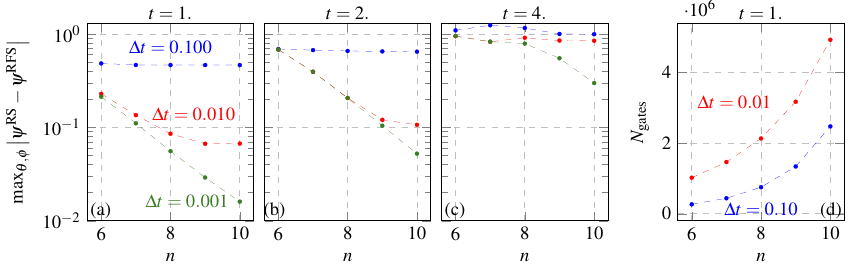}
\caption{
    \label{fig:scan-scaling}
    The maximum absolute difference between $\psi(\ytheta, \yphi, t)$ computed by the classical simulations in real space (RS) and in real-Fourier space (RFS) for various $\Delta t$ and for $t=1$ (a), $t=2$ (b), and $t=4$ (c).
    (d): The dependence of the number of gates in circuit~\ref{circ:scheme-mix} on $n$ for various $\Delta t = 0.100$ (blue markers) and $\Delta t = 0.010$ (red markers) for $t = 1$. 
}
\end{figure*}

Figure~\ref{fig:scan-scaling} shows how the precision of the QHS  algorithm changes with $\Delta t$, $t$, and $n$. 
It is clear that these computations require a sufficiently large $N$ to reduce the numerical error for a given time interval.
In fact, due to the exponential stretching induced by the Lyapunov exponent, which controls the butterfly effect, for a classically mixing or chaotic flow, one expects that the number of Fourier modes must increase exponentially in time to maintain a given accuracy.
This is clearly illustrated by the the nonlinear transfer approach in the third row, where numerical ``digitization'' artifacts  become visible for $t=4$.
In this case, the trajectories are still accurate, but the coarse-graining induced by interpolation of the values to grid of $N^2=64\times 64$ points generates the numerical artifacts.
Since, in practice, exponentially growing resolution is not achievable beyond a few Lyapunov times, one must consider how to properly consider the effects of coarse-graining on the solution.

\subsubsection{Complexity scaling}
The complexity scaling of the circuit~\ref{circ:scheme-mix} essentially rests on the QSP complexity which is 
\begin{equation}\label{eq:qsp-complexity}
    C_{\rm QSP} = \oO\left(\|A\| t + \log_2(\varepsilon_{\rm QSP}^{-1})\right).
\end{equation}
In our case, the spectral norm $\|A\| = 1$ for both $\yMM{\yTheta}$ and $\yMM{\yPhi}$, but the factor $k\omega$ in Eq.~\eqref{eq:exp-w-A} scales as $\oO(\alpha 2^n)$, which results in poor scaling with $N$.
The resulting Hamiltonian simulation with $N_t$ trotterization time intervals scales as 
\begin{equation}\label{eq:QHS-scaling}
    C_{\rm QHS} = \oO\left[n^2 N_t \left(N \alpha \Delta t + n \log_2(\varepsilon_{\rm QSP}^{-1})\right) + C_{\rm AA, init}\right].
\end{equation}
Here, the scaling has the multiplicative factor $n^2$ for two reasons.
The first reason is that the block-encoding oracles~\ref{circ:be-qhs} scale as $\oO(n)$ and they are addressed $C_{\rm QSP}$ times by each QSP circuit in Fig.~\ref{circ:enc-diag-M1}.
The second reason is that each $\Delta t$ time interval requires $\oO(n)$ QSP circuits. 

The additive factor $C_{\rm AA, init}$ is the cost of amplitude amplifying the initialization step,
\begin{equation}
    C_{\rm AA, init} = \oO\left[N_{\rm AA, init} f\left(N, \kappa, \varepsilon_{\rm init}^{-1},\dots\right) \right].
\end{equation}
The function $f\left(N, \kappa, \varepsilon_{\rm init}^{-1}, \dots\right)$, where $\varepsilon_{\rm init}$ is the error in the initialization, strongly depends on the form of the initial condition.
In the case of function~\eqref{eq:init-for Chebyschev} with $\kappa = 1$, the dependence $f\left(N, \kappa, \varepsilon_{\rm init}^{-1}, \dots\right)$ is
\begin{equation}\label{eq:init-complexity}
    f\left(N, \kappa, \varepsilon_{\rm init}^{-1}, \dots\right) = \oO\left(\log_2(N) \log_2(\varepsilon_{\rm init}^{-1})\right).
\end{equation}
This means that the numerical error of the QSVT computation of function~\eqref{eq:init-for Chebyschev} drops exponentially with the number of QSVT angles.
In our case, the number of QSVT angles required for computing the initial condition in a single spatial dimension with the error $\varepsilon_{\rm init}\approx 10^{-11}$ is $19$.

The QSP circuits are $1 - \oO(\epsilon_{\rm QSP})$ coherent but since we have $N_t n$ consequent QSP circuits, there is error accumulation.
As a result, the total error is $\oO(n N_t \epsilon_{\rm QSP})$ without taking into account the trotterization error which decreases quadratically with $\Delta t$.
One can avoid the multiplicative factor $n N_t$ by using a compression gadget~\cite{Fang23}, but this will require $\oO(\log_2{(N_t)})$ ancillary qubits.

In Fig.~\ref{fig:scan-scaling}d, one can see the scaling of circuit~\ref{circ:scheme-mix} with $n$ observed in numerical simulations.
The number of gates grows exponentially with $n$ which is consistent with the theoretical scaling~\eqref{eq:QHS-scaling}.

The number of the QSP angles required to perform Hamiltonian evolution with the generator $\yMM{\yTheta}$ within the time interval $\Delta t = 0.10$ is around $20$.
It should be noted that for $\Delta t = 0.01$, the number of the QSP angles does not decrease by a factor of $10$ but is approximately equal to $15$.
This happens because of the influence of the additive factor $\log_2(\varepsilon_{\rm QSP}^{-1})$ in the QSP complexity~\eqref{eq:qsp-complexity}.

The scaling with $N$ in Eq.~\eqref{eq:QHS-scaling} is a consequence of the Courant-Friedrichs-Lewy condition.
A high number of modes in Fourier space is needed to achieve sufficiently high spatial resolution, which in turn requires a small time step.
In the case of circuit~\ref{circ:enc-diag-M1}, this leads to a large factor $k\omega$.

\subsection{Amplitude Estimation of Physical Observables}\label{sec:ae}

\begin{figure*}[!t]
\centering
\includegraphics[width=0.80\textwidth]{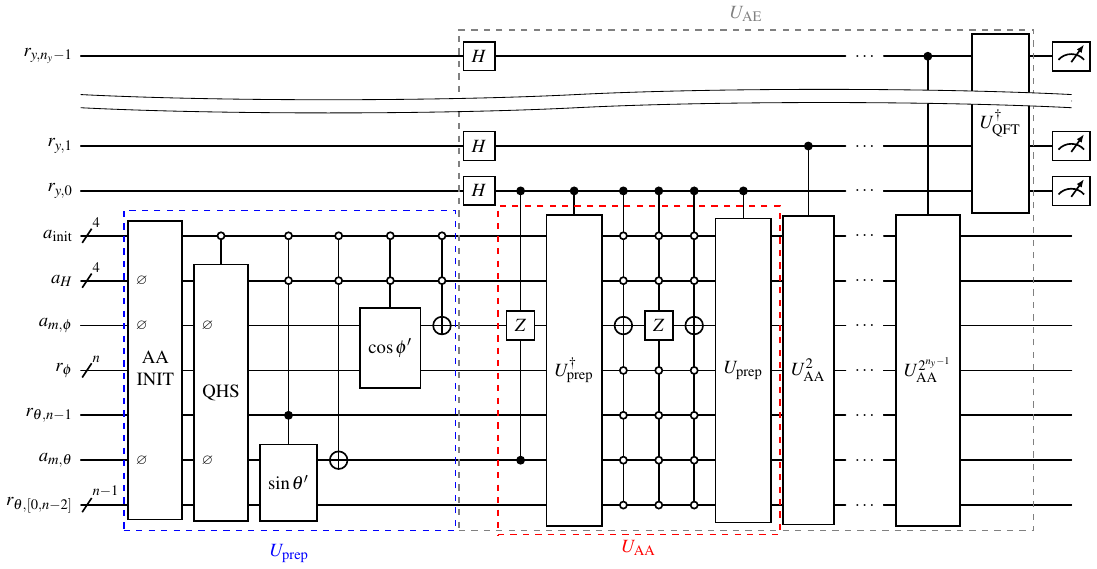}
\caption{
    \label{circ:AE}
    Schematic circuit performing AE for measuring $Q_2$, Eq.~\eqref{eq:AE-Q2}.
    The functions $\sin\yTheta$ and $\cos\yPhi$ are computed using circuit~\ref{circ:sin}.
    All qubits are initialized in the zero state.
    If $Q_1$ is computed, the $\sin\theta'$ block and the subsequent $X$ gate are excluded.
}
\end{figure*}

\newcommand{\UnitStep}[1]{ {\rm \Theta}\,{#1} }
We use the standard amplitude estimation (AE) procedure\cite{Brassard02} for measuring state probabilities computed by circuit~\ref{circ:scheme-mix}.
As test cases, we measure the observables:
\begin{align}\label{eq:Q-contin}
    Q_1 &= 
        \left\langle (\cos\phi)^2 \right\rangle 
\end{align}
and
\begin{multline}
    Q_2 = 
           \left\langle (\cos\phi)^2    \UnitStep{(\pi-\theta)}        \right\rangle 
          \\
        + \left\langle (\sin\theta \cos\phi)^2 \UnitStep{(\theta-\pi)} \right\rangle 
\end{multline}
where the angles are taken to have the range $[0,2\pi)$ and $\UnitStep(x)$ is the unit step function.
In discretized form, they become
\begin{equation}\label{eq:AE-Q1}
    Q_1 = \frac{1}{N^2} \sum_{j_{\yphi} = 0}^{N-1} \sum_{j_\ytheta = 0}^{N-1} |\psi_{j_\ytheta, j_{\yphi}}\cos\yPhi_{j_{\yphi}}|^2,
\end{equation}
and
\begin{align}\label{eq:AE-Q2}
    Q_2 
    &= \frac{1}{N^2} \sum_{j_{\yphi} = 0}^{N-1} \sum_{j_\ytheta = 0}^{N/2-1}   |\psi_{j_\ytheta, j_{\yphi}}\cos\yPhi_{j_{\yphi}}|^2 \nonumber\\
    &+ \frac{1}{N^2} \sum_{j_{\yphi} = 0}^{N-1} \sum_{j_\ytheta = N/2}^{N-1} |\psi_{j_\ytheta, j_{\yphi}}\sin\yTheta_{j_{\ytheta}}\cos\yPhi_{j_{\yphi}}|^2.
\end{align}

The quantum circuit for performing AE is shown in Fig.~\ref{circ:AE}.
The subcircuit shown in the blue box marks the preparation operator $U_{\rm prep}$, which consists of the initialization subcircuit described in Sec.~\ref{sec:be-init}, the circuit performing QHS described in Sec.~\ref{sec:enc-Hamiltonian}, and the subcircuits computing the trigonometric functions used in $Q_2$.
To compute $Q_1$ instead of $Q_2$, one needs to remove the ancillary qubit $a_{m, \ytheta}$, the subcircuit $\sin\yTheta$, and the subsequent controlled $X$ gate.
After the action of the preparation operator, the quantum state becomes
\begin{align}
   \ket{\psi_{\rm after\ prep}} &\sim \ket{0}_{r_y}\ket{0}_{a_{\rm init}, a_{\rm H}}\left[
        \ket{1}_{a_{m,\phi}}\ket{1}_{a_{m,\theta}}\ket{\psi_{\rm data}} + \dots
    \right] \nonumber\\
    &+ \ket{0}_{r_y}\ket{\neq 0}_{a_{\rm init}, a_{\rm H}}\ket{\dots},
\end{align}
where the state $\psi_{\rm data}$ is 
\begin{align}\label{eq:psi-data}
    \ket{\psi_{\rm data}} 
        &= \cos\phi_{j_\phi}'\ket{0}_{r_{\theta,n-1}}\ket{j_\phi}_{r_\phi}\ket{j_\theta}_{r_{\theta,[0,n-2]}}\nonumber\\
        &+ \cos\phi_{j_\phi}'\sin\theta_{j_\theta}'\ket{1}_{r_{\theta,n-1}}\ket{j_\phi}_{r_\phi}\ket{j_\theta}_{r_{\theta,[0,n-2]}}.
\end{align}
Circuit~\ref{circ:AE} allows estimation of the success probability
\begin{equation}\label{eq:prob}
    p_{\rm data} = \itimes{\psi_{\rm data}}{\psi_{\rm data}}.
\end{equation}
By comparing Eqs.~\eqref{eq:psi-data},~\eqref{eq:prob} with Eq.~\eqref{eq:AE-Q2} and taking into account the normalization coefficient~\eqref{eq:norm-init}, one obtains
\begin{equation}\label{eq:Q-from-p}
    Q = \frac{p_{\rm data}}{(N\eta_{\rm init})^2} ,
\end{equation}
where $Q$ is either $Q_1$ or $Q_2$.

Similarly to Eq.~\eqref{eq:AA-init}, the operator indicated by the red box is the AA operator $U_{\rm AA}$,
\begin{equation}
    U_{\rm AA} = U_{\rm prep} U_{R,{\rm init}}  U_{\rm prep}^\dagger U_{R,{\rm data}},
\end{equation}
where $U_{R,{\rm data}}$ is the reflector on the state of interest, i.e., the state entangled with $\ket{1}_{a_{m,\phi}}\ket{1}_{a_{m,\theta}}$, and $U_{R,{\rm init}}$ is the reflector on the initial state, which is the zero state.
The reflector $U_{R,{\rm data}}$ is implemented by a single Pauli gate $Z$ acting on the ancilla $a_{m,\phi}$ controlled by the unit state of the qubit $a_{m,\theta}$.
The reflector $U_{R,{\rm init}}$ is a combination of Pauli $Z$ and $X$ gates.

The AE operator, $U_{\rm AE}$, is indicated with the gray box.
It includes $N_y - 1$ repetitions of the AA oracle $U_{\rm AA}$, where $N_y = 2^{n_y}$.
The AE operator is a phase estimation circuit that estimates the eigenphase of the operator $U_{\rm AA}$, where the eigenphase corresponds to the eigenstate prepared by the operator $U_{\rm prep}$.
The probability $p_{\rm data}$ is estimated by measuring the bitstring of the register $r_y$:
\begin{equation}\label{eq:prob-est}
    \tilde{p}_{\rm data} = 1 - \sin^2\left(\frac{\pi i_y}{N_y}\right).
\end{equation}
Here, $\tilde{p}_{\rm data}$ is the estimation of $p_{\rm data}$, and $i_y$ is the integer representation of the bitstring measured in the register $r_y$.
With a probability more than $0.81$, measurement estimates $p_{\rm data}$ with the absolute error\cite{Brassard02}
\begin{equation}\label{eq:AE-error}
    \delta_{\rm data} \equiv |p_{\rm data} - \tilde{p}_{\rm data}| \leq 2\pi \frac{\sqrt{\tilde{p}_{\rm data}(1 - \tilde{p}_{\rm data})}}{N_y} + \frac{\pi^2}{N_y^2}.
\end{equation}
The estimated probability $\tilde{p}_{\rm data}$ is used instead of the exact value $p_{\rm data}$ in Eq.~\eqref{eq:Q-from-p} to obtain an estimated value of $Q$:
\begin{equation}\label{eq:Q-AE}
    \widetilde{Q} = \frac{\widetilde{p}_{\rm data}}{(N\eta_{\rm init})^2}.
\end{equation}
If one takes into account the normalization~\eqref{eq:Q-from-p}, then the error in the estimation $\widetilde{Q}$ is 
\begin{equation}\label{eq:err-norm}
    \delta_{{\rm data}, Q} = \frac{\delta_{\rm data}}{(N\eta_{\rm init})^2} .
\end{equation}

When we account for AE, the scaling~\eqref{eq:QHS-scaling} is modified to 
\begin{equation}\label{eq:AE-scaling}
    C_{\rm AE} = \oO\left(\delta_{\rm data}^{-1} C_{\rm QHS} \right),
\end{equation}
because the QHS circuit is repeated $\oO(N_y) = \oO(\delta_{\rm data}^{-1})$ times during the AE procedure.

\begin{figure}[!t]
\centering
\includegraphics[width=0.49\textwidth]{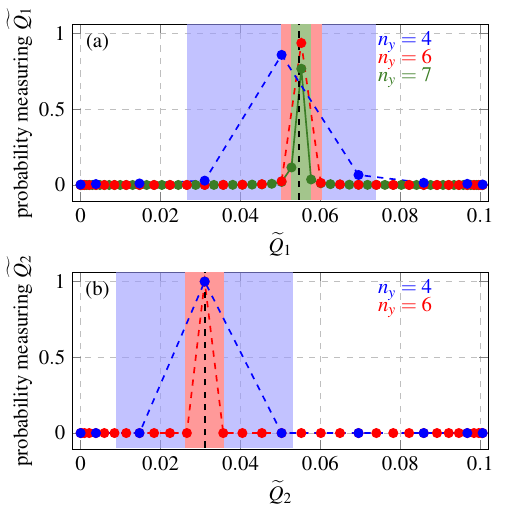}
\caption{
    \label{fig:AE-results}
    (a) Probability distribution of measuring different values of $\widetilde{Q}_1$ in circuit~\ref{circ:AE}.
    (b) Probability distribution of measuring different values of $\widetilde{Q}_2$.
    The black vertical lines indicate classically computed $Q_1$ and $Q_2$, Eqs.~\eqref{eq:AE-Q1} and~\eqref{eq:AE-Q2}, respectively.
    The filled colored regions indicate the intervals $[\widetilde{Q} - \delta_{{\rm data}, Q}, \widetilde{Q} + \delta_{{\rm data}, Q}]$ for various $n_y$.
}
\end{figure}

Measurement outcomes for the values $Q_1$ and $Q_2$ estimated by the AE circuit~\ref{circ:AE} are shown in Fig.~\ref{fig:AE-results}.
The emulator QuCF allows computation of the probabilities of all possible integer outcomes $i_y = 0, 1, \dots N_y - 1$.
By using Eq.~\eqref{eq:prob-est}, one computes the probability amplitudes $\widetilde{p}_{\rm data}$ for each possible $i_y$. 
The corresponding estimations of $Q$ are computed using Eq.~\eqref{eq:Q-AE}.
The integer $i_y$ with the highest probability corresponds to the most probable estimation of $Q$, with the absolute analytical error~\eqref{eq:err-norm}.

\section{Conclusion}\label{sec:conclusion}
In this work, we introduced the \emph{integrable Lie-Trotter-Suzuki} (I-LTS) decomposition for non-integrable dynamical systems and proved that this technique can be applied to arbitrary vector fields.
We also showed that the I-LTS decomposition has an explicit tensor product structure that is relatively easy for tensor network and quantum algorithms to exploit.

We then derived an explicit quantum algorithm for simulating the Stepanoff flow on the torus that is based on a natural I-LTS decomposition for the flow. 
A detailed procedure was presented for encoding the two-dimensional problem into a quantum circuit using modified angular coordinates that restore the symmetry in the Hamiltonian. 
The I-LTS decomposition then naturally encodes the system as a mixed combination of real space and Fourier space representations. 
This approach yielded two disjoint diagonal generators and significantly simplified the encoding of the QHS evolution operator.
The measurement of two prototypical examples of physical observables was also considered and explicit circuits for amplitude estimation of these observables were also provided.

The problem was initialized in real space using a combination of two QSVT circuits to compute the initial condition along the two spatial dimensions. 
Hamiltonian simulation was carried out over a sequence of short time intervals. 
Within each interval, the QHS operation was performed in the Fourier space of one coordinate, while keeping the other coordinate in real space.
The QHS was constructed as a sequence of controlled QSP circuits to encode the dependence of the generators on the Fourier indices. 
The QSP circuits queried simple trigonometric functions to encode the dependence on the coordinates in real space.

The proposed circuit was simulated using a numerical emulator of fault-tolerant quantum computers, and the results were found to be consistent with several alternative classical calculations of the Koopman operator. 
We analyzed the complexity scaling of the proposed QHS circuit and demonstrated that it has linear dependence on the number of points in Fourier space. 
This scaling arises from the CFL condition, which necessitates an increase in the number of time steps as the spatial resolution is refined.

A possible direction for subsequent investigation of this ergodic system is to simulate the Stepanoff flow, Eq.~\eqref{eq:sys-lagrangian}, using the Carleman linearization~\cite{Carleman32, Liu21} technique and whether the system requires the standard local version of the embedding or the globalized Carleman embedding first proposed in \onlinecite{Novikau25global}.
While the local embedding can be implemented using either LCHS or quantum linear system solver algorithms, the global version may require more sophisticated, and most likely hybrid, classical-quantum techniques.
An interesting research question is whether using the I-LTS decomposition relaxes the need for additional Carleman linearization charts required by the global Carleman embedding methods.

The simplicity of the I-LTS evolution operator enables the use of intrinsically linear methods, such as tensor networks and quantum algorithms, to handle nonlinear dynamics with a series of simple integrable steps.
There are a vast uncountable set of possible choices of I-LTS decompositions connected to the number of different atlases of coordinate charts that one can provide to represent a given manifold, and, hence, to represent a given flow on the manifold.
Since the notion of which I-LTS decomposition is most ``efficient'' in any given context can depend on which set of coordinates is most useful for the problem at hand, this reflects a rather large set of possible I-LTS decompositions that can be optimal in any given context.
The enormous flexibility in the choice of the I-LTS decomposition implies that different decompositions can be optimized for different purposes.

This realization may provide an explanation for why it has been found that multiple AI/ML methods can be useful for learning nonlinear dynamics from data~\cite{Lusch18, Brunton20, Rajendra2020modeling, Froyland2021spectral}
and are able to reproduce complex time-dependent behavior without a detailed understanding of the physical models or differential equations that determine the individual trajectories.
In fact, the I-LTS method is equivalent to a composition of rectified flow-matching steps~\cite{Liu2022flow,Hertrich2026relation} for transporting probability distributions  that constructs the map from initial to final states with a series of simple evolutions in curvilinear coordinates that change at each stage.
Perhaps AI/ML methodologies based on deep learning, kernel-based, and other methods are simply searching for optimal I-LTS decompositions, where the precise notion of optimality is defined by the loss (cost) function that is being minimized.
Ultimately, we hope that a clearer understanding of the mathematical structure behind the I-LTS decomposition will allow us to more easily understand and improve the way that AI/ML methods, tensor networks, and quantum algorithms can be applied to nonlinear dynamics.

\section*{Acknowledgments}
Work at Lawrence Livermore National Laboratory (LLNL) was performed under the auspices of the the U.S. Department of Energy (DOE) under Contract DE-AC52–07NA27344 and was supported by the DOE Office of Fusion Energy Sciences “Quantum Leap for Fusion Energy Sciences” project DE-SCL000070 (FWP SCW1680) and “Quantum Information Science and Koopman Operator Theory" DE-SC0025101 (FWP SCW1877).
DG and JS acknowledge support from the U.S. Department of Energy Office of Fusion Energy Sciences under grant DE-SC0025101. 
DG and MM acknowledge support from the U.S. Department of Defense, Basic Research Office under grant N00014-21-1-2946.
This research used resources of the National Energy Research Scientific Computing Center (NERSC), a DOE Office of Science User Facility supported by the Office of Science of the U.S. Department of Energy under Contract No. DE-AC02-05CH11231 using NERSC award FES-ERCAP0028618.

\appendix 
\section{Integrable Decompositions for Lorenz 1963 \label{sec:Lorenz}}
In this appendix, we consider several integrable decompositions for the Lorenz 1963 system.~\cite{Lorenz63}
The Lorenz 1963 system is an interesting example of a relatively simple low-dimensional (3D) dissipative system that hosts a strange attractor for certain values of the parameters and was the first example of a strange attractor that was discovered.
The Lorenz 1963 equations of motion are given by the velocity field $\BV=(V^x,V^y,V^z)^T$ where
\begin{align}
\dot x &= V^x =\sigma( y-x),\\
\dot y &= V^y = x(\rho-z)-y,\\
\dot z &= V^z = xy -\beta z.
\end{align}
Since $\nabla\cdot \BV=-1-\sigma-\beta$, this system is dissipative for $\beta+\sigma>-1.$
When $\beta>0$, the fixed point at $\{x,y,z\}=0$ is stable until it undergoes a pitchfork bifurcation once $\rho>1$.
In this case, the fixed point at the origin becomes unstable and there are two new fixed points at $x_\pm=y_\pm=\pm\sqrt{\beta (\rho-1)}$ and $z_\pm=\rho-1$.

There are multiple integrable LTS decompositions for this flow.
The canonical decomposition in Lemma~\ref{thm:local_I-LTS} is simple because the equations of motion have the form $V^i\propto (x^i_0-x)$, where for each $i$, the $x_0^i(x^j,x^k)$ are nonlinear functions of the other two coordinates, i.e. $j=\mod{(i+1,3)}$ and $k=\mod{(i+2,3)}$.
Hence, one can define angle-like coordinates
\begin{align}
\xi^i=\log{(x^i-x^i_0(x^j,x^k))},
\end{align}
where, for this form of the solution,  
\begin{align}
 x_0^1 &=y,\\
 x_0^2 &=x(\rho-z),\\
 x_0^3 &=x y/\beta.
\end{align}
This requires dividing the space into 8 different regions depending on the signs of the 3 components of $\{x-x^1_0,y-x^2_0,z-x^3_0\}$.

For this system, there is also a simpler decomposition into 2 flows, $\BV=\BV_0+\BV_1$.
For example, one can separate into the  flow that is linear at the origin and a remaining nonlinear flow
\begin{align}
\BV_0  &=\{\sigma(y-x),\ x\rho-y,\ -\beta z\},\\
\BV_1 &= \{0,\ -x z,\ x y\}.
\end{align}
This requires one region for $\BV_0$ and 8 regions for $\BV_1$, depending on the signs of the components of  $\BV_1$.

Alternatively, one can decompose into a flow at constant $z$ and a flow at constant $x$
\begin{align} \label{eq:L64_V0}
\BV_0 &=\{ \sigma(y-x),\ x(\rho-\gamma z)-y,\ 0\},\\
\BV_1 &= \{0,\ (\gamma-1)xz-y,\ x y-\beta z\}.
\end{align}
This simplifies even further for the choice $\gamma=1$.
At the origin, the linearization of $\BV_0$ has eigenvalues $\{-1-\sqrt{\rho-\gamma z},-1+\sqrt{\rho-\gamma z} ,0\}$ and the linearization of $\BV_1$ has  eigenvalues $\{0,-1,-\beta\}$.
Note that both flows have fixed points at the origin:  $\BV_1$ is stable (unstable) for $\beta>0$ ($<0$), while $\BV_0$ is oscillatory for $\rho-\gamma z<0$, stable for  $ 0\leq\rho-\gamma z\leq 1$,  and unstable for $1<\rho-\gamma z $.  

Yet another interesting decomposition arises from considering the two other fixed points for $x_\pm=y_\pm=\pm\sqrt{\beta (\rho-1)}$ and $z_\pm=\rho-1$.
In fact, one can choose a 3-component decomposition $\BV=\BV_++\BV_0+\BV_-$
\begin{align}
\BV_+ &=\tfrac{1}{2}\{\sigma(y-x),\ (x-x_+)(z_+ - z)+x-y,\ 0\},\\
\BV_- &=\tfrac{1}{2}\{\sigma(y-x),\ (x-x_-)(z_- - z)+x-y,\ 0\},\\
\BV_0&=\{0,\ 0,\ xy- \beta z\},
\end{align}
where the linearization of $\BV_\pm$ near the respective fixed point, $\Bx_\pm$, has a simple form 
for both $\BV_\pm$, the linearized motion near each fixed point is stable in $x-y$ and neutrally stable in $x+y$.
Again, we need 8 regions to cover the possible signs of the nonlinear terms.

\bibliography{main}

\end{document}